%% file: main.tex
\documentclass[acmsmall, nonacm]{acmart}

\acmJournal{PACMPL}
\acmVolume{1}
\acmNumber{CONF} 
\acmArticle{1}
\acmYear{2022}
\acmMonth{10}
\acmDOI{} 
\startPage{1}

\setcopyright{none}

\usepackage{booktabs}   
\usepackage{caption} 
\usepackage{wrapfig} 
\usepackage{subcaption} 
\usepackage{setspace}
\usepackage{newunicodechar}
\usepackage{listings}
\usepackage{placeins}
\usepackage{xspace}
\usepackage{plstx, listproc}
\usepackage{svg}
\input{definitions}
\usepackage{pdfpages}
\usepackage{multirow}
\usepackage{multicol}
\usepackage{array} 
\usepackage{savesym} 
\savesymbol{newlist}
\savesymbol{renewlist}
\usepackage{enumitem}
\restoresymbol{enumi}{newlist}
\restoresymbol{enumi}{renewlist}

\usepackage{tikz}
\newcommand{\cbox}[1]{\tikz[baseline=1pt]{\path[draw=#1,fill=#1] (0,0) rectangle (.25cm,.25cm);}}
\definecolor{legend_green}{HTML}{33A02C}
\definecolor{legend_purple}{HTML}{6A3D9A}
\definecolor{legend_red}{HTML}{DE2D26}
\definecolor{legend_yellow}{HTML}{FFD700}
\definecolor{legend_green_light}{HTML}{c8e4c4}
\definecolor{legend_purple_light}{HTML}{c4c4e8}

\usepackage{algorithmicx}
\usepackage{algpseudocode,algorithm,algorithmicx}

\newcommand*\Let[2]{\State #1 $\gets$ #2}
\algrenewcommand\algorithmicrequire{\textbf{Precondition:}}
\algrenewcommand\algorithmicensure{\textbf{Postcondition:}}

\newlength\myindent
\newcommand{\sepconj}{$*$}

\begin{document}

\title[\gco]{\gco: Symbolic Execution for Gradual Verification}         


\author{Jenna DiVincenzo}
\affiliation{
  \institution{Purdue University}            
  \country{USA}                    
}
\email{jennad@purdue.edu}          

\author{Ian McCormack}
\affiliation{
  \institution{Carnegie Mellon University}           
  \country{USA}                   
}
 \email{icmccorm@cs.cmu.edu}         

\author{Hemant Gouni}
\affiliation{
  \institution{Carnegie Mellon University} 
  \country{USA}
}
 \email{hsgouni@cs.cmu.edu}
 
\author{Jacob Gorenburg}
\affiliation{
  \institution{Haverford College} 
  \country{USA}
}
 \email{jgorenburg@haverford.edu}

\author{Jan-Paul Ramos-Dávila}
\affiliation{
  \institution{Cornell University} 
  \country{USA}
}
 \email{jvr34@cornell.edu}
 
\author{Mona Zhang}
\affiliation{
  \institution{Columbia University} 
  \country{USA}
}
 \email{mz2781@columbia.edu}
 
 \author{Conrad Zimmerman}
\affiliation{
  \institution{Brown University} 
  \country{USA}
}
 \email{conrad_zimmerman@brown.edu}

\author{Joshua Sunshine}
\affiliation{
  \institution{Carnegie Mellon University} 
  \country{USA}
}
 \email{sunshine@cs.cmu.edu}

\author{\'{E}ric Tanter}
\affiliation{
  \institution{University of Chile} 
  \country{Chile}
}
 \email{etanter@dcc.uchile.cl}
 
 \author{Jonathan Aldrich}
\affiliation{
  \institution{Carnegie Mellon University} 
  \country{USA}
}
 \email{jonathan.aldrich@cs.cmu.edu}

\renewcommand{\shortauthors}{J. DiVincenzo, I. McCormack, H. Gouni, J. Gorenburg, J. Ramos-Dávila, M. Zhang, C. Zimmerman, J. Sunshine, \'{E}. Tanter,  J. Aldrich}

\renewcommand{\footnotemark}{\mbox{}} 
\titlenote{
  This material is based upon work supported by a Google PhD Fellowship award and the
  \grantsponsor{GS100000001}{National Science
    Foundation}{http://dx.doi.org/10.13039/100000001} under Grant
  Nos.~\grantnum{GS100000001}{CCF-1901033}, \grantnum{GS100000001}{DGE1745016}, and \grantnum{GS100000001}{DGE2140739}.
  \'{E}. Tanter is partially funded by the ANID FONDECYT Regular Project 1190058 and the Millennium Science Initiative Program: code ICN17\_002.
  Any opinions, findings, and conclusions or recommendations expressed in this material are those of the authors and do not necessarily reflect the views of the National Science Foundation, Google, ANID, or the Millennium Science Initiative.
}

\begin{abstract}
Current static verification techniques such as separation logic support a wide range of programs. However, such techniques only support complete and detailed specifications, which places an undue burden on users. To solve this problem, prior work proposed gradual verification, which handles complete, partial, or missing specifications by soundly combining static and dynamic checking. Gradual verification has also been extended to programs that manipulate recursive, mutable data structures on the heap. Unfortunately, this extension does not reward users with decreased dynamic checking as specifications are refined. In fact, all properties are checked dynamically regardless of any static guarantees. Additionally, no full-fledged implementation of gradual verification exists so far, which prevents studying its performance and applicability in practice.

We present \gco, the first practicable gradual verifier for recursive heap data structures, which targets C0, a safe subset of C designed for education. Static verifiers supporting separation logic or implicit dynamic frames use symbolic execution for reasoning; so \gco, which extends one such verifier, adopts symbolic execution at its core instead of the weakest liberal precondition approach used in prior work. Our approach addresses technical challenges related to symbolic execution with imprecise specifications, heap ownership, and branching in both program statements and specification formulas. We also deal with challenges related to minimizing insertion of dynamic checks and extensibility to other programming languages beyond C0. Finally, we provide the first empirical performance evaluation of a gradual verifier, and found that on average, \gco decreases run-time overhead between 11-34\% compared to the fully-dynamic approach used in prior work. Further, the worst-case scenarios for performance are predictable and avoidable.
This work paves the way towards evaluating gradual verification at scale.
\end{abstract}



\maketitle

\section{Introduction}
\label{sec:intro}
Separation logic \cite{reynolds2002separation} supports the modular static verification of heap-manipulating programs. Its variant \emph{Implicit dynamic frames} (IDF) \cite{smans2009implicit} and extension with \emph{recursive abstract predicates} \cite{parkinson2005separation,smans2009implicit} further support verifying recursive heap data structures, such as trees, lists, and graphs. While these techniques allow users to specify and verify more code than ever before, tools implementing them (\eg Viper \cite{MuellerSchwerhoffSummers16}, VeriFast \cite{jacobs2011verifast}, Chalice \cite{leino2009verification}, JStar \cite{distefano2008jstar}, and SmallFoot \cite{berdine2006smallfoot}) are still largely unused due to the burden they place on their users. Such tools poorly support partial specifications, and thus require users to provide a number of auxiliary specifications (such as folds, unfolds, loop invariants, and inductive lemmas) in an all or nothing fashion to support inductive proofs of correctness. The tools also require many of these auxiliary specifications to be written before they can provide feedback on the correctness of specifications for important functional properties. For example, to prove that a simple insertion function preserves list acyclicity, static verifers need 1.5 times as many lines of auxiliary specifications to program code (\S\ref{sec:probs-static-verif}). They also need a significant number of these auxiliary specifications to uncover problems with specifications of the acyclic property (\S\ref{sec:probs-static-verif}).

Inspired by gradual typing~\cite{siek2006gradual}, \citet{bader2018gradual} proposed \emph{gradual verification} to support the incremental specification and verification of software. Users can write {\em imprecise} (i.e. partial) specifications backed by run-time checking where necessary. An imprecise formula can be fully unknown, 
written $\qm$, or combine a {\em static part} with the unknown, as in $\qm \:\ast\: x.f == 2$. 
\citet{wise2020gradual} extend \citet{bader2018gradual}'s initial system by designing and formalizing the first gradual verifier for recursive heap data structures. It supports imprecise specifications with accessibility predicates from IDF and abstract predicates; and thus, also (in theory) the run-time verification of these constructs.
During static verification, an imprecise specification can be optimistically strengthened (in non-contradictory ways) by the verifier to support proof goals. Wherever such strengthenings occur, dynamic checks are inserted to preserve soundness. Gradual verification smoothly supports the spectrum between static and dynamic verification. This is captured by properties adapted from gradual typing~\cite{siek2015refined}, namely the \emph{gradual guarantee}, stating that the verifier will not flag static or dynamic errors for specifications that are correct but imprecise, and the fact that gradual verification {\em conservatively extends} static verification, \ie~they coincide on fully-precise programs.

While promising, \citet{wise2020gradual}'s gradual verifier has neither been implemented nor validated in practice. Furthermore, their design relies on \emph{weakest liberal preconditions}~\cite{dijkstra75} for static reasoning rather than {\em symbolic execution}~\cite{king:cacm1976}, which is the ideal reasoning technique for tools based on separation logic or IDF. Indeed, Viper \cite{MuellerSchwerhoffSummers16}, VeriFast \cite{jacobs2011verifast}, JStar \cite{distefano2008jstar}, and SmallFoot \cite{berdine2006smallfoot} all support these permission logics with symbolic execution, not weakest liberal preconditions. Finally, \citet{wise2020gradual}'s gradual verifier is not efficient in the sense that it checks all memory safety and functional properties dynamically regardless of the precision of specifications.

This paper presents the design, implementation, and validation of \gco\footnote{\gco is hosted on Github: \href{https://github.com/gradual-verification/gvc0}{https://github.com/gradual-verification/gvc0}.}---the first gradual verifier for imperative programs manipulating recursive heap data structures that is based on symbolic execution. \gco targets C0, a safe subset of C designed for education, with appropriate support (and pedagogical material) for dynamic verification.
Technically, \gco is built on top of the Viper static verification infrastructure~\cite{MuellerSchwerhoffSummers16}, which facilitates the development of program verifiers supporting IDF and recursive abstract predicates. \gco's back-end leverages this infrastructure to simplify the implementation of gradual verifiers for other programming languages, and \gco's front-end demonstrates how this is done for C0.
Further, \gco minimizes the insertion of dynamic checks using statically available information and optimizes the checks' overhead at run time. 

Overall, we address new technical challenges in gradual verification related to symbolic execution, extensibility to multiple programming languages, and minimizing run-time checks and their overhead:

\begin{itemize}[leftmargin=14pt, topsep=2pt]
    \item \gco's symbolic execution algorithm is responsible for statically verifying programs with imprecise specifications and producing minimized run-time checks. In particular, \gco tracks the branch conditions created by program statements and specifications to produce run-time checks for corresponding execution paths. At run time, branch conditions are assigned to variables at the branch point that introduced them, which are then used to coordinate the successive checks as required. Further, \gco creates run-time checks by translating symbolic expressions into specifications---reversing the symbolic execution process.
    \item The run-time checks produced by \gco contain branch conditions, simple logical expressions, accessibility predicates, separating conjunctions, and predicates. Each of these constructs is specially translated into source code that can be executed at run time for dynamic verification.
    Logical expressions are turned into assertions. Accessibility predicates and separating conjunctions are checked by tracking and updating a set of owned heap locations. Finally, predicates are translated into recursive boolean functions. By encoding run-time checks into C0 source code, we avoid complexities from augmenting the C0 compiler to support dynamic verification. We also design these encodings to be performance friendly, \eg owned heap locations are tracked in a dynamic hash table.
\end{itemize}

Work on gradual typing performance has shown that minimizing the insertion of dynamic checks does not trivially correlate with overall execution performance; the nature of the inserted checks (such as higher-order function wrappers) as well as their location in the overall execution flow of a program can have drastic and hard-to-predict consequences~\cite{takikawa16,10.1145/3276503,camporaAl:icfp2018}. 
%
Therefore, our validation of \gco aims to empirically evaluate the relationship between minimizing check insertion and observed run-time performance in gradual verification. We evaluate the performance of \gco by adapting \citet{takikawa16}'s performance lattice method to gradual verification, exploring the performance characteristics for partial specifications of four common data structures. This method models the gradual verification process as a series of steps of partial specifications from an unspecified program (containing all $\qm$s) to a statically verifiable specification (not containing any $\qm$s) where, at each step, an atomic conjunct is added to the current, partial specification. Statically, we observe that as more specifications are added, more verification conditions can be statically discharged. Though imprecision introduces unavoidable run-time checks, gradual verification decreases run-time overhead by an average of 11-34\% compared to dynamic verification (and thus \citet{wise2020gradual}'s approach). Sources of run-time overhead correspond to the predictions made in prior work, and our study shows that the gradual guarantee holds empirically for our tool across thousands of sampled imprecise specifications. 


\section{\gco Improves the Static Specification Process}
\label{sec:probs-static-verif}

\begin{wrapfigure}{L}{2.70in}
\vspace{-0.1in}     
\begin{minipage}{\linewidth}
\begin{lstlisting}[xleftmargin=2.5em, name=ll-insert]
struct Node { int val; struct Node *next; };
typedef struct Node Node;

Node* insertLast(Node* list, int val)
{
    Node* y = list;
    while (y->next != NULL)
      { y = y->next; }
    y->next = alloc(struct Node);
    y->next->val = val;
    y->next->next = NULL;
    return list;
}
\end{lstlisting}
\end{minipage}
\footnotesize
\disableTttResize
\caption{Non-empty linked list insertion in C0}
\label{ex:ll-insert}
\end{wrapfigure}

Static verification tools, like Viper \cite{MuellerSchwerhoffSummers16}, VeriFast \cite{jacobs2011verifast}, Chalice \cite{leino2009verification}, JStar \cite{distefano2008jstar}, and SmallFoot \cite{berdine2006smallfoot}, require a number of user-provided auxiliary specifications, such as folds, unfolds, lemmas, and loop invariants, to prove properties about recursive heap data structures. Worse, they also require users to write many of these auxiliary specifications before the tools can provide useful feedback on the correctness of other specifications, including ones containing important functional properties. Therefore, users are burdened by writing many detailed and extraneous specifications with inadequate static feedback through the process. In this section, we illustrate this burden with a simple list insertion example (inspired by a similar introductory example and discussion from \citet{wise2020gradual}) and output from Viper. Then, we show in \S\ref{sec:gco-saves-the-day} how \gco overcomes this burden by smoothly supporting the spectrum between static and dynamic checking. Users can avoid writing auxiliary specifications and still get sound verification of their code with increased run-time checking. Users can also receive run-time feedback on the correctness of their specifications very early in the specification process, and the resulting error messages closely align with inherent problems in the specifications or in the program, making debugging them easier.

\fig~\ref{ex:ll-insert} implements a linked list and function that inserts a new node at the end of a given list, called \ttt{insertLast}, in C0 \cite{arnold2010c0}. The \ttt{insertLast} function traverses the list to its end with a \ttt{while} loop starting from the root. That is, \ttt{insertLast} implicitly assumes the list is non-empty (non-null) and acyclic; and that for multiple successive calls to \ttt{insertLast} the list remains acyclic and non-empty after insertion. These facts can be proven explicitly with static verification; the complete static specification is given in \fig~\ref{fig:ll-insert-sv}, highlighted in grey.

List acyclicity is specified with two predicates \ttt{acyclicSeg} and \ttt{acyclic}:
\begin{flalign*}
&\ttt{predicate acyclicSeg(Node* s, Node* e) =} &\\
&\qquad\phiCond{\ttt{s == e}}{\phiTrue}{\phiAcc{\ttt{s->val}} \phiAnd \phiAcc{\ttt{s->next}} \phiAnd \ttt{acyclicSeg(s->next, e)}} \\
&\ttt{predicate acyclic(Node* n) = } \ttt{acyclicSeg(Node* n, NULL)}
\end{flalign*}
The \ttt{acyclicSeg} predicate uses  \emph{accessibility predicates} and the \emph{separating conjunction} from IDF. Ownership is ensured through \emph{accessibility predicates} such as $\phiAcc{\ttt{s->val}}$; and, $\phiAcc{\ttt{s->val}}$ $\phiAnd$ $\ttt{s->val} == 2$ states that \ttt{s->val} is uniquely owned and contains the value 2. The {\em separating conjunction}, denoted by  $\&\&$,\footnote{Most IDF papers use \sepconj{} to denote the separating conjuction. We follow Viper and use $\&\&$ instead.} ensures memory disjointness: $\phiAcc{\ttt{s->next}} \phiAnd \phiAcc{\ttt{s->next->next}}$ states that the heap locations \ttt{s->next} and \ttt{s->next->next} are distinct (\ie $s \neq \ttt{s->next}$) and are each owned.

\begin{figure}[t]
\centering
\begin{lstlisting}[xleftmargin=2.5em, name=insertLast-sv]
(*@\label{sc-acyclicSeg-pred1}\lightgray{/*@~predicate acyclicSeg(Node* s, Node* e) = }@*) 
    (*@\lightgray{(s == e) ? true : acc(s->val) \phiAnd acc(s->next) \phiAnd acyclicSeg(s->next,e); @*/}@*)
(*@\label{sc-acyclic-pred}\lightgray{//@~predicate acyclic(Node* n) = acyclicSeg(n, NULL);}@*)
\end{lstlisting}
\begin{minipage}{0.45\linewidth}
\begin{lstlisting}[xleftmargin=2.5em, name=insertLast-sv]
Node* insertLast(Node* list, int val)
  //@ requires (*@\label{sv-pre}\lightgray{acyclic(list) \&\& list != NULL}@*);
  /*@ ensures (*@\label{sv-post1}\lightgray{acyclic(\textbackslash result) \&\&}@*)
                    (*@\label{sv-post2}\lightgray{\textbackslash result != NULL}@*); @*/
{
   (*@\label{sv-unfold-acyclic}\lightgray{//@~unfold acyclic(list);}@*)
   (*@\label{sv-unfold-acyclicSeg}\lightgray{//@~unfold acyclicSeg(list, NULL);}@*)
   (*@\label{sv-varassign-stmt}@*)Node* y = list;
   (*@\label{sv-fold-listyseg}\lightgray{//@~fold acyclicSeg(list, y);}@*)
   (*@\label{sv-loophead}@*)while (y->next != NULL)
   (*@\label{sv-loopinv-stmt1}@*)/*@ loop_invariant (*@\lightgray{acyclicSeg(list, y) \phiAnd }@*)
        (*@\label{sv-loopinv-stmt2}\lightgray{acc(y->next) \phiAnd acc(y->val) \phiAnd}@*)
        (*@\label{sv-loopinv-stmt3}\lightgray{acyclicSeg(y->next, NULL)}@*); @*/
   {
     (*@\label{sv-tmp-stmt}@*)Node* tmp = y;
     (*@\label{sv-loop-asmt}@*)y = y->next;
     (*@\label{sv-unfold-nextnullseg}\lightgray{//@~unfold acyclicSeg(y, NULL);}@*)
     (*@\label{sv-fold-next}\lightgray{//@~fold acyclicSeg(tmp->next, y);}@*)
     (*@\label{sv-fold-prevnextseg}\lightgray{//@~fold acyclicSeg(tmp, y);}@*)
     (*@\label{sv-mergelemma-loop}\lightgray{mergeLemma(list, tmp, y);}@*)
   }
(*@~@*)
(*@~@*)
\end{lstlisting}
\end{minipage}\hfill
\begin{minipage}{0.45\linewidth}
\begin{lstlisting} [xleftmargin=1em, name=insertLast-sv]
    (*@\label{sv-newnode-stmt-start}@*)y->next = alloc(struct Node);
    y->next->val = val;
    (*@\label{sv-newnode-stmt-end}@*)y->next->next = NULL;
    (*@\label{sv-fold-end}\lightgray{//@~fold acyclicSeg(y->next->next, NULL);}@*)
    (*@\label{sv-fold-newnode}\lightgray{//@~fold acyclicSeg(y->next, NULL);}@*)
    (*@\label{sv-fold-prevend}\lightgray{//@~fold acyclicSeg(y, NULL);}@*)
    (*@\label{sv-mergelemma-afterloop}\lightgray{mergeLemma(list, y, NULL);}@*)
    (*@\label{sv-fold-acyclic}\lightgray{//@~fold acyclic(list);}@*)
    (*@\label{sv-return}@*)return list;
}

(*@\label{sv-mergelemma-start}\lightgray{void mergeLemma (Node* a, Node* b, Node* c)}@*)
  (*@\lightgray{//@~requires~acyclicSeg(a, b) \&\& acyclicSeg(b, c);}@*)
  (*@\lightgray{//@~ensures~acyclicSeg(a, c);}@*)
(*@ \lightgray{\{}@*)
    (*@ \lightgray{if (a == b) \{}@*)
    (*@ \lightgray{\} else \{}@*)
        (*@ \lightgray{//@~unfold acyclicSeg(a, b);}@*)
        (*@ \lightgray{mergeLemma(a->next, b, c);}@*)
        (*@ \lightgray{//@~fold acyclicSeg(a, c);}@*)
    (*@ \lightgray{\}}@*)
(*@\label{sv-mergelemma-end}\lightgray{\}}@*)
\end{lstlisting}
\end{minipage}

\footnotesize
\disableTttResize
\fbox{\begin{tabular}{llll}
$\square$ & \small{Program code} &
\textcolor{light-gray}{$\blacksquare$} & \small{Static specification}
\end{tabular}}

\caption{The static verification of \ttt{insertLast} from \fig~\ref{ex:ll-insert}}
\label{fig:ll-insert-sv}
\end{figure}

Thus, the abstract recursive predicate \ttt{acyclicSeg}, which can be thought of as a pure boolean function, specifies that a list segment is acyclic. That is, \ttt{acyclicSeg(s, e)} denotes that all heap locations in list \ttt{s} are distinct up to node \ttt{e} by recursively generating accessibility predicates for each node in \ttt{s} up to \ttt{e}, joined with the separating conjunction. Further, \ttt{acyclicSeg(n, NULL)} denotes that all heap locations in list \ttt{n} are distinct and so \ttt{n} is acyclic, as specified with \ttt{acyclic(n)}.

Now that we have specified \ttt{acyclic}, we use it in \ttt{insertLast}'s precondition (line \ref{sv-pre}) and postcondition (lines \ref{sv-post1}-\ref{sv-post2}) to denote preservation of list acyclicity. We also specify that \ttt{insertLast} preserves list non-nullness with simple comparison logic (\ie \ttt{list != NULL} and \ttt{\textbackslash result != NULL}). Ideally, we would stop here and static verifiers would be able to prove \ttt{insertLast}'s implementation is correct with respect to this specification; however, as you can see in \fig~\ref{fig:ll-insert-sv} such tools require many more specifications. In fact, there are 27 lines of auxiliary specifications (comprised of folds, unfolds, loop invariants, and inductive lemmas); in contrast to 18 lines of wanted specifications (the predicates and pre- and postconditions) and program code. Furthermore, these auxiliary specifications are complex, as discussed next.

\subsection{Auxiliary Specifications}
\label{sec:aux-specs}
Static verifiers cannot reliably unroll recursive predicates during verification; so, such tools rely on explicit \emph{fold} and \emph{unfold} statements to control the availability of predicate information in the verifier. This treats predicates \emph{iso-recursively}; while an \emph{equi-recursive} interpretation treats predicates as their complete unrolling \cite{summers2013formal}. Consequently, the \ttt{acyclic} and \ttt{acyclicSeg} predicates are unfolded and folded often in \fig~\ref{fig:ll-insert-sv} (lines \ref{sv-unfold-acyclic}-\ref{sv-unfold-acyclicSeg}, \ref{sv-fold-listyseg}, \ref{sv-unfold-nextnullseg}-\ref{sv-fold-prevnextseg}, \ref{sv-fold-end}-\ref{sv-fold-prevend}, and \ref{sv-fold-acyclic}). Looking closely, we see that \ttt{acyclic(list)}, which is assumed true from the precondition, is unfolded on line \ref{sv-unfold-acyclic}. This consumes \ttt{acyclic(list)} and produces its body \ttt{acyclicSeg(list, NULL)}, which is subsequently unfolded on line \ref{sv-unfold-acyclicSeg}. Then, at the fold on line \ref{sv-fold-acyclic}, the body of \ttt{acyclic(list)} is packed up into the predicate itself to prove the list remains acyclic after insertion. 

Additionally, static verifiers cannot tell if or when a loop will end (in our example the verifier cannot tell when the list being iterated over ends), but must verify all paths through the program. Therefore, static verifiers reason about loops using specifications called \emph{loop invariants}, which are properties that are preserved for each execution of the loop including at entry and exit. Further, loop invariants must also provide information necessary for proof obligations after the loop, \eg that the list in \ttt{insertLast} is acyclic after insertion. In \fig~\ref{fig:ll-insert-sv}, these constraints result in the loop invariant on lines \ref{sv-loopinv-stmt1}-\ref{sv-loopinv-stmt3} that segments the list into three disjoint and acyclic parts: from the root up to the current node \ttt{y} ($\ttt{acyclicSeg(list, y)}$), the current node \ttt{y} ($\phiAcc{\ttt{y->val}} \phiAnd \phiAcc{\ttt{y->next}}$), and from the node after \ttt{y} to the end ($\ttt{acyclicSeg(y->next, NULL)}$). Exposing \ttt{y} via its accessibility predicates provides access to \ttt{y->next} on line \ref{sv-loop-asmt} in the loop body; and, \ttt{acyclicSeg(list, y)} helps prove \ttt{acyclic(list)} holds after the loop, as we will see next.  

To prove \ttt{acyclic(list)} holds at the end of \ttt{insertLast} (line \ref{sv-return}), it is sufficient to prove instead that \ttt{acyclicSeg(list, NULL)} holds (line \ref{sv-fold-acyclic}). After inserting a new node at the end of the list (lines \ref{sv-newnode-stmt-start}-\ref{sv-newnode-stmt-end}), we can build up an inductive proof with folds (lines \ref{sv-fold-end}-\ref{sv-fold-prevend}) that the list is acyclic from the insertion point \ttt{y} to the new end, \ie \ttt{acyclicSeg(y, NULL)} holds. We also have that the list is acyclic from the root to \ttt{y} (\ttt{acyclicSeg(list, y)}) from to the loop invariant, and so, we are done after proving transitivity of acyclic list segments, \eg \ie \ttt{acyclicSeg(list, y)} and \ttt{acyclicSeg(y, NULL)} implies \ttt{acyclicSeg(list, NULL)}. Sadly, static verifiers cannot automatically discharge such inductive proofs, and so we specify the proof steps in \ttt{mergeLemma} on lines \ref{sv-mergelemma-start}-\ref{sv-mergelemma-end}. Then, after using the lemma on line \ref{sv-mergelemma-afterloop}, we achieve our proof goal.

As we can see, not only do users of static verifiers need to write a number of auxiliary specifications in support of proof goals, the specifications are often more complex compared to the program code itself even for simple examples like \ttt{insertLast}. Worse even, is that while users are developing these complex specifications static tools provide limited feedback on their correctness as demonstrated next (\S\ref{sec:bad-feedback}). 

\subsection{Lack of Early Specification Feedback}
\label{sec:bad-feedback}

\begin{figure}[t]
\centering
\begin{minipage}{0.45\linewidth}
\begin{lstlisting}[xleftmargin=2.5em, name=insertLast-iv]
Node* insertLast(Node* list, int val)
  //@ requires (*@\label{iv-pre}\lightgreen{acyclic(list) \&\& list != NULL}@*);
  /*@ ensures (*@\label{iv-post1}\lightgreen{acyclic(\textbackslash result) \&\&}@*)
                    (*@\label{iv-post2}\lightgreen{\textbackslash result != NULL}@*); @*/
{
   (*@\label{iv-unfold-acyclic}\lightpink{//@~unfold acyclic(list);}@*)
   (*@\label{iv-unfold-acyclicSeg}\lightpink{//@~unfold acyclicSeg(list, NULL);}@*)
   (*@\label{iv-varassign-stmt}@*)Node* y = list;
   (*@\label{iv-fold-listyseg}\lightred{//@~fold acyclicSeg(list, y);}@*)
   (*@\label{iv-loophead}@*)while (y->next != NULL)
   (*@\label{iv-loopinv-stmt1}@*)/*@ loop_invariant (*@\lightred{acyclicSeg(list, y) \phiAnd }@*)
        (*@\label{iv-loopinv-stmt2}\lightpurple{acc(y->next)} \neonblue{\phiAnd acc(y->val) \phiAnd}@*)
        (*@\label{iv-loopinv-stmt3}\neonblue{acyclicSeg(y->next, NULL)}@*); @*/
   {
     (*@\label{iv-tmp-stmt}@*)Node* tmp = y;
     (*@\label{iv-loop-asmt}@*)y = y->next;
\end{lstlisting}
\end{minipage}\hfill
\begin{minipage}{0.45\linewidth}
\begin{lstlisting} [xleftmargin=1em, name=insertLast-iv]
     (*@\label{iv-unfold-nextnullseg}\neonblue{//@~unfold acyclicSeg(y, NULL);}@*)
     (*@\label{iv-fold-next}\lightred{//@~fold acyclicSeg(tmp->next, y);}@*)
     (*@\label{iv-fold-prevnextseg}\lightred{//@~fold acyclicSeg(tmp, y);}@*)
     (*@\label{iv-mergelemma-loop}\neonyellow{mergeLemma(list, tmp, y);}@*)
   }
   
   (*@\label{iv-newnode-stmt-start}@*)y->next = alloc(struct Node);
   (*@\label{iv-newnode-stmt-val}@*)y->next->val = val;
   (*@\label{iv-newnode-stmt-end}@*)y->next->next = NULL;
   (*@\label{iv-fold-end}\lightorange{//@~fold acyclicSeg(y->next->next, NULL);}@*)
   (*@\label{iv-fold-newnode}\lightorange{//@~fold acyclicSeg(y->next, NULL);}@*)
   (*@\label{iv-fold-prevend}\lightorange{//@~fold acyclicSeg(y, NULL);}@*)
   (*@\label{iv-mergelemma-afterloop}\neonyellow{mergeLemma(list, y, NULL);}@*)
   (*@\label{iv-fold-acyclic}\lightorange{//@~fold acyclic(list);}@*)
   (*@\label{iv-return}@*)return list;
}
\end{lstlisting}
\end{minipage}

\footnotesize
\disableTttResize
\fbox{\begin{tabular}{llllllll}
\textcolor{light-green}{$\blacksquare$} & \small{$1^{st}$ increment} (least precise) &
\textcolor{light-purple}{$\blacksquare$} & \small{$2^{nd}$ increment} &
\textcolor{light-pink}{$\blacksquare$} & \small{$3^{rd}$ increment} &
\textcolor{neon-blue}{$\blacksquare$} & \small{$4^{th}$ increment} \\
\textcolor{light-orange}{$\blacksquare$} & \small{$5^{th}$ increment} &
\textcolor{light-red}{$\blacksquare$} & \small{$6^{th}$ increment} &
\textcolor{neon-yellow}{$\blacksquare$} & \small{$7^{th}$ increment} (full spec) & &
\end{tabular}}

\caption{The incremental verification of \ttt{insertLast} from \fig~\ref{ex:ll-insert}}
\label{fig:ll-insertlast-incremental}
\end{figure}

\noindent Since static verifiers, like Viper, limit themselves to reasoning about predicates iso-recursively and rely on loop invariants to prove properties about loops, feedback on the correctness of specifications early in the specification process is limited. For example, consider that a user named Daisy incorrectly specifies the body of \ttt{acyclicSeg} (our recursive predicate) as $\ttt{(s == e) ?}$ $\phiAcc{\ttt{s->val}} \phiAnd \phiAcc{\ttt{s->next}} \phiAnd$ $\ttt{acyclicSeg(s->next, e)}$ : $\phiTrue$, which swaps the branches of the ternary in the correct specification from \fig~\ref{fig:ll-insert-sv}. Let's see how Daisy comes across this error while using Viper to incrementally specify \ttt{insertLast} in \fig~\ref{fig:ll-insertlast-incremental}. Each increment from the first to the last (seventh) is highlighted in a different color. The first increment, highlighted in green, specifies the precondition and postcondition of \ttt{insertLast} with \ttt{acyclic} and \ttt{acyclicSeg} (lines \ref{iv-pre}-\ref{iv-post2}). Since predicates are black boxes in static verifiers, Viper only tells Daisy that there is insufficient permission to access \ttt{y->next} in the loop condition on line \ref{iv-loophead}. So, Daisy specifies the required $\phiAcc{\ttt{y->next}}$ permission (as highlighted in purple for the second increment on line \ref{iv-loopinv-stmt2}) in the loop invariant, which must frame the loop condition. But, alas Viper cannot prove that $\phiAcc{\ttt{y->next}}$ holds on entry to the loop. Without realizing the branches of \ttt{acyclicSeg} are out of order, Daisy expects \ttt{acyclicSeg(list, NULL)}'s body to provide $\phiAcc{\ttt{y->next}}$ at loop entry as \ttt{list != NULL} and \ttt{y == list}; and so, she unfolds \ttt{acyclic(list)} and \ttt{acyclicSeg(list, NULL)} on lines \ref{iv-unfold-acyclic}-\ref{iv-unfold-acyclicSeg} making up the third specification increment highlighted in pink. Unfortunately, Viper still reports that $\phiAcc{\ttt{y->next}}$ does not hold on entry to the loop, which alerts Daisy to the problem with \ttt{acyclicSeg}.

With Viper, Daisy required three specification increments to detect a bug in the first unfolding of \ttt{acyclicSeg}, and this problem gets worse the deeper the bug is in the recursive predicate. 
For example, consider now that \ttt{acyclicSeg}'s body is incorrectly specified as $\ttt{(s == e) ?}~ \phiTrue :$ $\phiAcc{\ttt{s->val}} \phiAnd \phiAcc{\ttt{s->next}} \phiAnd \ttt{acyclicSeg(e, s->next)}$, which swaps \ttt{s->next} and \ttt{e} in the recursive call to \ttt{acyclicSeg}. As a result, \ttt{acyclicSeg} asserts in lock-step that the nodes in lists \ttt{s} and \ttt{e} are accessible and separated---which is not the intended behavior of \ttt{acyclicSeg}---and \ttt{acyclicSeg} now always fails when \ttt{e} reaches its end, \ie is \ttt{NULL}, as it tries to assert $\phiAcc{\ttt{e->val}}$ and $\phiAcc{\ttt{e->next}}$. It takes until the fourth specification increment highlighted in blue to discover that \ttt{acyclicSeg} is incorrectly specified using Viper. As before, Daisy is led to specifying the first three increments by Viper's error messages that first require $\phiAcc{\ttt{y->next}}$ in the loop invariant and then require $\phiAcc{\ttt{y->next}}$ to hold on entry to the loop. This time, however, \ttt{acyclicSeg(list, NULL)}'s body contains $\phiAcc{\ttt{y->next}}$ when \ttt{list != NULL}, so Viper can prove $\phiAcc{\ttt{y->next}}$ holds on loop entry and instead reports that the loop invariant $\phiAcc{\ttt{y->next}}$ might not be preserved by the loop body. Daisy recalls the loop iterates over all nodes in the list with the current node being \ttt{y}, so preserving $\phiAcc{\ttt{y->next}}$ in the loop is the same as showing that Viper has accessibility predicates for every node in the list. As a result, she specifies the fourth increment (lines \ref{iv-loopinv-stmt2}-\ref{iv-loopinv-stmt3} and \ref{iv-unfold-nextnullseg}), which continuously unfolds the \ttt{acyclicSeg(list, NULL)} predicate on every iteration of the loop and captures the information in its body in the loop invariant. Alas, Viper reports that the new loop invariant does not hold on entry as it cannot prove \ttt{acyclicSeg(y->next, NULL)} holds here. Since this information should come from unfolding \ttt{acyclicSeg(list, NULL)} on line \ref{iv-unfold-acyclicSeg}, Daisy takes another look at \ttt{acyclicSeg}'s body and discovers her specification error. That is, it takes four specification increments for Daisy to realize her mistake and the fourth increment required her to think deeply about her while loop.

Clearly, static verifiers burden their users, like Daisy, by requiring them to write a number of complex auxiliary specifications both for proofs and to receive useful feedback on the correctness of their specifications. Fortunately, as we will show next in \S\ref{sec:gco-saves-the-day}, \gco overcomes this burden by smoothly supporting the spectrum between static and dynamic checking.

\section{\gco to the Rescue}
\label{sec:gco-saves-the-day}

In this section, we show how \gco's ability to smoothly integrate static and dynamic checking allows users to overcome specification burdens inherent to static verification, \eg that users have to write complex auxiliary specifications in support of proofs and to receive useful feedback on the correctness of their specifications.

\subsection{Ignore Auxiliary Specifications with \gco}
\label{sec:ignore-specs-gco}

In \S\ref{sec:aux-specs}, we saw how static verifiers force users to write complex auxiliary specifications (like folds and unfolds, loop invariants, and inductive lemmas) in support of proof goals. In contrast, \gco allows users to write as many or as few auxiliary specifications as they want, and instead utilizes dynamic verification to check proof obligations not discharged statically due to missing specifications. For example, consider \fig~\ref{fig:ll-insertlast-incremental}, in which our user Daisy incrementally specifies \ttt{insertLast} for preservation of list acyclicity and non-nullness. With \gco, Daisy only needs to specify the first increment in green, which contains the pre- and postcondition of \ttt{insertLast} on lines \ref{iv-pre}-\ref{iv-post2}. She can completely avoid specifying the auxiliary specifications in the rest of the increments by instead specifying $\qm$ in the loop invariant on line \ref{iv-loophead}. Then, \gco uses the allocation statement on line \ref{iv-newnode-stmt-start} to statically validate the write accesses of \ttt{y->next->val} and \ttt{y->next->next} on lines \ref{iv-newnode-stmt-val}-\ref{iv-newnode-stmt-end}. It can also prove statically that the list after insertion is non-null. All other proof obligations, which ensure memory safety of the while loop and the list after insertion is acyclic, are checked dynamically.

Run-time checking memory safety of the while loop involves three parts: 1) verifying access to \ttt{y->next} in the loop condition (line \ref{iv-loophead}), 2) verifying access to \ttt{y->next} in the loop body (line \ref{iv-loop-asmt}), and 3) verifying access to \ttt{y->next} in the alloc statement directly after the loop (line \ref{iv-newnode-stmt-start}). That is, \gco asserts ownership of all heap locations in the given list, which are being iterated over by the loop. Consequently, the larger the list the higher the run-time cost of verification for \ttt{insertLast}. This cost is unacceptable to Daisy, so she statically specifies memory safety of the while loop in the first four increments in \fig\ref{fig:ll-insertlast-incremental} (lines \ref{iv-pre}-\ref{iv-post2}, \ref{iv-unfold-acyclic}-\ref{iv-unfold-acyclicSeg}, \ref{iv-loopinv-stmt2}-\ref{iv-loopinv-stmt3}, and \ref{iv-unfold-nextnullseg}). The new loop invariant (lines \ref{iv-loopinv-stmt2}-\ref{iv-loopinv-stmt3}) uses \ttt{acyclicSeg} to expose $\phiAcc{\ttt{y->next}}$ for verifying access to \ttt{y->next} in the loop condition, loop body, and after the loop. The unfolds on lines \ref{iv-unfold-acyclic}-\ref{iv-unfold-acyclicSeg} and \ref{iv-unfold-nextnullseg} are used to prove that the loop invariant is preserved by the loop given the precondition on line \ref{iv-pre}. After all this work, Daisy is not interested in also statically specifying the list after insertion is acyclic, and so, she joins $\qm$ with her newly specified loop invariant. As a result, \gco now additionally checks memory safety of the while loop statically reducing run-time cost, and only checks \ttt{acyclic(\textbackslash result)} from the postcondition of \ttt{insertLast} (line \ref{iv-post1}) dynamically (\ie the list returned after insertion is acyclic). By allowing \gco to check \ttt{acyclic(\textbackslash result)} dynamically, Daisy saved herself a lot of specification effort. She avoided building up \ttt{acyclicSeg} from the previous end of the list to the new one (increment five in orange, lines \ref{iv-fold-end}-\ref{iv-fold-prevend}), specifying a more complex loop invariant (increment six in red, lines \ref{iv-fold-listyseg}, \ref{iv-loopinv-stmt1}, and \ref{iv-fold-next}-\ref{iv-fold-prevnextseg}), and stating and proving transitivity of acyclic list segments (increment seven in yellow, lines \ref{iv-mergelemma-loop} and \ref{iv-mergelemma-afterloop}). Daisy is very happy to make this human-effort vs run-time cost trade-off.

\subsection{\gco Provides Earlier Feedback with Run-Time Checks}
As we saw in \S\ref{sec:bad-feedback}, static verifiers struggle to provide early feedback on specification errors in predicates. Using Viper, it took until the third specification increment in \fig\ref{fig:ll-insertlast-incremental} for Daisy to discover the simple error in \ttt{acyclicSeg}'s body, $\ttt{(s == e) ?}$ $\phiAcc{\ttt{s->val}} \phiAnd \phiAcc{\ttt{s->next}} \phiAnd$ $\ttt{acyclicSeg(s->next, e)}$ : $\phiTrue$, which swaps the branches of the ternary in the correct specification from \fig~\ref{fig:ll-insert-sv}. In contrast, with \gco, Daisy easily discovers this error on the first specification increment (in green, lines \ref{iv-pre}-\ref{iv-post2}), which specifies the pre- and postcondition of \ttt{insertLast}. She additionally specifies $\qm$ on the loop invariant and provides a simple test case that calls \ttt{insertLast} on a list with one node. Then, \gco alerts Daisy to the error in \ttt{acyclicSeg} by reporting at run time that $\phiAcc{\ttt{s->val}}$ from \ttt{acyclicSeg} does not hold at the end of the list where \ttt{s == NULL}. Clearly, $\phiAcc{\ttt{s->val}}$ will never hold when \ttt{s == NULL}, so Daisy realizes she swapped the branches in \ttt{acyclicSeg}. Similarly, the second example of an error in \ttt{acyclicSeg}'s specification, which swaps \ttt{s->next} and \ttt{e} in the recursive call to \ttt{acyclicSeg} ($\ttt{(s == e) ?}~ \phiTrue :$ $\phiAcc{\ttt{s->val}} \phiAnd \phiAcc{\ttt{s->next}} \phiAnd \ttt{acyclicSeg(e, s->next)}$), took four specification increments in \fig\ref{fig:ll-insertlast-incremental} to be exposed by Viper. Again with \gco, Daisy can detect the error by the first increment as long as she specifies $\qm$ on her loop invariant and supplies a simple test case with a list containing two elements. In this case, \gco reports at run time that $\phiAcc{\ttt{s->val}}$ from \ttt{acyclicSeg} does not hold for \ttt{s} which is \ttt{NULL}. Since $\phiAcc{\ttt{s->val}}$ will never hold when \ttt{s == NULL}, Daisy takes another look at \ttt{acyclicSeg}'s body and realizes her error. 
That is, \gco's dynamic checking of partial specifications is helpful for detecting errors in recursive predicates much earlier in the specification process than static verification alone and the errors better capture the inherent problems in the specifications.

To summarize, users of \gco may write as many or as little auxiliary specifications as they want and still get sound verification of their code by trading off between human-effort and run-time cost. Users can also receive feedback on the correctness of their specifications much earlier in the specification process than if they used static verification alone and the run-time errors reported often closely match the inherent problems with the specification.

\section{\gco's Design and Implementation}
\label{sec:gradual-c0}

\begin{wrapfigure}{L}{2.50in}
\centering
\includegraphics[width=0.46\textwidth]{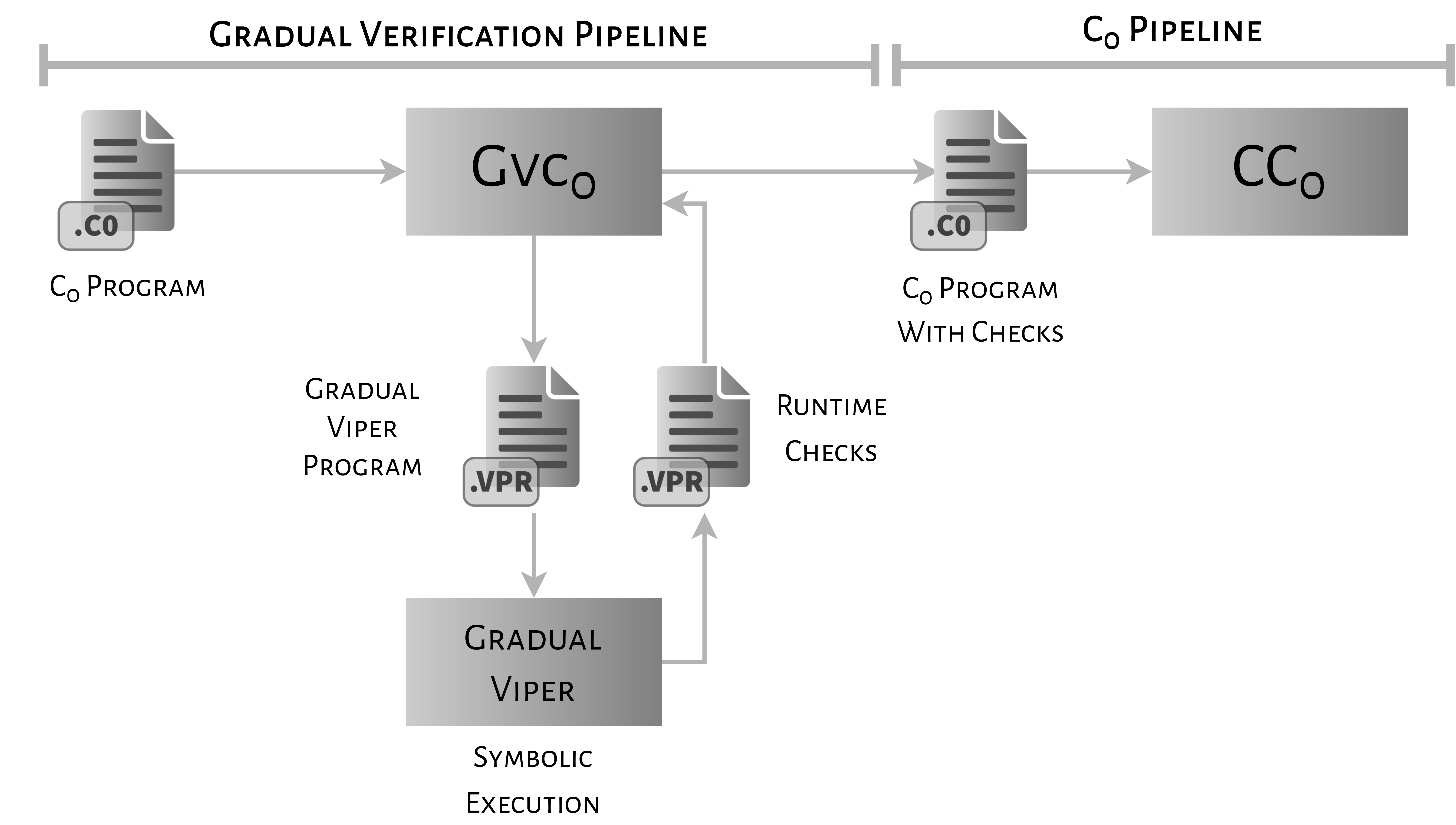}
\disableTttResize
\caption{System design of \gco}
\label{fig:gco-design}
\end{wrapfigure}

\gco is a working gradual verifier for the C0 programming language \cite{arnold2010c0} that is built as extension of the Viper static verifier~\cite{MuellerSchwerhoffSummers16}. Our goals for the design and implementation of \gco are to: 
\begin{itemize}
    \item be easily extensible to other programming languages beyond C0, 
    \item minimize run-time overhead from verification as much as possible without introducing highly complex algorithms, and
    \item use symbolic execution for static reasoning.
\end{itemize}

Consequently, we settled on the design illustrated in \fig\ref{fig:gco-design}. \gco is structured in two major sub-systems: 1) the gradual verification pipeline and 2) the C0 pipeline. 
Within the gradual verification pipeline, a C0 program is first translated AST-to-AST into a \gviper program by \gco's frontend module, \gvc. The \gvc module implements a simple parser, abstract syntax, and type checker for C0 programs to facilitate the translation. The \gviper module is the backend of \gco and implements its own parser, abstract syntax, and type checker for its own imperative language called the \gviper language. This language is the Viper language plus imprecise formulas and allows \gviper to support multiple frontend languages, not just C0. We chose to build a C0 frontend first because C0 is a pedagogical version of C designed with dynamic verification in mind, and we plan to use it in the classroom.  

Once the C0 program is translated into a \gviper program, it is optimistically statically verified by the \gviper module. \gviper extends Viper's symbolic execution based verifier to support imprecise formulas and resulting holes in static reasoning as inspired by \citet{wise2020gradual} and gradual typing \cite{siek2015refined, siek2006gradual, hermanAl:hosc10}. Consequently, by construction \gviper supports full static verification of programs when specifications are complete. Differing from \citet{wise2020gradual}'s work, \gviper also extends the symbolic execution algorithm to create a description of needed run-time checks in support of static holes. The run-time checks are minimized with statically available information during reasoning. Finally, \gvc takes these run-time checks and encodes them in the original C0 program to produce a sound, gradually-verified program. The C0 pipeline takes this C0 program and feeds it to the C0 compiler, CC0, which is used to execute the program. Note, the encoding of checks into C0 source code optimizes for run-time performance and simplifies extending C0 with dynamic verification in our domain.  

The rest of this section describes the implementation of \gviper and \gvc's designs in more detail and illustrates the concepts via example. We also highlight design and implementation choices influenced by our goals. 
\S\ref{sec:c0-to-viper} discusses how C0 programs are translated to \gviper programs, along with modifications made to both C0 and Viper for gradual verification. 
Then, \S\ref{sec:gviper-sv-highlevel} and \S\ref{sec:gviper-sv} detail \gviper's symbolic execution approach and how it produces minimized run-time checks. Finally, \S\ref{sec:c0-runtime-checks} focuses on how \gvc turns run-time checks from \gviper into C0 code for dynamic verification.

\subsection{Translating C0 Source Code to \gviper Source Code for Verification}
\label{sec:c0-to-viper}
The C0 language, with its minimal set of language features and its existing support for specifications, serves well as the target language for our implementation. As its name suggests, C0 borrows heavily from C, but its feature set is reduced to better suit its intended purpose as a tool in computer science education \cite{arnold2010c0}. It is a memory-safe subset of C that forbids casts, pointer arithmetic, and pointers to stack-allocated memory. All pointers are created with heap allocation, and de-allocation is handled by a garbage collector. 

\begin{figure}[t]
\begin{minipage}[t]{.45\linewidth}
\centering\scriptsize\ttfamily\disableTttResize
\begin{plstx}
*(variables): x [\in] \mathit{VAR} \\
*(values): v [\in] \mathit{VAL} \\
*(expressions): e [\in] \mathit{EXPR} \\
*(statements): s [\in] \mathit{STMT} \\
*(operators): op [\in] \ttt{+},~ \ttt{-},~ \ttt{/},~ \ttt{*},~ \ttt{==},~ \ttt{!=},~ \ttt{<=},~ \ttt{>=},~ \ttt{<},~ \ttt{>} \\
\end{plstx}
\end{minipage}\hspace{3em}
\begin{minipage}[t]{.45\linewidth}
\centering\scriptsize\ttfamily\disableTttResize
\begin{plstx}
*(struct names): S [\in] \mathit{STRUCTNAME} \\
*(field names): f [\in] \mathit{FIELDNAME} \\
*(predicate names): p [\in] \mathit{PREDNAME} \\
*(method names): m [\in] \mathit{METHODNAME} \\
\end{plstx}
\end{minipage}
{\footnotesize
\disableTttResize
\caption{Shared abstract syntax definitions}
\label{fig:shared-syntax}}
\par\medskip
\begin{minipage}[t]{0.45\linewidth}
\centering\scriptsize\ttfamily\disableTttResize
\begin{plstx}
: \mathtt{P} ::= \overline{struct}\ \overline{predicate}\ \overline{method} \\
: struct ::= \mlightorange{\ttt{struct}~ S~ \{\ \overline{T~ f}\ \}} \\
: predicate ::= \ttt{//@predicate}~ p(\overline{T~ x})~ \mathtt{=}~ \gphi \\
: method ::= \tilde{T}~ m(\overline{T~ x})\ contract \ \{\ s \ \} \\
: contract ::= \ttt{//@requires}~ \gphi;\ \ttt{//@ensures}~ \gphi; \\
: T ::= \mlightorange{\ttt{struct}~ S} | \ttt{int} | \ttt{bool} | \mlightorange{\ttt{char}} | \mlightpink{T*} \\
: \tilde{T} ::= \mlightorange{\mathtt{void}} | T \\
: s ::= s;\ s | T~ x | \mlightorange{T~ x~ \ttt{=}~ e} | x~ \ttt{=}~ e | l~ \ttt{=}~ e | e | \mlightpink{\ttt{assert}(e)} | \ttt{//@assert}~ \phi | \ttt{//@fold}~ p(\overline{\tilde{e}}) | \ttt{//@unfold}~ p(\overline{\tilde{e}}) | \sIf{e}{s}{s} | \sWhilegvc{e}{\gphi}{s} | \mlightpink{\sForgvc{s}{e}{s}{\gphi}{s}} \\
: e ::= v | x | op(\overline{e}) | e \ttt{->} f | \mlightpink{\mathtt{*}e} | \mlightpink{m(\overline{e})} | \mlightpink{\ttt{alloc}(T)} | \mlightpink{\phiCond{e}{e}{e}} \\
: \tilde{e} ::= v | x | op(\overline{\tilde{e}}) | \tilde{e} \ttt{->} f | \mlightpink{\ttt{*}\tilde{e}} \\
: l ::= x \ttt{->} f | \mlightpink{\ttt{*} x} | \mlightpink{l \ttt{->} f} | \mlightpink{\ttt{*} l} \\
: x ::= \ttt{\textbackslash result} | id \\
: v ::= n | \mlightorange{c} | \ttt{NULL} | \phiTrue | \phiFalse \\
: \gphi ::= \withqm{\phi} | \theta \\
: \theta ::= \ttt{self-framed } \phi \\
: \phi ::= \tilde{e} | p(\overline{\tilde{e}}) | \phiAcc{l} | \phi \ttt{ \&\& } \phi | \phiCond{\tilde{e}}{\phi}{\phi} \\
\end{plstx}
\captionof{figure}{\gvc abstract syntax}
\label{fig:gvc-syntax}
\end{minipage}\hspace{2.5em}
\begin{minipage}[t]{.45\linewidth}
\centering\scriptsize\ttfamily\disableTttResize
\begin{plstx}
: \mathtt{P} ::= \overline{\mathit{field}}\ \overline{predicate}\ \overline{method} \\
: \mathit{field} ::= \mlightorange{\ttt{field}~f:T} \\
: \mathit{predicate} ::= \ttt{predicate } \pred{p}{\overline{x:T}}{\gphi} \\
*: \mathit{method} [::=,] \ttt{method}~ m(\overline{x:T})~ \ttt{returns}~(\overline{y:T})~ | \mathit{contract}~ \{~s~\} \\
: \mathit{contract} ::= \contract{\gphi}{\gphi} \\
: T ::= \Tint | \Tbool | \mlightpink{\Tref} \\
: s ::= \sSeq{s}{s} | \mlightorange{\sDeclare{T}{x}} | \mlightorange{\sVarAssign{x}{e}} | \mlightpink{\sFieldAssign{x}{f}{e}} | \mlightpink{\sAlloc{x}{\overline{f}}} | \mlightpink{\sCall{\overline{x}}{m}{\overline{e}}} | \sAssert{\phi} | \sFold{\phiAcc{p(\overline{e})}} | \sUnfold{\phiAcc{p(\overline{e})}} | \sIf{e}{s}{s} | \sWhile{e}{\gphi}{s} \\
: e ::= v | x | op(\overline{e}) | e\ttt{.}f \\
: x ::= \eresult | id \\
: v ::= n | \enull | \phiTrue | \phiFalse \\
: \gphi ::= \withqm{\phi} | \sphi \\
: \sphi ::= \text{self-framed } \phi \\
: \phi ::= e | \phiAcc{p(\overline{e})} | \phiAcc{e.f} | \phiCons{\phi}{\phi} | \phiCond{e}{\phi}{\phi} \\
\end{plstx}
\caption{\gviper abstract syntax}
\label{fig:abstract-syntax}
\end{minipage}
\footnotesize
\disableTttResize
\par\smallskip
\fbox{
    \begin{tabular}{llll}
        \textcolor{light-orange}{$\blacksquare$} &
        \begin{minipage}{2in}
            \small{Representation differs slightly in \gvc vs. \gviper}
        \end{minipage} &
        \textcolor{light-pink}{$\blacksquare$} &
        \begin{minipage}{2in}
            \small{Functionality in \gvc that requires non-trivial translation to \gviper}
        \end{minipage}
    \end{tabular}
}
\caption{Abstract syntax comparison for \gvc and \gviper}
\label{fig:syntax}
\end{figure}

The abstract syntax for C0 programs supported by \gco is given in \fig~\ref{fig:gvc-syntax}, \ie \gvc's abstract syntax. \gvc programs are made of struct and method declarations that largely follow C syntax. 
What differs from both C and C0 is \gvc's specification language. Methods may specify constraints on their input and output values as side-effect-free gradual formulas $\gphi$, usually in \ttt{//@requires} or \ttt{//@ensures} clauses in the method header. 
Loops and abstract predicates contain invariants and bodies respectively that are made of gradual formulas. Such formulas $\gphi$ are imprecise formulas $\withqm{\phi}$ or complete boolean formulas $\phi$ (Note, in this case, $\phi$ must be self-framed as defined in IDF). A formula $\phi$ joins boolean values, boolean operators, predicate instances, accessibility predicates, and conditionals via the separating conjunction $\phiAnd$.
\gvc programs also contain \ttt{//@fold p($\overline{\tilde{e}}$)} and \ttt{//@unfold p($\overline{\tilde{e}}$)} statements for predicates and \ttt{//@assert $\phi$} statements for convenience.

To support the gradual verification of many different imperative programming languages, \gviper verifies programs written in its own custom imperative language, which is designed to ease the translation from other imperative languages into it. The \gviper language's abstract syntax is given in \fig~\ref{fig:abstract-syntax}. The \gvc and \gviper languages are roughly 1-to-1, including their specification languages, so translation is mostly straightforward, but there are some differences as highlighted in yellow (trivial) and blue (non-trivial) in \fig~\ref{fig:syntax}. 
For example, \ttt{for} loops in \gvc are rewritten as \ttt{while} loops in \gviper, and \ttt{alloc(struct T)} expressions are translated to \ttt{new} statements containing \ttt{struct T}'s fields. Additionally, \gvc allows method calls, \ttt{alloc}s, and ternaries in arbitrary expressions, while \gviper only allows such constructs in corresponding program statements\footnote{Note, ternaries correspond to if statements}. Therefore, \gvc uses fresh temporary variables to version expressions containing the aformentioned constructs into program statements in \gviper. The temporary variables are then used in the original expression in place of the corresponding method call, \ttt{alloc}, or ternary.
Nested field assignments, such as \ttt{x->y->z = a}, are similarly expanded into multiple program statements using temporary variables. Value type pointers in \gvc are rewritten as pointers to single-value structs that can be easily translated into \gviper syntax. 
Finally, \ttt{assert($e$)} statements are essentially ignored in \gviper; $e$ is translated into \gviper syntax to verify its heap accesses, but $e$ is not asserted. Instead, the assert is always kept in the original C0 program and is checked exclusively at run time. \fig~\ref{fig:bank-gv} provides a simple example program written in both the \gvc language (\fig~\ref{fig:ex-bank-c0}) and \gviper language (\fig~\ref{fig:ex-bank-gv}) for reference.

Note that \gvc does not support array and string values since 
gradually verifying any interesting properties about such constructs requires non-trivial extensions to current gradual verification theory. Similarly, the \gviper language, in contrast to the Viper language, does not support the aforementioned constructs and fractional permissions.

\subsection{\gviper: Symbolic Execution for Gradual Verification}
\label{sec:gviper-sv-highlevel}

In this section, we describe the design and implementation of \gviper's symbolic execution based algorithm supporting the static verification of imprecise formulas. Gradual verification of recursive heap data structures was formalized with weakest liberal preconditions~\cite{wise2020gradual}. All of the static verifiers supporting separation logic or IDF---such as Viper \cite{MuellerSchwerhoffSummers16}, VeriFast \cite{jacobs2011verifast}, JStar \cite{distefano2008jstar}, and SmallFoot \cite{berdine2006smallfoot}---reason with symbolic execution. Therefore, this work serves as a first guide for building gradual verifiers from static verifiers using symbolic execution for reasoning. Additionally, unlike in \citet{wise2020gradual}'s work, we use our static reasoning algorithm not just for optimistic static verification but also for soundly reducing the number of run-time checks required during dynamic verification. That is, during a single execution of \gviper a program is statically verified and a set of minimized run-time checks is produced for program points where the algorithm is optimistic during verification due to imprecision.

Before formalizing \gviper's implementation in \S\ref{sec:gviper-sv}, we first demonstrate at a high-level with examples how symbolic execution is used to perform optimistic static verification of programs containing recursive heap data structures and how minimized run-time checks are produced during this process. We also point out novel technical challenges faced and solutions developed thanks to relying on symbolic execution both for static verification and minimizing run-time checks.

\subsubsection{\textbf{Optimistic static verification in \gviper by example}}

\begin{figure}[p]
\centering
\begin{subfigure}[t]{0.45\linewidth}
\footnotesize
\disableTttResize
\begin{lstlisting}[xleftmargin=2.5em, name=bank-c0]
struct Account { int balance; }; 
typedef struct Account Account;
(*@\label{bank-pred1}\drkgray{/*@~predicate geqTo(Account* a1, Account* a2) = }@*) 
  (*@\drkgray{? \phiAnd a1->balance >= a2->balance \phiAnd}@*)
  (*@\drkgray{a2->balance >= 0; @*/}@*)
(*@\lightgray{/*@~predicate positive(Account* a) = }@*)
  (*@\label{bank-pred2}\lightgray{acc(a->balance) \phiAnd a->balance >= 0; @*/}@*)

Account* withdraw(Account* a1, Account* a2)
  //@ requires (*@\label{bank-pre}\lightgray{geqTo(a1,a2)}@*);
  //@ ensures (*@\label{bank-post1}\drkgray{? \phiAnd positive(a2) \phiAnd}@*)
              (*@\label{bank-post2}\drkgray{positive(\textbackslash result)}@*);
(*@\label{bank-withdraw-start}@*){
  (*@\label{bank-unfold}\lightgray{//@~unfold geqTo(a1,a2);}@*)
  (*@\label{bank-if}@*)if (a1 == NULL || a2 == NULL) {
    (*@\label{bank-null}@*)return a1;
  (*@\label{bank-else}@*)} else {
    (*@\label{bank-newB}@*)int newB = a1->balance - a2->balance;
    (*@\label{bank-fieldassign}@*)a1->balance = newB;
    (*@\label{bank-folda1}\lightgray{//@~fold positive(a1);}@*)
    (*@\label{bank-folda2}\lightgray{//@~fold positive(a2);}@*)
    (*@\label{bank-nonnull}@*)return a1;
  (*@\label{bank-if-end}@*)}
}
(*@~@*)
\end{lstlisting}
\caption{A simple bank withdraw example written in the \gco language}
\label{fig:ex-bank-c0}
\end{subfigure}\hspace{0.7cm}
\begin{subfigure}[t]{0.45\linewidth}
\footnotesize
\disableTttResize
\begin{lstlisting}[xleftmargin=2.5em, name=bank-gv]
field balance: Int

(*@\label{bankgv-pred1}\drkgray{predicate geqTo(a1: Ref, a2: Ref) }@*) 
  (*@\label{bankgv-pred1-body1}\drkgray{\{ ? \phiAnd a1.balance >= a2.balance \phiAnd}@*)
  (*@\label{bankgv-pred1-body2}\drkgray{a2.balance >= 0 \} }@*)
(*@\lightgray{predicate positive(a: Ref)}@*)
  (*@\label{bankgv-pred2}\lightgray{\{ acc(a.balance) \phiAnd a.balance >= 0 \} }@*)

method withdraw(a1: Ref, a2: Ref) 
returns (res: Ref)
  requires (*@\label{bankgv-pre}\lightgray{acc(geqTo(a1,a2))}@*)
  ensures (*@\label{bankgv-post1}\drkgray{? \phiAnd acc(positive(a2)) \phiAnd}@*)
          (*@\label{bankgv-post2}\drkgray{acc(positive(res))}@*)
(*@\label{bankgv-withdraw-start}@*){
  (*@\label{bankgv-unfold}\lightgray{unfold acc(geqTo(a1,a2))}@*)
  (*@\label{bankgv-if}@*)if (a1 == null || a2 == null) {
    (*@\label{bankgv-null}@*)res := a1
  (*@\label{bankgv-else}@*)} else {
    (*@\label{bankgv-newB}@*)var newB:Int = a1.balance-a2.balance
    (*@\label{bankgv-fieldassign}@*)a1.balance := newB
    (*@\label{bankgv-folda1}\lightgray{fold acc(positive(a1))}@*)
    (*@\label{bankgv-folda2}\lightgray{fold acc(positive(a2))}@*)
    (*@\label{bankgv-nonnull}@*)res := a1
  (*@\label{bankgv-if-end}@*)}
}
\end{lstlisting}
\caption{A simple bank withdraw program written in the \gviper language}
\label{fig:ex-bank-gv}
\end{subfigure}\\[1ex]
{\footnotesize
\disableTttResize
\fbox{\begin{tabular}{llllll}
$\square$ & \small{Program code} & \textcolor{light-gray}{$\blacksquare$} & \small{Static specification} &
\textcolor{drk-gray}{$\blacksquare$} & \small{Imprecise specification}
\end{tabular}}}\\[1ex]
\begin{subfigure}[t]{\linewidth}
\scriptsize\centering
\begin{tabular}{|p{0.015\linewidth}|p{0.04\linewidth}|p{0.14\linewidth}|l|p{0.14\linewidth}|p{0.15\linewidth}|p{0.2\linewidth}|}
 \hline
 \textbf{Ln} & \textbf{Impr-ecise} & \textbf{Opt. Heap} & \textbf{Heap} & \textbf{Var Store} & \textbf{Path Condition} & \textbf{Run-time Checks} \\ \hline
 \ref{bankgv-withdraw-start}-\ref{bankgv-unfold} & No & $\emptyset$ & geqTo(t1,t2) & a1$\rightarrow$t1 ; a2$\rightarrow$t2 ; res$\rightarrow$t3 & $\emptyset$ & $\emptyset$\\ \hline
 \ref{bankgv-unfold}-\ref{bankgv-if} & Yes & acc(t1,balance,p1) ; acc(t2,balance,p2) & $\emptyset$ & a1$\rightarrow$t1 ; a2$\rightarrow$t2 ; res$\rightarrow$t3 & t1 != null ; t2 != null ; p1 >= p2 ; p2 >= 0 & $\emptyset$ \\ \hline
 \ref{bankgv-if}-\ref{bankgv-null} &  - & - & - & - & - & - \\ \hline
 \ref{bankgv-null}-\ref{bankgv-else} & - & - & - & - & - & - \\ \hline
 \ref{bankgv-else}-\ref{bankgv-newB} & Yes & acc(t1,balance,p1) ; acc(t2,balance,p2) & $\emptyset$ & a1$\rightarrow$t1 ; a2$\rightarrow$t2 ; res$\rightarrow$t3 & t1 != null ; t2 != null ; p1 >= p2 ; p2 >= 0 & $\emptyset$ \\ \hline
 \ref{bankgv-newB}-\ref{bankgv-fieldassign} & Yes & acc(t1,balance,p1) ; acc(t2,balance,p2) & $\emptyset$ & a1$\rightarrow$t1 ; a2$\rightarrow$t2 ; res$\rightarrow$t3 ; newB$\rightarrow$t4 & t1 != null ; t2 != null ; p1 >= p2 ; p2 >= 0 ; t4 = p1 - p2 & $\emptyset$ \\ \hline
 \ref{bankgv-fieldassign}-\ref{bankgv-folda1} & Yes & $\emptyset$ & acc(t1,balance,p3) & a1$\rightarrow$t1 ; a2$\rightarrow$t2 ; res$\rightarrow$t3 ; newB$\rightarrow$t4 & t1 != null ; t2 != null ; p1 >= p2 ; p2 >= 0 ; t4 = p1 - p2 ; p3 = t4 & $\emptyset$ \\ \hline
 \ref{bankgv-folda1}-\ref{bankgv-folda2} & Yes & $\emptyset$ & positive(t1) & a1$\rightarrow$t1 ; a2$\rightarrow$t2 ; res$\rightarrow$t3 ; newB$\rightarrow$t4 & t1 != null ; t2 != null ; p1 >= p2 ; p2 >= 0 ; t4 = p1 - p2 ; p3 = t4 & $\emptyset$ \\ \hline
 \ref{bankgv-folda2}-\ref{bankgv-nonnull} & Yes & $\emptyset$ & positive(t2) & a1$\rightarrow$t1 ; a2$\rightarrow$t2 ; res$\rightarrow$t3 ; newB$\rightarrow$t4 & t1 != null ; t2 != null ; p1 >= p2 ; p2 >= 0 ; t4 = p1 - p2 ; p3 = t4 & $l_{bc}$, $\neg$(a1 = null || a2 = null) $\longrightarrow$ $l_{c1}$, acc(a2.balance) \hspace{0.5em} ; \hspace{0.5em} $l_{bc}$, $\neg$(a1 = null || a2 = null) $\longrightarrow$ $l_{c2}$, a2.balance >= 0 \\ \hline
 \ref{bankgv-nonnull}-\ref{bankgv-if-end} & Yes & $\emptyset$ & positive(t2) & a1$\rightarrow$t1 ; a2$\rightarrow$t2 ; res$\rightarrow$t3 ; newB$\rightarrow$t4 & t1 != null ; t2 != null ; p1 >= p2 ; p2 >= 0 ; t4 = p1 - p2 ; p3 = t4 ; t3 = t1 & $l_{bc}$, $\neg$(a1 = null || a2 = null) $\longrightarrow$ $l_{c1}$, acc(a2.balance) \hspace{0.5em} ; \hspace{0.5em} $l_{bc}$, $\neg$(a1 = null || a2 = null) $\longrightarrow$ $l_{c2}$, a2.balance >= 0 \\ \hline
 \ref{bankgv-if-end} & \multicolumn{6}{p{0.8\linewidth}|}{$l_{bc}$, $\neg$(a1 = null || a2 = null) $\longrightarrow$ $l_{c1}$, acc(a2.balance) \hspace{0.5em} ; \hspace{0.5em} $l_{bc}$, $\neg$(a1 = null || a2 = null) $\longrightarrow$ $l_{c2}$, a2.balance >= 0 \hspace{0.5em} ; \hspace{0.5em} $l_{bc}$, $\neg$(a1 = null || a2 = null) $\longrightarrow$ $l_{c3}$, acc(positive(res))} \\ \hline
\end{tabular}
\footnotesize
\disableTttResize
\caption{Contents of the symbolic state at each program point during \gviper's static verification of \ttt{withdraw} in \fig~\ref{fig:ex-bank-gv}}
\label{fig:ex-bank-gv-table}
\end{subfigure}\\

\caption{A gradually verified, bank withdraw program that is contrived to illustrate how \gviper works}
\label{fig:bank-gv}
\end{figure}

The simple program given in \fig~\ref{fig:ex-bank-gv} implements a \ttt{withdraw} function (method), which subtracts the balance in one bank account (the subtrahend) from the balance in another account (the minuend) returning the result.\footnote{Note, we refer to the version of the withdraw program written in the \gviper language rather than its \gco counterpart in \fig~\ref{fig:ex-bank-c0} as we will be discussing how \gviper works in this section.} Any client program of \ttt{withdraw} must ensure the subtrahend's balance is less than or equal to the minuend's balance and that both balances are positive as specified by \ttt{withdraw}'s precondition (line \ref{bankgv-pre}). Then, \ttt{withdraw} will return an account with a positive balance as specified by \ttt{withdraw}'s postcondition (lines \ref{bankgv-post1}-\ref{bankgv-post2}). Additionally, \ttt{withdraw}'s postcondition ensures the subtrahend's balance remains positive as well. Note, the \ttt{withdraw} example is contrived to better illustrate how \gviper works and its interesting aspects.

\paragraph{Well-formedness of user written specifications}
\gviper begins static verification by first checking user written specifications, like predicate bodies, preconditions, and postconditions, for well-formedness. That is, user specifications must be self-framed and cannot contain duplicate accessibility predicates or predicates joined by the separating conjunction $\&\&$. Self-framing from IDF \cite{smans2009implicit} simply means that a formula must contain accessibility predicates for any heap locations accessed in the formula. In gradual verification \cite{wise2020gradual}, $\qm$ can represent these accessibility predicates. For example, in the \ttt{withdraw} program \ttt{geqTo}'s body (lines \ref{bankgv-pred1-body1}-\ref{bankgv-pred1-body2}) is self-framed, because $\qm$ can represent $\phiAcc{\ttt{a1.balance}}$ and $\phiAcc{\ttt{a2.balance}}$ to frame \ttt{a1.balance} and \ttt{a2.balance}. On the other hand, \ttt{positive}'s body (line \ref{bankgv-pred2}) is classically self-framed as it explicitly contains $\phiAcc{\ttt{a.balance}}$ to frame \ttt{a.balance}. All of the user written formulas, which are \ttt{geqTo}'s body (lines \ref{bankgv-pred1-body1}-\ref{bankgv-pred1-body2}), \ttt{positive}'s body (line \ref{bankgv-pred2}), \ttt{withdraw}'s precondition (line \ref{bankgv-pre}), and its postcondition (lines \ref{bankgv-post1}-\ref{bankgv-post2}), are well-formed.

Next, \gviper optimistically statically verifies each function in the given program, \eg the \ttt{withdraw} function in our running example. This involves symbolically executing the function from top to bottom and tracking information in a symbolic state. Information is gathered from the execution of both specifications and code, and proof obligations are established by the symbolic state. If any obligations are established optimistically, corresponding run-time checks are stored in the symbolic state. \fig~\ref{fig:ex-bank-gv-table} displays the contents of the symbolic state at every program point (marked by program lines) during the verification of \ttt{withdraw}. In general, a symbolic state can be thought of as writing an intermediate logical formula in a special form. We will discuss the contents of a symbolic state in more detail as we work through the \ttt{withdraw} example.

\paragraph{Producing a precondition}
At the start of \ttt{withdraw} (lines \ref{bankgv-withdraw-start}-\ref{bankgv-unfold}), information in the precondition, \eg \ttt{geqTo(a1,a2)}, is \emph{produced} or translated into an empty symbolic state resulting in the first state in the table in \fig~\ref{fig:ex-bank-gv-table}. As with formulas, symbolic states may be imprecise or not, meaning information may be missing from the state due to imprecision. In fact, you can think of an imprecise symbolic state as representing an imprecise intermediate formula. Here, the precondition \ttt{geqTo(a1,a2)} is precise,\footnote{{\disableTttResize Note, precision in a static context as in \gviper means the formula does not contain $\qm$ at the top-level. Predicates are treated as black-boxes, so even if their bodies are imprecise, as with \ttt{geqTo(a1,a2)}, a formula containing them, such as \ttt{geqTo(a1,a2)}, can be precise.}} so the state also remains precise. Had the precondition been imprecise, then the state would also become imprecise. Local variables are mapped to symbolic values in a symbolic, variable store. Since \ttt{a1} and \ttt{a2} are arguments to withdraw and \ttt{res} the return value, they are all assigned fresh symbolic values \ttt{t1}, \ttt{t2}, and \ttt{t3} respectively in the store. Then, permissions like accessibility predicates and predicates can be stored in a symbolic heap (either the optimistic heap or heap) in terms of the symbolic values. We call symbolic versions of permissions \emph{heap chunks}. Since \ttt{geqTo(a1,a2)} is concretely known it is added directly to the heap as the heap chunk \ttt{geqTo(t1,t2)}. An important invariant of the heap is that permissions in it are guaranteed to be separated in memory, \ie when they are joined by the separating conjunction they return true. The optimistic heap contains heap chunks for accessibility predicates that are optimistically assumed during verification and is introduced in this work to reduce the number of run-time checks produced by \gviper. We will see how this works as we continue to discuss the \ttt{withdraw} example. For now, the optimistic heap is empty. Similarly, both the path condition and set of run-time checks both remain empty. The path condition contains constraints on symbolic values that have been collected on the current verification path. The precondition \ttt{geqTo(t1,t2)} only contains permission information, so the path condition is empty. Further, producing a formula into the symbolic state does not introduce any run-time checks.

\paragraph{Unfolding a predicate}
Next, \gviper executes the unfold statement on line \ref{bankgv-unfold} causing the predicate \ttt{geqTo(a1,a2)} to be \emph{consumed} and then its body to be \emph{produced} into the state on lines \ref{bankgv-unfold}-\ref{bankgv-if}. In general, consuming a formula 1) checks whether the formula is established by the symbolic state, 2) generates minimized run-time checks for the state to establish the formula soundly, and 3) removes permissions asserted in the formula from the symbolic state. That is, consume is \gviper's mechanism for checking proof obligations; and as we will see, is used a few different times throughout the verification of \ttt{withdraw}. Here, since \ttt{geqTo(t1,t2)} is in the heap, \ttt{geqTo(a1,a2)} is established by the symbolic state and no run-time checks are required. It is then removed from the heap as it is ``consumed". After consumption, \ttt{geqTo(a1,a2)}'s body (lines \ref{bankgv-pred1-body1}-\ref{bankgv-pred1-body2}) is produced into the current state (the one without \ttt{geqTo(t1,t2)}). The body of \ttt{geqTo} is imprecise, so the symbolic state is made imprecise (as seen in \fig~\ref{fig:ex-bank-gv-table} at lines \ref{bankgv-unfold}-\ref{bankgv-if}). The rest of \ttt{geqTo}'s body is a boolean expression constraining \ttt{a1} and \ttt{a2}'s account balances: \ttt{a1}'s balance is greater than or equal to \ttt{a2}'s balance and \ttt{a2}'s balance is positive. Before adding these constraints to the path condition, \gviper first looks for heap chunks in the current symbolic state corresponding to accessibility predicates that frame \ttt{a1.balance} and \ttt{a2.balance} in \ttt{geqTo}'s body. Both the heap and optimistic heap are empty, but the state is imprecise so the missing heap chunks are optimistically assumed to be in the state. In fact, it is sound to make this assumption without any run-time checks, because we are producing (rather than consuming) the predicate body. As a result, fresh symbolic values \ttt{p1} and \ttt{p2} for \ttt{a1.balance} and \ttt{a2.balance} respectively are generated and used to record constraints on the balances in the path condition. \gviper also records that the receivers \ttt{a1} and \ttt{a2} are non-null in the path condition, because \gviper assumed they can be safely de-referenced. Finally, \gviper adds heap chunks $\phiAcc{\ttt{t1,balance,p1}}$ and $\phiAcc{\ttt{t2,balance,p2}}$ to the optimistic heap to 1) record the mappings of locations to their values and 2) avoid producing run-time checks for the accessibility predicates later in the program. These heap chunks are added to the optimistic heap rather than the heap, because \ttt{getTo}'s body does not specify whether or not \ttt{a1} (\ttt{t1}) or \ttt{a2} (\ttt{t2}) alias. So adding them to the heap would break the heap's invariant. The optimistic heap, however, does not maintain any invariants. It is also convenient to store optimistic heap chunks in their own heap signaling that they are available due to optimism in the verification. The final symbolic state after consuming \ttt{geqTo(a1,a2)} and producing its body is given in \fig~\ref{fig:ex-bank-gv-table} lines \ref{bankgv-unfold}-\ref{bankgv-if}.

\paragraph{Branching} After the unfold on line \ref{bankgv-unfold}, \gviper reaches the start of the if statement on the following line \ref{bankgv-if}. As is common with static verifiers based on symbolic execution, \gviper's execution branches at if statements. Execution also branches at other conditioned points, such as logical conditionals and loops. \gviper analyzes the \emph{then} branch (lines \ref{bankgv-if}-\ref{bankgv-else}) under the assumption the condition \ttt{a1 == \enull || a2 == \enull} is true, and the \emph{else} branch (lines \ref{bankgv-else}-\ref{bankgv-if-end}) under the assumption \ttt{a1 == $\enull$ || a2 == $\enull$} is false. These assumptions are added to the path condition for each execution path respectively. However, in our example the symbolic state going into the if statement (\fig~\ref{fig:ex-bank-gv-table} lines \ref{bankgv-unfold}-\ref{bankgv-if}) states that both \ttt{a1} and \ttt{a2} are non-null. So the \emph{then} branch is infeasible, and \gviper prunes this execution path resulting in the blank symbolic states in \fig~\ref{fig:ex-bank-gv-table} from lines \ref{bankgv-if}-\ref{bankgv-else}. Both accounts being non-null means the \emph{else} branch condition for sure holds and so execution proceeds down this branch without any changes to the symbolic state. That is, for \ttt{withdraw} to be statically verified, this one execution path must successfully verify. If \gviper executed both branches, then determining verification success is a bit more complicated:
\begin{itemize}[leftmargin=14pt, topsep=2pt]
    \item If the current symbolic state is precise, then both execution paths must successfully verify. This is the default functionality in static verifiers.
    \item If the current symbolic state is imprecise, then verification succeeds when one or both paths successfully verify. When only one path succeeds and the state is imprecise, \gviper optimistically assumes the state contains information that forces program execution down the success path only at run time. This more permissive static functionality is critical for adhering to the gradual guarantee at branch points. To ensure the program will never actually execute the failing branch (\ie, to ensure soundness), \gviper adds a run-time check for the success path's condition at the branch point.
\end{itemize}

\paragraph{Variable assignment}
Let's look now at how \gviper verifies the \ttt{else} branch (lines \ref{bankgv-else}-\ref{bankgv-if-end}). At the variable assignment on line \ref{bankgv-newB}, \gviper first evaluates the right-hand expression to the symbolic value \ttt{p1 - p2}. To do this, \gviper first looks for heap chunks for \ttt{a1.balance} and \ttt{a2.balance} in the current symbolic heaps (\fig~\ref{fig:ex-bank-gv-table} lines \ref{bankgv-else}-\ref{bankgv-newB}) to both frame the locations and get their values. 
Both heap chunks are in the optimistic heap, so no run-time checks are required for framing and \ttt{p1} and \ttt{p2} are used in the evaluation of the right-hand expression. Note, if \gviper did not add the aforementioned heap chunks to the optimistic heap when producing the body of \ttt{geqTo(a1,a2)}, then \gviper would create run-time checks for them here in the program. However, these checks would be duplicates, because \gviper also checks that these heap chunks are available when ensuring the precondition \ttt{getTo(a1,a2)} holds in client contexts at calls to \ttt{withdraw}. So sound tracking of heap chunks in an optimistic heap has helped us avoid duplicating run-time checks! Finally, \gviper adds a new mapping to the variable store for \ttt{newB} and its fresh symbolic value \ttt{t4}; and then, adds the constraint \ttt{t4 = p1 - p2} to the path condition to record information from the assignment in the symbolic state (\fig~\ref{fig:ex-bank-gv-table} lines \ref{bankgv-newB}-\ref{bankgv-fieldassign}).

\paragraph{Field assignment}
Next, \ttt{a1.balance} is assigned \ttt{newB}'s value in the field assignment on line \ref{bankgv-fieldassign}. \gviper mimics this behavior symbolically by first pulling \ttt{newB}'s value \ttt{t4} from the symbolic state (\fig~\ref{fig:ex-bank-gv-table} lines \ref{bankgv-newB}-\ref{bankgv-fieldassign}). Then, \gviper looks for \ttt{a1.balance}'s heap chunk in the state for framing, and if there, removes the chunk as \ttt{a1.balance}'s value may change in the write, \ie $\phiAcc{\ttt{a1.balance}}$ is \emph{consumed}. \gviper also asserts that \ttt{a1} is non-null. The heap chunk for \ttt{a1.balance} is in the optimistic heap and \ttt{a1 != NULL} is in the path condition, so no run-time checks are needed here. Then, \ttt{a1.balance}'s heap chunk is removed from the state; and, unfortunately, this action causes \ttt{a2.balance}'s heap chunk to be removed from the state as well. \gviper does not know whether or not \ttt{a1} and \ttt{a2} alias, because this information does not appear in the current path condition and the optimistic heap does not maintain the separation invariant. Then, since the state is imprecise \gviper could assume that \ttt{a1.balance} and \ttt{a2.balance} refer to the same heap location, \ie \ttt{a1} and \ttt{a2} are aliased. In this case, removing \ttt{a1.balance}'s heap chunk requires also removing \ttt{a2.balance}'s heap chunk as \ttt{a2.balance} may have also changed with the write. On the other hand, the case where they do not alias and \ttt{a2.balance}'s heap chunk can stay in the optimistic heap is also possible from \gviper's perspective. To simply and soundly cover both cases \gviper removes \ttt{a2.balance}'s heap chunk by default when alias information is unknown. As we will see next, this comes at the cost of additional run-time checks later in the verification of \ttt{withdraw}. Finally, \gviper \emph{produces} a new heap chunk for \ttt{a1.balance} into the state to track its new, fresh symbolic value \ttt{p3} after the write and updates the path condition with the assignment information \ttt{p3 = t4} (\fig~\ref{fig:ex-bank-gv-table} lines \ref{bankgv-fieldassign}-\ref{bankgv-folda1}).

\paragraph{Folding a predicate}
After \ttt{a1.balance} is assigned a new balance, \gviper executes the fold statement on line \ref{bankgv-folda1}. Folding a predicate is similar to unfolding a predicate except that the functionality is reversed: \ttt{positive(a1)}'s body is \emph{consumed} from the current state (\fig~\ref{fig:ex-bank-gv-table} lines \ref{bankgv-fieldassign}-\ref{bankgv-folda1}) and then \ttt{positive(a1)} is \emph{produced} into the state after consumption (\fig~\ref{fig:ex-bank-gv-table} lines \ref{bankgv-folda1}-\ref{bankgv-folda2}). The body of \ttt{positive(a1)} is $\phiAcc{\ttt{a1.balance}} \phiAnd \ttt{a1.balance} >= 0$, so $\phiAcc{\ttt{a1.balance}}$ is consumed first then \ttt{a1.balance >= 0} second. The heap chunk $\phiAcc{\ttt{t1,balance,p3}}$ corresponding to $\phiAcc{\ttt{a1.balance}}$ is in the heap and \ttt{a1 != null} holds in the path condition, so no run-time check is required for consuming $\phiAcc{\ttt{a1.balance}}$. Then, \ttt{a1.balance}'s heap chunk is removed from the heap as seen in \fig~\ref{fig:ex-bank-gv-table} lines \ref{bankgv-folda1}-\ref{bankgv-folda2}. Next, the boolean expression \ttt{a1.balance >= 0} is evaluated to its symbolic value \ttt{p3 >= 0}. Recall that to do this, \gviper must look up a heap chunk for \ttt{a1.balance} in the symbolic state to both frame the heap location and get its value. However, \gviper just removed this heap chunk from the current state due to the left-to-right execution of consume. To solve this issue, \gviper looks for framing and value information in the state before the fold, \ie the state before consumption at lines \ref{bankgv-fieldassign}-\ref{bankgv-folda1} in \fig~\ref{fig:ex-bank-gv-table}. As we know, \ttt{a1.balance}'s heap chunk is in this state and maps \ttt{a1.balance} to the value \ttt{p3}, so no run-time check is needed for framing. Then, \ttt{p3 >= 0} is asserted against the current path condition. Since \ttt{p1 >= p2 >= 0} and \ttt{p3 = t4 = p1 - p2} are in the path condition, \ttt{p3} is clearly greater than or equal to 0 and is proven directly by the path condition. That is, no run-time check is needed for \ttt{p3 >= 0}. Finally, \ttt{positive(a1)} is produced into the current state, which adds \ttt{positive(t1)} to the heap resulting in the final version of the state in \fig~\ref{fig:ex-bank-gv-table} lines \ref{bankgv-folda1}-\ref{bankgv-folda2}.

Next, \gviper executes the second fold statement on line \ref{bankgv-folda2} with the aforementioned state. This fold statement consumes \ttt{positive(a2)}'s body and then produces \ttt{positive(a2)} into the state. So as before, \gviper first consumes $\phiAcc{\ttt{a2.balance}}$ and then \ttt{a2.balance >= 0}. The receiver \ttt{a2} is proved to be non-null by the path condition; however this time, the heap chunk for $\phiAcc{\ttt{a2.balance}}$ is not in either of the heaps. Fortunately for us, the state is imprecise and can optimistically contain this heap chunk, so a run-time check is produced for $\phiAcc{\ttt{a2.balance}}$ as seen in the state after folding \ttt{positive(a2)} in \fig~\ref{fig:ex-bank-gv-table} lines \ref{bankgv-folda2}-\ref{bankgv-nonnull}. Since this run-time check occurs down the \ttt{else} branch of the if statement in \ttt{withdraw}, branch information, \eg $l_{bc}, \neg\ttt{(a1 = null || a2 = null)}$, is included with the check. The location $l_{bc}$ specifies where the branch point originated in the program, \eg line \ref{bankgv-if}, and $\neg\ttt{(a1 = null || a2 = null)}$ is the assumption made at the branch point for the current execution path. Additionally, $l_{c1}$ contains the location where the check itself is required in the program, \eg line \ref{bankgv-folda2}. While it does not happen in the \ttt{withdraw} example, sometimes different checks are required at the same program point down different execution paths. So, \gviper attaches branch information for the entire execution path to each run-time check to allow \gvc (or other frontends) to apply checks only on the execution path they are required. This ensures soundness and reduces run-time checking. Now, \gviper removes $\phiAcc{\ttt{a2.balance}}$ from the current state (\fig~\ref{fig:ex-bank-gv-table} lines \ref{bankgv-folda1}-\ref{bankgv-folda2}), which actually causes \ttt{positive(t1)} to be removed from the heap as well. Predicates are treated as black boxes in \gviper; so unless told otherwise, \gviper conservatively assumes $\phiAcc{\ttt{a2.balance}}$ is in \ttt{positive(t1)} and removes \ttt{positive(t1)} from the heap alongside $\phiAcc{\ttt{a2.balance}}$. The only way \gviper can guarantee $\phiAcc{\ttt{a2.balance}}$ is not in $\ttt{positive(t1)}$ is if a heap chunk for \ttt{a2.balance} and \ttt{positive(t1)} both exist in the heap, as the heap maintains the separation invariant. In this case, \ttt{positive(t1)} can remain in the heap while only $\phiAcc{\ttt{a2.balance}}$ is removed. Of course, in our example the heap chunk for \ttt{a2.balance} is definitely not in the heap, so \ttt{postive(t1)} is removed.
 
Continuing, \gviper consumes \ttt{a2.balance >= 0}, which first looks for a heap hunk to frame \ttt{a2.balance} in the state before the consume (\fig~\ref{fig:ex-bank-gv-table} lines \ref{bankgv-folda1}-\ref{bankgv-folda2}). However, neither of the heaps contain a heap chunk for \ttt{a2.balance}. As before, \gviper uses imprecision to optimistically assume the heap chunk is in the state and produces a run-time check for $\phiAcc{\ttt{a2.balance}}$. Since this run-time check for the same location already exists in the state, the two checks are condensed into the first one. 
Then, \gviper returns a fresh symbolic value for \ttt{a2.balance}, say \ttt{p4}, to evaluate \ttt{a2.balance >= 0} down to \ttt{p4 >= 0}. Note, \gviper can only return a fresh value here, because the heaps do not contain a heap chunk recording \ttt{a2.balance}'s value in the state. Unfortunately, this means \gviper cannot prove \ttt{a2.balance >= 0} holds as no constraints exist for \ttt{p4} in the path condition. But, this also means \ttt{p4 >= 0} does not contradict existing information in the path condition. So, imprecision in the state can optimistically represent \ttt{p4 >= 0} and a run-time check for \ttt{a2.balance >= 0} is generated as seen in \fig~\ref{fig:ex-bank-gv-table} lines \ref{bankgv-folda2}-\ref{bankgv-nonnull}. A few things of note here:
\begin{itemize}[leftmargin=14pt, topsep=2pt]
    \item Run-time checks are originally computed in terms of symbolic values, \eg \ttt{p4 >= 0}, but are ultimately replaced with counterparts written in terms of program variables, \eg \ttt{a2.balance >= 0}. This replacement by the \emph{translate} function in \gviper simplifies the implementation of run-time checks for frontends like \gvc, which operate on program variables and concrete values not symbolic values. The \ttt{translate} function uses mappings in the symbolic heaps and store to reverse the symbolic execution. Special considerations are made for fresh symbolic values like \ttt{p4}, aliasing between object values, and different variable contexts.
    \item On another note, if consuming $\phiAcc{\ttt{a1.balance}}$ at the field assignment on line \ref{bankgv-fieldassign} did not also consume the heap chunk for \ttt{a2.balance}, then the run-time checks for $\phiAcc{\ttt{a2.balance}}$ and \ttt{a2.balance >= 0} would not be necessary. \gviper conservatively assumed \ttt{a1} and \ttt{a2} were aliased at the consume, so it removed both chunks from the state. However, in practice \ttt{a1} and \ttt{a2} are likely to be distinct objects; and in fact, folding \ttt{positive(a1)} then \ttt{positive(a2)} is a good sign the developer of \ttt{withdraw} expects \ttt{a1} and \ttt{a2} to be distinct. In this case, \ttt{a2.balance}'s heap chunk does not need to be removed making the aforementioned run-time checks unnecessary. Unfortunately, since we designed \gviper to be conservative for simplicity, these run-time checks are only eliminated when the developer explicitly specifies that \ttt{a1} and \ttt{a2} are not aliased, such as in the precondition of \ttt{withdraw}. So, we are trading more optimal run-time checks for simplicity in our consume algorithm.
    \item While it does not happen here in our \ttt{withdraw} example, there may be times where parts of a symbolic, boolean expression are proven statically and the rest optimistically. In this case, \gviper re-writes the expression into conjunctive normal form and computes the conjuncts in this form that cannot be proven statically by the path condition. These conjuncts (after translation) will then be checked at run time. We call this process computing the \emph{difference} between the expression and the path condition, and it results in minimized run-time checks given statically available information.
\end{itemize}
Finally, a heap chunk for \ttt{positive(a2)} (\eg \ttt{positive(t2)}) is produced into the heap resulting in the final form of the state in \fig~\ref{fig:ex-bank-gv-table} lines \ref{bankgv-folda2}-\ref{bankgv-nonnull}.

\paragraph{Return value assignment} Then, \gviper reaches the variable assignment on line \ref{bankgv-nonnull}, which assigns \ttt{a1} to \ttt{res}---the return value of \ttt{withdraw}. \gviper first looks up the symbolic value \ttt{t1} for \ttt{a1} and then the symbolic value \ttt{t3} for \ttt{res} in the variable store. \gviper stores the information \ttt{t3 = t1} from the assignment in the path condition resulting in the next symbolic state in \fig~\ref{fig:ex-bank-gv-table} lines \ref{bankgv-nonnull}-\ref{bankgv-if-end}.

\paragraph{Consuming a postcondition}
Finally, \gviper reaches the end of \ttt{withdraw} down its one and only execution path on line \ref{bankgv-if-end}. So the last thing \gviper must do to verify the function, is to \emph{consume} the postcondition $\qm \phiAnd \phiAcc{\ttt{positive(a2)}} \phiAnd \phiAcc{\ttt{positive(res)}}$ (lines \ref{bankgv-post1}-\ref{bankgv-post2}) in the current symbolic state (\fig~\ref{fig:ex-bank-gv-table} lines \ref{bankgv-nonnull}-\ref{bankgv-if-end}). \gviper begins by first consuming \ttt{positive(a2)} then \ttt{positive(res)}. The heap chunk for \ttt{positive(a2)} is in the heap, so no run-time check is needed for it. Then \ttt{positive(t2)} is removed from the heap leaving both symbolic heaps empty. As a result (and because the state is imprecise), consuming \ttt{positive(res)} in the next step results in a run-time check for the predicate as seen in the final set of run-time checks required for \ttt{withdraw} given in \fig~\ref{fig:ex-bank-gv-table} line \ref{bankgv-if-end}. Note, consuming $\phiAcc{\ttt{positive(a2)}} \phiAnd \phiAcc{\ttt{positive(res)}}$ requires both consuming the predicates individually (which we've done) and ensuring that one predicate does not access heap locations overlapping with the other (in adherence with the separating conjunction \ttt{\&\&}). Unfortunately, the state does not contain enough information to prove this fact statically, \eg only the heap chunk for \ttt{positive(a2)} appears in the heap, but the state is imprecise! So when \gviper optimistically assumes \ttt{positive(res)} holds, it also assumes \ttt{positive(res)} is separated from \ttt{positive(a2)}. \gviper flags \ttt{positive(res)}'s run-time check with this additional check for \gvc to handle. Additionally, after consuming the static part of an imprecise formula, \eg $\phiAcc{\ttt{positive(a2)}} \phiAnd \phiAcc{\ttt{positive(res)}}$ in \ttt{withdraw}'s postcondition, \gviper makes the state imprecise and empties both symbolic heaps. The $\qm$ in the imprecise formula can represent any permission available in the state, so they must be removed by \emph{consume}.

\paragraph{Takeaways} To summarize, \gviper statically verifies the \ttt{withdraw} function successfully, and produces run-time checks for $\phiAcc{\ttt{a2.balance}}$ before line \ref{bankgv-fieldassign}, \ttt{a2.balance >= 0} also before line \ref{bankgv-fieldassign}, and \ttt{positive(res)} at the end of \ttt{withdraw} (line \ref{bankgv-if-end}). The \ttt{withdraw} function will be completely verified if these checks succeed at run time. During our discussion of the \ttt{withdraw} function, we highlighted a number of technical challenges addressed and solutions developed related to designing and implementing \gviper. One of our goals was for \gviper to minimize run-time checks as much as possible without using highly complex algorithms. For this we introduced the optimistic heap, which tracks heap chunks that are optimistically assumed during static verification and can be soundly used to reduce run-time checking in successive program statements from where they originated. In \ttt{withdraw}, we saw the heap chunks for \ttt{a1.balance} and \ttt{a2.balance}, which were added to the optimistic heap during the production of \ttt{geqTo(a1,a2)}'s body (line \ref{bankgv-unfold}), be used to eliminate duplicate run-time checks at the assignment on line \ref{bankgv-newB}. We had to make careful considerations for the separating conjunction and removal of heap chunks at consumes to ensure sound tracking of heap chunks in the optimistic heap. We also defined and implemented the \ttt{diff} function, which utilizes conjunctive normal form to optimize run-time checks for boolean expressions. Finally, \gviper conservatively removes heap chunks from the symbolic heaps that may alias with other heap chunks removed at a \emph{consume}. We saw in \ttt{withdraw} that this comes at the cost of additional run-time checks: consuming \ttt{a1.balance}'s heap chunk at the field assignment on line \ref{bankgv-fieldassign} also consumed \ttt{a2.balance}'s heap chunk resulting in run-time checks for $\phiAcc{\ttt{a2.balance}}$ and \ttt{a2.balance >= 0} before line \ref{bankgv-fieldassign}. That is, we are trading more optimal run-time checks for simplicity in our consume algorithm.

Another goal for \gviper, is for it to use symbolic execution for static reasoning. We accomplished this goal, but not without dealing with some technical challenges. Symbolic execution based static verifiers generate and discharge proof obligations written in terms of symbolic values, causing \gviper, which extends this system, to follow suit. As a result, \gviper naturally generates run-time checks written in terms of symbolic values as well. Unfortunately, dynamic verifiers only operate on program variables and concrete values not symbolic ones. To bridge this gap between the static and dynamic systems, we implemented a \ttt{translate} function in \gviper that re-writes run-time checks containing symbolic values to ones containing program variables and concrete values while being careful about aliases. Finally, execution splitting at branch points led to some trickiness in gradual verification. Different run-time checks may appear at the same program point along different execution paths, so we augmented \gviper to attach branching information to run-time checks. We also augmented \gviper to be more optimistic about verification success when dealing with failing execution paths in the presence of imprecision. This was done in compliance with the gradual guarantee \cite{wise2020gradual}.

\subsection{\gviper: Implemented Algorithm}
\label{sec:gviper-sv}
In this section, we formalize the symbolic execution algorithm implemented by \gviper. A high-level description of how it works is given in \S\ref{sec:gviper-sv-highlevel}. Our algorithm extends Viper's symbolic execution algorithm, and so \gviper's design is heavily influenced by \citet{MuellerSchwerhoffSummers16}'s work.
Like Viper, \gviper's algorithm consists of 4 major functions: \predicate{eval}, \predicate{produce}, \predicate{consume}, and \predicate{exec}. The functions evaluate expressions, produce (inhale) and consume (exhale) formulas, and execute program statements respectively. Following Viper's lead, our 4 functions are defined in continuation-passing style, where the last argument of each of the aforementioned functions is a continuation $Q$. The continuation is a function that represents the remaining symbolic execution that still needs to be performed. Note that the last continuation returns a boolean ($\lambda~\_~.~\success$ or $\lambda~\_~.~\failure$), indicating whether or not symbolic execution was successful.

The rest of this section is outlined as follows. Run-time checks and the collections that hold them are described in \S\ref{sec:gviper-runtimechecks}. We define symbolic states in \S\ref{sec:gviper-symstate} and preliminaries in \S\ref{sec:gviper-prelims}. Finally, the 4 major functions of our algorithm are given in their own sections: \predicate{eval} \S\ref{sec:gviper-eval}, \predicate{produce} \S\ref{sec:gviper-produce}, \predicate{consume} \S\ref{sec:gviper-consume}, and \predicate{exec} \S\ref{sec:gviper-exec}. 
Throughout this section, we make clear where Viper has been extended to support imprecise formulas with yellow highlighting in figures. We also use blue highlighting to indicate extensions for run-time check generation and collection.

\subsubsection{\textbf{Run-time checks}} 
\label{sec:gviper-runtimechecks}

Run-time checks produced by \gviper are collected in the $\rchecksall$ set. A run-time check is a 4-tuple $(\bcs_c,~ \origin_c,~ \location_c,~ \phi_c)$, where $\bcs_c$ is a set of branch conditions, $\origin_c$ and $\location_c$ denote where the run-time check is required in the program, and $\phi_c$ is what must be checked. A branch condition in $\bcs_c$ is also a tuple of $(\origin_e,~ \location_e,~ e)$, where $\origin_e$ and $\location_e$ define the program location at which \gviper's execution branches on the condition $e$. A $\location$ is the AST element in the program where the branch or check occurs, denoted as a formula $\phi_l$. Sometimes, the condition being checked is defined elsewhere in the program (\eg in the precondition of a method) but we need to relate it to the method being verified.  The $\origin$ is used to do this.  It is $\none$ when the condition is in the method being verified; otherwise, it contains a method call, fold, unfold, or special loop statement from the method being verified that referenced the check specified in the $\location$.  An example run-time check is: $(\{ (\none,~ x > 2,~ \neg (x > 2)) \},~ \sCall{z}{m}{y},~ \phiAcc{y.f},~ \phiAcc{y.f})$. The check is for accessing $y.f$, and it is required for $m$'s precondition element $\phiAcc{y.f}$ at the method call statement $\sCall{z}{m}{y}$. The check is only required when $\neg (x > 2)$, which is evaluated at the program point where the AST element $x > 2$ exists. Since $\neg (x > 2)$'s $\origin$ is $\none$, it comes from an if or assert statement.

Further, $\rchecks$ is used to collect run-time checks down a particular execution path in \gviper. $\rchecks$ is a 3-tuple $(\bcs_p,~ \origin_p,~ \rcs_p)$ where $\bcs_p$ is the set of branch conditions collected down the execution path $p$, $\origin_p$ is the current origin that is set and reset during execution, and $\rcs$ is the set of run-time checks collected down $p$.
Two auxiliary functions are used to modify $\rchecks$: \predicate{addcheck} and \predicate{addbc}. The \predicate{addcheck} function takes an $\rchecks$ collection $\rchecks_{arg}$, a $\location$ $\phi_l$ for a check, and the check itself, and returns a copy of $\rchecks_{arg}$ with the run-time check added to $\rchecks_{arg}.\rcs$. If necessary, \predicate{addcheck} uses $\rchecks_{arg}.\origin$ and substitution to ensure $\phi_l$ and the check refer to the correct context. For example, let $\phi_l$ and check $\phi_c$ come from asserting a precondition for $\sCall{z}{m}{y}$. Then, \predicate{addcheck} performs the substitutions: $\phi_l[t \mapsto m_{arg}]$ (precondition declaration context) and $\phi_c[t \mapsto y]$ (method call context) where $t$ is the symbolic value for $y$. The \predicate{addbc} function operates similarly to \predicate{addcheck} but for branch conditions.

\subsubsection{\textbf{Symbolic State}} 
\label{sec:gviper-symstate}
We use $\sigma \in \Sigma$ to denote a symbolic state, which is a 6-tuple\\ $(\isimp,~\oh,~h,~\gamma,~\pi,~ \rchecks)$ consisting of a boolean $\isimp$, a symbolic heap $\oh$, another symbolic heap $h$, a symbolic store $\gamma$, a path condition $\pi$, and a collection $\rchecks$ (defined in \S\ref{sec:gviper-runtimechecks}).
The boolean $\isimp$ records whether or not the state is imprecise, the symbolic store $\gamma$ maps local variables to their symbolic values, and the path condition $\pi$ (defined in \S\ref{sec:gviper-prelims}) contains constraints on symbolic values that have been collected on the current verification path.

A symbolic heap is a multiset of heap chunks for fields or predicates that are currently accessible. A field chunk $\chunk{id}{r}{\delta}$ (representing expression $r.id$) consists of the field name $id$, the receiver's symbolic value $r$, and the field's symbolic value $\delta$---also referred to as the \emph{snapshot} of a heap chunk. For a predicate chunk $\chunk{id}{args}{\delta}$, $id$ is the predicate name, $args$ is a list of symbolic values that are arguments to the predicate, and $\delta$ is the snapshot of the predicate. A predicate's snapshot represents the values of the heap locations abstracted over by the predicate. The symbolic, \emph{optimistic} heap $\oh$ contains heap chunks that are accessible due to optimism in the symbolic execution, while $h$ contains heap chunks that are statically accessible. Further, only $h$ maintains the invariant that its heap chunks are separated in memory, and thus, can be joined successfully by the separating conjunction.
The \emph{empty symbolic state} is\\ 
$\sigma_{0} = ( \isimp \assign \phiFalse,~ \oh \assign \emptyset,~ h \assign \emptyset,~ \gamma \assign \emptyset,~ \pi \assign \emptyset,$ $\rchecks \assign (\emptyset,~ \none,~ \emptyset) )$.

\subsubsection{\textbf{Preliminaries}}
\label{sec:gviper-prelims}
We introduce a few preliminary definitions here that will be helpful later.
A path condition $\pi$ is a stack of tuples $(id,~ bc,~ pcs)$. An $id$ is a unique identifier that determines the constraints on symbolic values that have been collected between two branch points in execution. The $bc$ entry is the symbolic value for the branch condition from the first of two branch points, and $pcs$ is the set of constraints that have been collected. Branch points can be from if statements and logical conditionals in formulas. Functions \predicate{pc-all}, \predicate{pc-add}, and \predicate{pc-push} manipulate path conditions and are formally defined in Appendix \fig~\ref{fig:pc-funcs}. The \predicate{pc-all} function collects and returns all the constraints in $\pi$, \predicate{pc-add} adds a new constraint to $\pi$, and \predicate{pc-push} adds a new stack entry to $\pi$. 
Similarly, snapshots for heap chunks have their own related functions: \emph{unit}, \emph{pair}, \emph{first}, and \emph{second}. The constant \emph{unit} is the empty snapshot, \emph{pair} constructs pairs of snapshots, and \emph{first} and \emph{second} deconstruct pairs of snapshots into their sub-parts.
Further, $\fresh$ is used to create fresh snapshots, symbolic values, and other identifiers depending on the context. The \predicate{havoc} function similarly updates a symbolic store by assigning a fresh symbolic value to each variable in a given collection of variables.
Finally, $\checkv{\pi}{t} = \phiImplies{\pcall{\pi}}{t}$ queries the underlying SAT solver to see if the given constraint $t$ is valid in a given path condition $\pi$ (\ie $\pi$ proves or implies $t$).

\subsubsection{\textbf{Symbolic execution of expressions}}
\label{sec:gviper-eval}
The symbolic execution of expressions by the eval function is defined in \fig~\ref{fig:eval-rules}. Using the current symbolic state, eval evaluates an expression to a symbolic value $t$ and returns $t$ and the current state to the continuation $Q$. Variable values are looked up in the symbolic store and returned. For $op(\overline{e})$, its arguments $\overline{e}$ are each evaluated to their symbolic values $\overline{t}$. A symbolic value $op'(\overline{t})$ is then created and returned with the state after evaluation. Each $op$ has a corresponding symbolic value $op'$ of the same arity. For example, $e_1 + e_2$ results in the symbolic value $add(t_1,t_2)$ where $e_1$ and $e_2$ evaluate to $t_1$ and $t_2$ respectively. 

Finally, the most interesting rule is for fields $e.f$. The receiver $e$ is first evaluated to $t$ resulting in a new state $\sigma_2$. Then, eval looks for a heap chunk for $t.f$ first in the current heap $h$.\footnote{Heap lookup in eval also looks for heap chunks that are aliases (according to the path condition) to the chunk in question.} If a chunk exists, then the heap read succeeds and $\sigma_2$ and the chunk's snapshot $\delta$ is returned to the continuation. If a chunk does not exist in $h$, then eval looks for a chunk in the optimistic heap $\oh$, and if found the chunk's snapshot is returned with $\sigma_2$. If a heap chunk for $t.f$ is not found in either heap, then the heap read can still succeed when $\sigma_2$ is imprecise. As long as $t \neq \enull$ does not contradict the current path condition $\sigma_2.\pi$ (the call to \predicate{assert}, Appendix \fig~\ref{fig:check-assert-funcs}), $\sigma_2$'s imprecision optimistically provides access to $t.f$. Therefore, a run-time check for $\phiAcc{e_t.f}$ is created and added to $\sigma_2$'s set of run-time checks (highlighted in blue). 
Note that $e_t.f$ is used in the check rather than $t.f$, because---unlike $t$ which is a symbolic value---the expression $e_t$ can be evaluated at run time. Specifically, \predicate{translate} (described in Appendix \fig~\ref{fig:algorithm-translate}) is called on $t$ with the current state $\sigma_2$ to compute $e_t$.
Additionally, the AST element $e.f$ is used to denote the check's location.

Afterwards, a fresh snapshot $\delta$ is created for $t.f$'s value, and a heap chunk $\chunk{f}{t}{\delta}$ for $t.f$ and $\delta$ is created and added to $\sigma_2$'s optimistic heap passed to the continuation. Similarly, the constraint $t \neq \enull$ is added to $\sigma_2$'s path condition. By adding $\chunk{f}{t}{\delta}$ to the optimistic heap, the following accesses of $t.f$ are statically verified by the optimistic heap, which reduces the number of run-time checks produced. 
Finally, verification of the heap read for $t.f$ fails when none of the aforementioned cases are true.
\fig~\ref{fig:eval-p-rules} and \fig~\ref{fig:eval-c-rules} in the Appendix define variants of \predicate{eval}, called \predicate{eval-p} and \predicate{eval-c}, that are used in \predicate{produce} and \predicate{consume} respectively. The \predicate{eval-p} variant does not introduce run-time checks and \predicate{eval-c} does not extend the optimistic heap and path condition, because the aforementioned functionalities are not needed in these contexts.

\begin{figure}[t]
{\scriptsize\ttfamily
\disableTttResize
\begin{alignat*}{2}  
  &\eval{\sigma}{t}{Q}
    &&= Q(\sigma,~t) \\
  &\eval{\sigma}{x}{Q}
    &&= Q(\sigma,~\sigma.\gamma(x)) \\
  &\eval{\sigma_1}{op(\overline{e})}{Q}
    &&=  \eval{\sigma_1}{\overline{e}}{(\lambda~ \sigma_2, ~\overline{t} ~.~ Q(\sigma_2,~ op'(\overline{t})))}\\
  &\eval{\sigma_1}{e.f}{Q}
    &&=  \eval{\sigma_1}{e}{(\lambda~ \sigma_2, ~t ~.~ \\
      &&&\quad \texttt{if}~~ (\exists~ \chunk{f}{r}{\delta} \in \sigma_2.h ~.~ \checkv{\sigma_2.\pi}{r = t}) ~~\ttt{then}~~\\
      &&&\qquad Q(\sigma_2,~\delta)\\
      &&&\quad \mlightyellow{\texttt{else if}~~ (\exists~ \chunk{f}{r}{\delta} \in \sigma_2.\oh ~.~ \checkv{\sigma_2.\pi}{r = t}) ~~\ttt{then}~~}\\
      &&&\qquad \mlightyellow{Q(\sigma_2,~\delta)}\\
      &&&\quad \mlightyellow{\texttt{else} ~~\texttt{if}~~ (\sigma_2.\isimp) ~~\ttt{then}~~}\\
      &&&\qquad \mlightyellow{res,~\_ \assign \assertv{\sigma_2.\isimp}{\sigma_2.\pi}{t \neq \enull}} \\
      &&&\qquad \mlightblue{e_t \assign \translate{\sigma_2}{t}} \\
      &&&\qquad \mlightblue{\rchecks' \assign \addcheck{\sigma_2.\rchecks}{e.f}{\phiAcc{e_t.f}}}\\
      &&&\qquad \mlightyellow{\delta \assign \fresh}\\
      &&&\qquad \mlightyellow{res \wedge Q(\sigma_2 \{ \oh \assign \sigma_2.\oh \cup \chunk{f}{t}{\delta},~ \pi \assign \pcadd{\sigma_2.\pi}{\{t \neq \enull\}}, ~\mlightblue{\rchecks \assign \rchecks'} \},~\delta)}\\
      &&&\quad \texttt{else}~ \failure}
\end{alignat*}
}
\footnotesize
\disableTttResize
\fbox{\begin{tabular}{llll}
\textcolor{light-yellow}{$\blacksquare$} & \small{Handles imprecision} &
\textcolor{light-blue}{$\blacksquare$} & \small{Handles run-time check generation and collection}
\end{tabular}}

\caption{Rules for symbolically executing expressions}
\label{fig:eval-rules}
\end{figure}

\subsubsection{\textbf{Symbolic production of formulas}}
\label{sec:gviper-produce}
\begin{figure}
{\scriptsize\ttfamily
\disableTttResize
\begin{alignat*}{2}
  &\mlightyellow{\produce{\sigma}{\withqm{\phi}}{\delta}{Q}} 
    &&= \produce{\sigma \{\isimp \assign \phiTrue \}}{\phi}{\second{\delta}}{Q}\\
  &\produce{\sigma_1}{e}{\delta}{Q} 
    &&= \hevalp{\sigma_1}{e}{(\lambda~ \sigma_2,~t ~.~ Q(\sigma_2 \{ \pi \assign \pcadd{\sigma_2.\pi}{\{t,\delta=unit\}}\}))}\\
  &\produce{\sigma_1}{\phiAcc{p(\overline{e})}}{\delta}{Q} 
    &&= \hevalp{\sigma_1}{\overline{e}}{(\lambda~ \sigma_2,~\overline{t} ~.~ Q(\sigma_2 \{ h \assign \sigma_2.h \uplus \chunk{p}{\overline{t}}{\delta} \})) } \\
  &\produce{\sigma_1}{\phiAcc{e.f}}{\delta}{Q} 
    &&= \hevalp{\sigma_1}{e}{(\lambda~ \sigma_2,~t ~.~ \\
      &&&\quad Q(\sigma_2 \{ h \assign \sigma_2.h \uplus \chunk{f}{t}{\delta},~ \pi \assign \pcadd{\sigma_2.\pi}{\{t \neq \enull \}} \})) }\\
  &\produce{\sigma_1}{\phi_1 \phiAnd \phi_2}{\delta}{Q} 
    &&= \produce{\sigma_1}{\phi_1}{\first{\delta}}{(\lambda~ \sigma_2 ~.~ \produce{\sigma_2}{\phi_2}{\second{\delta}}{Q})} \\
  &\produce{\sigma_1}{\phiCond{e}{\phi_1}{\phi_2}}{\delta}{Q} 
    &&= \hevalp{\sigma_1}{e}{(\lambda~ \sigma_2,~t ~.~ \\
    &&&\quad \hbranch{\sigma_2}{e}{t}{(\lambda~ \sigma_3 ~.~ \produce{\sigma_3}{\phi_1}{\delta}{Q})}{(\lambda~ \sigma_3 ~.~ \produce{\sigma_3}{\phi_2}{\delta}{Q})}
    )}
\end{alignat*}
}

\footnotesize
\disableTttResize
\fbox{\begin{tabular}{llll}
\textcolor{light-yellow}{$\blacksquare$} & \small{Handles imprecision} &
\textcolor{light-blue}{$\blacksquare$} & \small{Handles run-time check generation and collection}
\end{tabular}}

\caption{Rules for symbolically producing formulas}
\label{fig:produce-rules}
\end{figure}
Produce (\fig~\ref{fig:produce-rules}) is responsible for adding information to the symbolic state, in particular, the path condition and the heap $h$. Producing an imprecise formula makes the symbolic state imprecise.  The \predicate{produce} rule for an expression $e$ evaluates $e$ to its symbolic value and produces it into the path condition. The produce rules for accessibility predicates containing fields and predicates are similar, so we focus on the rule for fields only. The field $e.f$ in $\phiAcc{e.f}$ first has its receiver $e$ evaluated to a symbolic value $t$. Then, using the parameter $\delta$ a fresh heap chunk $\chunk{f}{t}{\delta}$ is created and added to the heap before invoking the continuation. Note, the disjoint union $\uplus$ ensures $\chunk{f}{t}{\delta}$ is not already in the heap before $\chunk{f}{t}{\delta}$ is added; otherwise, verification fails. Further, $\phiAcc{e.f}$ implies $e \neq \enull$ and so that fact is recorded in the path condition as $t \neq \enull$. When the separating conjunction $\phiCons{\phi_1}{\phi_2}$ is produced, $\phi_1$ is first produced into the symbolic state, followed by $\phi_2$. 
Finally, to produce a conditional, \gviper branches on the symbolic value $t$ for the condition $e$ splitting execution along two different paths. Along one path $\phi_1$ is produced into the state under the assumption that $t$ is true, and along the other path $\phi_2$ is produced under the $\neg t$ assumption. Both paths follow the continuation to the end of its execution, and a branch condition corresponding to the $t$ assumption made is added to the symbolic state. Paths are pruned when they are infeasible (the assumption about $t$ would contradict the current path conditions). Overall verification success is computed from the results of the two execution paths, and an imprecise state allows this computation to be optimistic when one path successfully verifies and the other doesn't. In this case, \predicate{branch} optimistically marks verification a success when normally it should fail, because the state may optimistically contain information that prunes the failure case. A run-time check is then added for the success path's condition to ensure soundness. This functionality is important for adhering to the gradual guarantee \cite{wise2020gradual}. The formal definition of \predicate{branch} is in \fig~\ref{fig:branch-func}, and other details for \predicate{branch} and \predicate{produce} are given in Appendix \S\ref{sec:appendix-produce}. Note, \predicate{produce} only adds run-time checks for branching to the symbolic state.

\begin{figure}[t]
\scriptsize\ttfamily
\disableTttResize
\begin{alignat*}{2}
  &\branch{\sigma}{e}{t}{Q_t}{Q_{\neg t}} = \\
    &\quad (\pi_T,~\rchecks_T) \assign (\pcpush{\sigma.\pi}{\fresh}{t},~\mlightblue{\addbc{\sigma.\rchecks}{e}{e}}) \\
    &\quad (\pi_F,~\rchecks_F) \assign (\pcpush{\sigma.\pi}{\fresh}{\neg t},~\mlightblue{\addbc{\sigma.\rchecks}{e}{\neg e}}) \\
    &\quad \mlightyellow{\ttt{if}~~ (\sigma.\isimp) ~~\ttt{then}}~~\\
    &\qquad \mlightyellow{res_T \assign (\ttt{if}~~ \neg \checkv{\sigma.\pi}{\neg t} ~~\ttt{then}~~ Q_t(\sigma\{ \pi \assign \pi_T,~ \mlightblue{\rchecks \assign \rchecks_T} \}) ~~\ttt{else}~~\failure)} \\
    &\qquad \mlightyellow{res_F \assign (\ttt{if}~~ \neg \checkv{\sigma.\pi}{t} ~~\ttt{then}~~ Q_{\neg t}(\sigma\{ \pi \assign \pi_F,~ \mlightblue{\rchecks \assign \rchecks_F} \}) ~~\ttt{else}~~\failure)} \\
    &\qquad \mlightblue{\ttt{if}~~ ((res_T \wedge \neg res_F) \vee (\neg res_T \wedge res_F)) ~~\ttt{then}}\\
    &\qquad\quad \mlightblue{\rchecks' \assign \addcheck{\sigma.\rchecks}{e}{(\ttt{if}~~(res_T)~~\ttt{then}~~e~~\ttt{else}~~\neg e)}} \\
    &\qquad\quad \mlightblue{\rchecksall \assign \rchecksall \cup \rchecks'.\rcs.\predicate{last}} \\
    &\qquad \mlightyellow{res_T \vee res_F} \\
    &\quad \mlightyellow{\ttt{else}}\\
    &\qquad (\ttt{if}~~ \neg \checkv{\sigma.\pi}{\neg t} ~~\ttt{then}~~ Q_t(\sigma\{ \pi \assign \pi_T,~ \mlightblue{\rchecks \assign \rchecks_T} \}) ~~\ttt{else}~~\success) \qquad \wedge \\
    &\qquad (\ttt{if}~~ \neg \checkv{\sigma.\pi}{t} ~~\ttt{then}~~ Q_{\neg t}(\sigma\{ \pi \assign \pi_F,~ \mlightblue{\rchecks \assign \rchecks_F} \}) ~~\ttt{else}~~\success)
\end{alignat*}

\fbox{\begin{tabular}{llll}
\textcolor{light-yellow}{$\blacksquare$} & \small{Handles imprecision} &
\textcolor{light-blue}{$\blacksquare$} & \small{Handles run-time check generation and collection}
\end{tabular}}

\caption{Formally defining the \predicate{branch} function}
\label{fig:branch-func}
\end{figure}

\subsubsection{\textbf{Symbolic consumption of formulas}}
\label{sec:gviper-consume}

\begin{figure}[t]
\scriptsize\ttfamily
\disableTttResize
\begin{alignat*}{2}
  &\consume{\sigma_1}{\theta}{Q}
    &&= \mlightyellow{\sigma_2 \assign \sigma_1 \{~ h,~ \pi \assign \consolidate{\sigma_1.h}{\sigma_1.\pi} ~\} } \\
    &&&\quad \consumep{\sigma_2}{\mlightyellow{\sigma_2.\isimp}}{\mlightyellow{\sigma_2.\oh}}{\sigma_2.h}{\theta}{(\lambda~ \sigma_3,~\mlightyellow{\oh'},~h_1,~\delta_1 ~.~ \\
    &&&\qquad Q(\sigma_3 \{ \mlightyellow{\oh \assign \oh'},~ h \assign h_1 \},~\delta_1))} \\
  &\mlightyellow{\consume{\sigma_1}{\withqm{\phi}}{Q}}
    &&= \mlightyellow{\sigma_2 \assign \sigma_1 \{~ h,~ \pi \assign \consolidate{\sigma_1.h}{\sigma_1.\pi} ~\} } \\
    &&&\quad \consumep{\sigma_2}{\mlightyellow{\phiTrue}}{\mlightyellow{\sigma_2.\oh}}{\sigma_2.h}{\phi}{(\lambda~ \sigma_3,~\mlightyellow{\oh'},~h_1,~\delta_1 ~.~ \\
    &&&\qquad Q(\sigma_3 \{ \mlightyellow{\isimp \assign \phiTrue,~ \oh \assign \emptyset,~ h \assign \emptyset} \},~\mlightyellow{pair(unit,~\delta_1)}))}
\end{alignat*}

\begin{alignat*}{2}  
  &\consumep{\sigma}{\mlightyellow{f_{\qm}}}{\mlightyellow{\oh}}{h}{(\mlightblue{e},t)}{Q} 
    &&= res,~\mlightblue{\overline{t}} \assign \mlightyellow{\assertv{\sigma.\isimp}{\sigma.\pi}{t}} \\
      &&&\quad \mlightblue{\rchecks' \assign \addcheck{\sigma.\rchecks}{e}{\translate{\sigma}{\overline{t}}}} \\
      &&&\quad res \wedge Q(\sigma\{\mlightblue{\rchecks \assign \rchecks'}\},~\mlightyellow{\oh},~h,~unit) \\
  &\consumep{\sigma_1}{\mlightyellow{f_{\qm}}}{\mlightyellow{\oh}}{h}{e}{Q}
    &&= \hevalpc{\sigma_1 \{\mlightyellow{\isimp \assign f_{\qm}} \}}{e}{(\lambda~ \sigma_2,~t ~.~ \\
      &&&\hspace{-0.1cm} \consumep{\sigma_2 \{\mlightyellow{\isimp \assign \sigma_1.\isimp} \}}{\mlightyellow{f_{\qm}}}{\mlightyellow{\oh}}{h}{(\mlightblue{e},t)}{Q})}\\
  &\consumep{\sigma_1}{\mlightyellow{f_{\qm}}}{\mlightyellow{h_{\qm}}}{h}{\phiAcc{e.f}}{Q} 
    &&= \hevalpc{\sigma_1 \{\mlightyellow{\isimp \assign f_{\qm} }\}}{e}{(\lambda~ \sigma_2,~t ~.~ \\
      &&&\quad \mlightyellow{\sigma_3 \assign \sigma_2 \{\isimp \assign \sigma_1.\isimp \}} \\
      &&&\quad res,~\mlightblue{\overline{t}} \assign \mlightyellow{\assertv{\sigma_3.\isimp}{\sigma_3.\pi}{t \neq \enull}} \\
      &&&\quad res ~~\wedge~~ (\\
      &&&\quad \mlightblue{\rchecks' \assign \addcheck{\sigma_3.\rchecks}{\phiAcc{e.f}}{\translate{\sigma_3}{\overline{t}}}} \\
      &&&\quad \mlightyellow{(h_1,~\delta_1,~b_1) \assign \heaprema{\sigma_3.\isimp}{h}{\sigma_3.\pi}{f(t)}} \\
      &&&\quad \mlightyellow{\ttt{if}~~ (\sigma_3.\isimp) ~~\ttt{then}} \\
      &&&\qquad \mlightyellow{(\oh',~\delta_2,~b_2) \assign \heaprema{\sigma_3.\isimp}{\oh}{\sigma_3.\pi}{f(t)}} \\
      &&&\qquad \mlightblue{\ttt{if}~~(b_1 = b_2 = \phiFalse)~~\ttt{then}} \\
      &&&\qquad\quad \mlightblue{\rchecks'' \assign \addcheck{\rchecks'}{\phiAcc{e.f}}{\phiAcc{\translate{\sigma_3}{t}.f}}} \\
      &&&\qquad \mlightblue{\ttt{else}~~\rchecks'' \assign \rchecks'} \\
      &&&\qquad \mlightyellow{Q(\sigma_3\{\mlightblue{\rchecks \assign \rchecks''}\},~h_{\qm}',~h_1,~(\ttt{if}~~(b_1)~~\ttt{then}~~\delta_1~~\ttt{else}~~\delta_2))} \\
      &&&\quad \mlightyellow{\ttt{else}~~\ttt{if}~~ (b_1) ~~\ttt{then}~~ Q(\sigma_3\{\mlightblue{\rchecks \assign \rchecks'}\},~\sigma_3.\oh,~h_1,~\delta_1)} \\
      &&&\quad \mlightyellow{\ttt{else}~~ \failure}
      ))} 
\end{alignat*}

\fbox{\begin{tabular}{llll}
\textcolor{light-yellow}{$\blacksquare$} & \small{Handles imprecision} &
\textcolor{light-blue}{$\blacksquare$} & \small{Handles run-time check generation and collection}
\end{tabular}}

\caption{\emph{Select rules} for symbolically consuming formulas}
\label{fig:consume-rules-select}
\end{figure}

The goals of \predicate{consume} are 3-fold: 1) given a symbolic state $\sigma$ and formula $\gphi$, check whether $\gphi$ is established by $\sigma$, \ie $\gphiImplies{\gphi_{\sigma}}{\gphi}$ where $\gphi_{\sigma}$ is the formula which represents the state $\sigma$, 2) produce and collect run-time checks that are minimally sufficient for $\sigma$ to establish $\gphi$ soundly, and 3) remove accessibility predicates and predicates that are asserted in $\gphi$ from $\sigma$. The rules for \predicate{consume} are given in full and described in great detail in Appendix \S\ref{sec:appendix-consume}. We give select rules in \fig~\ref{fig:consume-rules-select} and an abstract description here.

The functionality of \predicate{consume} is split across two functions: \predicate{consume}, which is the interface to \predicate{consume} accepting only a state, formula, and continuation, and \predicate{consume'}, which is a helper function performing \predicate{consume}'s major functionality. Note, before calling \predicate{consume'}, \predicate{consume} first adds non-alias information from the heap to the path condition and checks that the heap and path condition are non-contradictory using \predicate{consolidate} \cite{schwerhoff2016advancing}. 

Then, the two functions work together to accomplish the aforementioned goals. For the first and second goals, heap chunks representing accessibility predicates and predicates in $\gphi$ are looked up in the heap $h$ and optimistic heap $\oh$ from $\sigma$. When $\sigma$ is precise, the heap chunks must be in $h$ or verification fails. If $\sigma$ is imprecise, then the heap chunks are always justified either by the heaps or imprecision. Run-time checks for heap chunks that are verified by imprecision are collected in $\sigma.\rchecks$.
The \predicate{consume'} rule for $\phiAcc{e.f}$ (and the rule for $\phiAcc{p(\overline{e})}$ which is similar) supports this functionality by calling \predicate{heap-rem-acc} (defined in Appendix \fig~\ref{fig:heap-add-rem}) for the look-up, assigning the boolean results to $b_1$ and $b_2$, and then using them in \ttt{if-then-else} and \ttt{else-if} casing. The blue highlighting in the $\isimp$ is \ttt{true} case in the aforementioned rule handles the run-time checks.
Clauses in $\gphi$ containing logical expressions are first evaluated to a symbolic value $t$, which is then checked against $\sigma$'s path condition $\pi$. If $\sigma$ is precise, then $\phiImplies{\pcall{\pi}}{t}$ must hold (\ie the constraints in $\pi$ prove $t$) or verification fails. In contrast, when $\sigma$ is imprecise, $\bigwedge \pcall{\pi} \wedge t$ must hold (\ie $t$ does not contradict constraints in $\pi$) otherwise verification fails. In this case, a run-time check is added to $\sigma.\rchecks$ for the set of residual symbolic values in $t$ that cannot be proved statically by $\pi$.
The \predicate{consume'} rules for expressions and symbolic values implement this behavior. The call to \predicate{assert} (defined in Appendix \fig~\ref{fig:check-assert-funcs}) checks $t$ against $\pi$ and returns the result and any residual symbolic values. Note, \predicate{assert} uses \predicate{diff} from Appendix \S\ref{sec:appendix-diff-translate} to compute the residuals. The part highlighted in blue adds the run-time check for the residuals to the state.
Finally, fields used in $\gphi$ must have corresponding heap chunks in $h$ when $\sigma$ and $\gphi$ are precise; otherwise when $\sigma$ or $\gphi$ are imprecise, field access can be justified by either the heaps or imprecision. A run-time check containing an accessibility predicate for the field is added to $\sigma.\rchecks$ when imprecision is relied on. This is all handled by the second argument $f_{\qm}$ to \predicate{consume'} and \predicate{eval-c} called by \predicate{consume'} on expressions.  

The third goal of \predicate{consume} is to remove heap chunks $\overline{hc_i}$ representing accessibility predicates and predicates in $\gphi$ from $\sigma$, and in particular, from heaps $h$ and $\oh$. When $\sigma$ and $\gphi$ are both precise, the heap chunks in $\overline{hc_i}$ are each removed from $h$ ($\oh$ is empty here). If $\gphi$ is imprecise, then all heap chunks in both heaps are removed as they may be in $\overline{hc_i}$ or $\gphi$ may represent them with imprecision. Finally, when $\sigma$ is imprecise and $\gphi$ is precise, any heap chunks in $h$ or $\oh$ that overlap with or may potentially overlap with (thanks to $\sigma$'s imprecision) heap chunks in $\overline{hc_i}$ are removed.
The calls to \predicate{heap-rem-acc} (and its counterpart \predicate{heap-rem-pred}) in \predicate{consume'}, the extra heaps tracked in \predicate{consume'}, and the heap assignments in the continuations from \predicate{consume} come together to implement heap chunk removal.

\subsubsection{\textbf{Symbolic execution of statements}}
\label{sec:gviper-exec}

\begin{figure}[t]
{\scriptsize\ttfamily
\disableTttResize
\begin{alignat*}{2}
  &\exec{\sigma_1}{\sFieldAssign{x}{f}{e}}{Q}
    &&= \heval{\sigma_1}{e}{(\lambda~ \sigma_2,~t ~.~
       \hconsume{\sigma_2}{\phiAcc{x.f}}{(\lambda~ \sigma_3,~\_ ~.~ \\
      &&&\qquad \hproduce{\sigma_3}{\phiCons{\phiAcc{x.f}}{x.f = t}}{pair(\fresh,~unit)}{Q})})}\\
  &\exec{\sigma_1}{\sCall{\overline{z}}{m}{\overline{e}}}{Q}
    &&=  \heval{\sigma_1}{\overline{e}}{(\lambda~ \sigma_2,~\overline{t} ~.~ \\
      &&&\qquad \mlightblue{\rchecks' \assign \sigma_2.\rchecks\{ \origin \assign (\sigma_2,~\sCall{\overline{z}}{m}{\overline{e}},~ \overline{t}) \}} \\
      &&&\qquad \hconsume{\sigma_2\{\mlightblue{\rchecks \assign \rchecks'}\}}{meth_{pre}[\overline{meth_{args} \mapsto t}]}{(\lambda~ \sigma_3,~\delta ~.~ \\
      &&&\qquad\quad \mlightyellow{\ttt{if}~~(\iseimp{meth_{pre}})~~\ttt{then}} \\
      &&&\qquad\qquad \mlightyellow{\sigma_4 \assign \sigma_3 \{ \isimp \assign \phiTrue,~ \oh \assign \emptyset,~ h \assign \emptyset,~ \gamma \assign \havoc{\sigma_3.\gamma}{\overline{z}} \}} \\
      &&&\qquad\quad \mlightyellow{\ttt{else}}~~\sigma_4 \assign \sigma_3 \{ \gamma \assign \havoc{\sigma_3.\gamma}{\overline{z}} \} \\
      &&&\qquad\quad \hproduce{\sigma_4}{meth_{post}[\overline{meth_{args} \mapsto t}][\overline{meth_{ret} \mapsto z}]}{\fresh}{ \\
      &&&\qquad\qquad (\lambda~ \sigma_5 ~.~ Q(\sigma_5\{\mlightblue{\rchecks \assign \sigma_5.\rchecks\{ \origin \assign \none \}} \})
      )})})}
\end{alignat*}
}

\footnotesize
\disableTttResize
\fbox{\begin{tabular}{llll}
\textcolor{light-yellow}{$\blacksquare$} & \small{Handles imprecision} &
\textcolor{light-blue}{$\blacksquare$} & \small{Handles run-time check generation and collection}
\end{tabular}}

\caption{\emph{Select rules} for symbolically executing program statements}
\label{fig:exec-rules-partial}
\end{figure}

The \predicate{exec} rules in \gviper, which symbolically execute program statements, are largely unchanged from Viper. The only differences are 1) the rules now utilize versions of \predicate{eval}, \predicate{produce}, \predicate{consume}, and \predicate{branch} defined previously in this paper and 2) the rules track $\origin$s where appropriate. To provide an intuition, select rules for \predicate{exec} are given in \fig~\ref{fig:exec-rules-partial}; the full set of rules are listed in the Appendix \S\ref{sec:appendix-exec}. The \predicate{exec} function takes a symbolic state $\sigma$, program statement $stmt$, and continuation $Q$. Then, \predicate{exec} symbolically executes $stmt$ using $\sigma$ to produce a potentially modified state $\sigma'$, which is passed to the continuation.

Symbolic execution of field assignments first evaluates the right-hand side expression $e$ to the symbolic value $t$. Any field reads in $e$ are either directly or optimistically verified using $\sigma_1$. Then, the resulting state $\sigma_2$ must establish write access to $x.f$ in \predicate{consume}, \ie $\gphiImplies{\sigma_2}{\phiAcc{x.f}}$. Calling \predicate{consume} also removes the field chunk for $\phiAcc{x.f}$ from $\sigma_2$ (if it is in there) resulting in $\sigma_3$. Therefore, the call to \predicate{produce} can safely add a fresh field chunk for $\phiAcc{x.f}$ alongside $x.f = t$ to $\sigma_3$ before it is passed to the continuation $Q$. Under the hood, run-time checks are collected and passed to $Q$.

The method call rule evaluates the arguments $\overline{e}$ to symbolic values $\overline{t}$, consumes the method precondition (substituting arguments with $\overline{t}$) while making sure the origin is set properly for check and branch condition insertion, havocs existing assumptions about the variables being assigned to, produces knowledge from the postcondition, and finally continues after resetting the origin to none. An in-depth explanation is in the Appendix, along with the other \predicate{exec} rules and \predicate{equi-imp} definition.

Note that while \gviper treats predicates iso-recursively in all other cases, it makes an exception when consuming preconditions at method calls (and loop invariants before entering loops), which can be seen in the \ttt{if-then} in the method call rule (\fig~\ref{fig:exec-rules-partial}). If \gviper determines the precondition (invariant) is equi-recursively imprecise, then it will conservatively remove all the heap chunks from both symbolic heaps and make the state imprecise after the consume. This exception ensures the static verification semantics in \gviper lines up with the equi-recursive, dynamic verification semantics encoded by \gvc (describe in \S\ref{sec:c0-runtime-checks}) such that \gco is sound. Interestingly, \citet{wise2020gradual}'s gradual verifier does not need this special case, because it does not optimize run-time checks with statically available information. Once optimization is introduced, the semantics across the two systems need to be more tightly integrated to ensure soundness. \citet{zimmerman2024sound} alerted us to this issue and proposed the aforementioned solution.
 

\subsubsection{Valid \gviper programs}
\label{sec:gviper-valid}
\begin{figure}[t]
\scriptsize\ttfamily
\disableTttResize
\begin{alignat*}{2}  
  &\verify{\mathtt{method}~ \ m(\overline{x:T})~ \mathtt{returns}~(\overline{y:T})} &&= 
    \hwellformed{\sigma_0\{ \gamma \assign \sigma_0.\gamma[\overline{x \mapsto \fresh }][\overline{y \mapsto \fresh}]\}}{meth_{pre}}{\fresh}{(\lambda~ \sigma_1 ~.~ \\
    &&&\qquad \hwellformed{\sigma_1\{\mlightyellow{\isimp \assign \phiFalse,~ h_{\qm} \assign \emptyset},~ h \assign \emptyset\}}{meth_{post}}{ \\
    &&&\qquad\quad \fresh}{(\lambda~ \_ ~.~ \success )} \\
    &&&\qquad \wedge~ \\
    &&&\qquad \hexec{\sigma_1}{meth_{body}}{(\lambda~ \sigma_2 ~.~ \\
    &&&\qquad\quad \hconsume{\sigma_2}{meth_{post}}{(\lambda~ \sigma_3,~ \_ ~.~ \\
    &&&\qquad\qquad \mlightblue{\rchecksall \assign \rchecksall \cup \sigma_3.\rchecks.\rcs} ~;~ \success )})})}\\
  &\verify{\texttt{predicate } p\ttt{(}\overline{x:T}\ttt{)}} 
  &&= 
    \hwellformed{\sigma_0\{ \gamma \assign \sigma_0.\gamma[\overline{x \mapsto \fresh}]\}}{pred_{body}}{\fresh}{(\lambda~ \_ ~.~ \success )}
\end{alignat*}

\fbox{\begin{tabular}{llll}
\textcolor{light-yellow}{$\blacksquare$} & \small{Handles imprecision} &
\textcolor{light-blue}{$\blacksquare$} & \small{Handles run-time check generation and collection}
\end{tabular}}

\caption{Rules defining a valid \gviper program}
\label{fig:valid-program}
\end{figure}

Finally, putting everything together, a \gviper program is checked by examining each of its method and predicate definitions to ensure they are well-formed (formally defined in Appendix \fig~\ref{fig:wellformed-func}). The formal definitions are given in \fig~\ref{fig:valid-program}, and a more detailed description of the rules is given in the Appendix, \S\ref{sec:gviper-validprogram}. Intuitively, for each method, we define symbolic values for the method arguments, and then create an initial symbolic state by calling the \predicate{produce} function on the method precondition\footnote{Note, \predicate{produce} is part of \predicate{well-formed}.}.  We then call the \predicate{exec} function on the method body, which symbolically executes the body and ensures that all operations are valid based on that precondition. Finally, we invoke the \predicate{consume} function on the final symbolic state and the postcondition, verifying that the former implies the latter.  Throughout these operations a set of run-time checks is built up, which (along with success or failure) is the ultimate result of gradual verification.

\subsection{Dynamic Verification: Encoding Run-time Checks into C0 Source Code}
\label{sec:c0-runtime-checks}
After static verification, \gviper returns a collection of run-time checks $\rchecksall$ that are required for soundness to \gvc. Then, \gvc creates a C0 program from the run-time checks in $\rchecksall$ and the original C0 program by encoding the checks in C0 source code. The C0 program is sent to the C0 compiler to be compiled, executed, and thus dynamically verified. We chose to encode the run-time checks directly in source code to avoid complexities from augmenting the C0 compiler with support for dynamic verification. Further, since C0 is a simple imperative language, any more expressive language should be able to encode the checks far more easily. That is, we hope this work serves as a guide to the developers of \gviper frontends for other languages on how to implement efficient dynamic verification for gradual verification---especially, when modifying the compiler for their language is difficult. The rest of this section illustrates \gvc's encoding of run-time checks into C0 source code via example. We also highlight design points in the encoding that minimize run-time overhead of the checks during execution.

Now, consider the C0 program in \fig~\ref{fig:ex-checks-gviper} that implements a method for inserting a new node at the end of a list, called \ttt{insertLastWrapper}. Note, when passed a non-empty list, \ttt{insertLastWrapper} calls \ttt{insertLast} from \fig~\ref{ex:ll-insert} to perform insertion (line \ref{ilw-gviper-methcall}). Here, \ttt{insertLast} is gradually verified with the simpler and fully specified (precise) \ttt{acyclic} predicate given on lines \ref{ilw-gviper-acyclic-start}-\ref{ilw-gviper-acyclic-end} in \fig~\ref{fig:ex-checks-gviper}. For our purposes, we only need to know that \ttt{insertLast}'s precondition is $\qm \phiAnd \ttt{acyclic(list)} \phiAnd$$\ttt{list != NULL}$ and its postcondition is $\ttt{acyclic(\textbackslash result)} \phiAnd$$\ttt{\textbackslash result != NULL}$. The \ttt{insertLastWrapper} method is also gradually specified: its precondition is $\qm$ (line \ref{ilw-gviper-precond})---requiring unknown information---and its postcondition is \ttt{acyclic(\textbackslash result)} (line \ref{ilw-gviper-postcond})---ensuring the list after insertion is acyclic. \fig~\ref{fig:ex-checks-gviper} also contains run-time checks generated by \gviper for \ttt{insertLastWrapper}, as highlighted in blue. The first check (lines \ref{ilw-gviper-check1-begin}-\ref{ilw-gviper-check1-end}) ensures the list \ttt{l} sent to \ttt{insertLast} (line \ref{ilw-gviper-methcall}) is acyclic, and is only required when \ttt{l} is non-empty (non-null). The second check (lines \ref{ilw-gviper-check2-begin}-\ref{ilw-gviper-check2-end}) ensures the list returned from \ttt{insertLastWrapper} is acyclic, and is only required when \ttt{insertLastWrapper}'s parameter \ttt{l} is empty (null).
These checks are not executable by the C0 compiler; therefore, \gvc takes the program and checks in \fig~\ref{fig:ex-checks-gviper} and returns the executable program in \fig~\ref{fig:ex-checks-gvc0}. That is, \gvc encodes branch conditions (lines \ref{ilw-gviper-check1-begin} and \ref{ilw-gviper-check2-begin}), predicates (lines \ref{ilw-gviper-check1-end} and \ref{ilw-gviper-check2-end}), accessibility predicates ($\phiAcc{\ttt{l->val}}$ and $\phiAcc{\ttt{l->next}}$ in acyclic's body, lines \ref{ilw-gviper-acyclic-start}-\ref{ilw-gviper-acyclic-end}), and separating conjunctions (also in acyclic's body) from \gviper into C0 source code. We discuss the aforementioned encodings in sections \S\ref{sec:gvc0-branchconds}, \S\ref{sec:gvc0-predicates}, and \S\ref{sec:gvc0-accpreds} respectively. While not in the \ttt{insertLastWrapper} example, \gvc translates checks of simple logical expressions into C0 assertions: \eg \ttt{assert(y >= 0);}.

\begin{figure}[t]
\begin{minipage}[t]{0.43\linewidth}
{\scriptsize\ttfamily
\disableTttResize
\begin{lstlisting}[xleftmargin=2.5em, name=ex-checks-gviper]
(*@\label{ilw-gviper-acyclic-start}@*)/*@ predicate acyclic(Node* l) = 
      l == NULL ? true : 
        acc(l->val) && acc(l->next) &&
        (*@\label{ilw-gviper-acyclic-end}@*)acyclic(l->next) ;@*/

Node* insertLastWrapper(Node* l, int val)
  (*@\label{ilw-gviper-precond}@*)//@ requires ?;
  (*@\label{ilw-gviper-postcond}@*)//@ ensures acyclic(\result);
{
  if (l == NULL) {
    l = alloc(struct Node);
    l->val = val;
    l->next = NULL;
  } else {
  (*@\label{ilw-gviper-check1-begin}\lightblue{(\{($\none$, l == NULL, $\neg$(l == NULL))\},}@*)
    (*@\label{ilw-gviper-check1-end}\lightblue{(l=insertLast(l,val), acyclic(l), acyclic(l))}@*)
    (*@\label{ilw-gviper-methcall}@*)l = insertLast(l, val);
  }
  return l;
  (*@\label{ilw-gviper-check2-begin}\lightblue{(\{($\none$, l == NULL, l == NULL)\},}@*)
    (*@\label{ilw-gviper-check2-end}\lightblue{($\none$, acyclic(\textbackslash result), acyclic(\textbackslash result))}@*)
}
\end{lstlisting}
}

\footnotesize
\disableTttResize
\begin{center}
\fbox{\begin{tabular}{llll}
\textcolor{light-blue}{$\blacksquare$} & \small{Run-time checks from \gviper}
\end{tabular}}
\end{center}

\caption{Original \ttt{insertLastWrapper} program with run-time checks from \gviper}
\label{fig:ex-checks-gviper}
\end{minipage}\hfill
\begin{minipage}[t]{0.47\linewidth}
{\scriptsize\ttfamily
\disableTttResize
\begin{lstlisting} [xleftmargin=1em, name=ex-checks-gvc0]
Node* insertLastWrapper(Node* l,int val,
  (*@\label{ilw-gvc0-ownedfields-param}\lightblue{OwnedFields* \_ownedFields}@*))
{
  (*@\label{ilw-gvc0-cond}\lightblue{bool \_cond\_1 = l == NULL;}@*)
  if (l == NULL) {
    (*@\label{ilw-gvc0-alloc}@*)l = alloc(struct Node);
    (*@\label{ilw-gvc0-accalloc}\lightblue{l->\_id = addStructAcc(\_ownedFields, 2);}@*)
    l->val = val; l->next = NULL;
  } else {
    (*@\label{ilw-gvc0-acyclicl}\lightblue{if (!\_cond\_1) \{ acyclic(l, \_ownedFields); \}}@*)
    (*@\label{ilw-gvc0-init1-begin}\lightblue{OwnedFields* \_tempFields =}@*)
        (*@\label{ilw-gvc0-init1-end}\lightblue{initOwnedFields(\_ownedFields->instCntr);}@*)
    (*@\label{ilw-gvc0-sep1}\lightblue{sep\_acyclic(l, \_tempFields);}@*)
    (*@\label{ilw-gvc0-ownedfields-pass}@*)l = insertLast(l,val,(*@\lightblue{\_ownedFields}@*));
  }
  (*@\label{ilw-gvc0-acyclicresult}\lightblue{if (\_cond\_1) \{ acyclic(l, \_ownedFields); \}}@*)
  (*@\label{ilw-gvc0-init2-begin}\lightblue{OwnedFields* \_tempFields1 =}@*)
      (*@\label{ilw-gvc0-init2-end}\lightblue{initOwnedFields(\_ownedFields->instCntr);}@*)
  (*@\label{ilw-gvc0-sep2}\lightblue{sep\_acyclic(l, \_tempFields1);}@*)
  return l;
}  
\end{lstlisting}
}

\footnotesize
\disableTttResize
\begin{center}
\fbox{\begin{tabular}{llll}
\textcolor{light-blue}{$\blacksquare$} & \small{Run-time checks from \gvc}
\end{tabular}}
\end{center}

\caption{\gvc generated \ttt{insertLastWrapper} program with run-time checks}
\label{fig:ex-checks-gvc0}
\end{minipage}
\end{figure}


\subsubsection{\textbf{Encoding branch conditions}}
\label{sec:gvc0-branchconds}
Run-time checks contain branch conditions that denote the execution path a check is required on. For example, in \fig~\ref{fig:ex-checks-gviper} \ttt{acyclic(\textbackslash result)} should only be checked at lines \ref{ilw-gviper-check2-begin}-\ref{ilw-gviper-check2-end} when \ttt{l == NULL}, as indicated by the branch condition \ttt{($\none$,l == NULL,l == NULL)}. Therefore, \gvc first encodes the condition \ttt{l == NULL} into C0 code. In general, conditions are encoded as logical expressions in C0 and assigned to fresh boolean variables at the program point where they originated---we call this \emph{versioning}. Then, the boolean variable is used in checks in place of the condition. For example, the $\origin$ and $\location$ pair \ttt{($\none$,l == NULL)} tells \gvc that \ttt{l == NULL} must be evaluated at the program point in \ttt{insertLastWrapper} containing the \ttt{l == NULL} AST element. As a result, in \fig~\ref{fig:ex-checks-gvc0} a boolean variable \ttt{\_cond\_1} is introduced on line \ref{ilw-gvc0-cond} to hold the value of \ttt{l == NULL}. The condition variable \ttt{\_cond\_1} is then used in the C0 run-time check for \ttt{acyclic(\textbackslash result)} later in the program (line \ref{ilw-gvc0-acyclicresult}). To reduce run-time overhead, \ttt{\_cond\_1} is also used in the check for \ttt{acyclic(l)} on line \ref{ilw-gvc0-acyclicl}, which relies on the same branch point \ttt{($\none$,l == NULL)}. Further, while not demonstrated here, \gvc supports short-circuit evaluation of conditions on the same execution path to reduce run-time overhead. Finally, note, \gvc carefully places versioning code after any run-time checks encoded at the program point where the condition originated to ensure this new assignment code (not verified by \gviper) is framed, \ie correct.

\subsubsection{\textbf{Encoding predicates}}
\label{sec:gvc0-predicates}
Now that \gvc has versioned the branch conditions in \fig~\ref{fig:ex-checks-gviper} into variables, \gvc can use the variables to develop C0 run-time checks. The \gviper check \ttt{(\{($\none$,l == NULL,$\neg$(l == NULL))\}, (l=insertLast(l,val),acyclic(l),acyclic(l)))} is translated into \ttt{if (!\_cond\_1) \{acyclic(l,\_ownedFields);\}} on line \ref{ilw-gvc0-acyclicl} in \fig~\ref{fig:ex-checks-gvc0}. \gvc places this C0 check according to the $\origin$, $\location$ pair \ttt{((l=insertLast(l, val),acyclic(l))}, which points to the program point just before the call to \ttt{insertLast} on line \ref{ilw-gvc0-ownedfields-pass}. The branch condition becomes the if statement with condition \ttt{!\_cond\_1} (\S\ref{sec:gvc0-branchconds}), and \ttt{acyclic(l)} is turned into the C0 function call \ttt{acyclic(l,\_ownedFields)}. The \ttt{acyclic} function implements acyclic's predicate body as C0 code: it asserts true for empty lists and recursively verifies accessibility predicates (using \ttt{\_ownedFields}) for nodes in non-empty lists. That is, predicates are encoded and treated equi-recursively by \gvc. For efficiency, separation of list nodes is encoded separately on lines \ref{ilw-gvc0-init1-begin}-\ref{ilw-gvc0-sep1}. We discuss the dynamic verification of accessibility predicates and the separating conjunction in C0 code next (\S\ref{sec:gvc0-accpreds}). Finally, a similar C0 check is created for \ttt{acyclic(\textbackslash result)} on lines \ref{ilw-gvc0-acyclicresult}-\ref{ilw-gvc0-sep2}.

\subsubsection{\textbf{Encoding accessibility predicates and separating conjunctions}}
\label{sec:gvc0-accpreds}
\gvc implements run-time tracking of owned heap locations in C0 programs to verify accessibility predicates and uses of the separating conjunction. 

\paragraph{Encoding owned fields in C0 source code.}
An owned field is a tuple $(id, \mathit{field})$ where $id$ is an integer identifying a struct instance (object in C0) and $\mathit{field}$ is an integer indexing a field in the struct. The \ttt{OwnedFields} struct, which is implemented as a dynamic hash table to improve check performance, contains currently owned fields. That is, hashed object identifiers ($id$) and then field identifiers ($\mathit{field}$) are used to index into \ttt{OwnedFields} where a boolean that determines whether or not the field is currently owned is stored. Since objects are tracked with integers, all struct definitions in a C0 program are modified to contain an additional \ttt{\_id} field.

\paragraph{Semantics of tracking owned fields (inspired by \citet{wise2020gradual}).} At the entry point to a C0 program (\eg \ttt{main}), an empty \ttt{OwnedFields} struct called \ttt{\_ownedFields} is allocated and initialized. This is not shown in \fig~\ref{fig:ex-checks-gvc0}. Then, when a new struct instance is created---such as allocating a new node on line \ref{ilw-gvc0-alloc} in \fig~\ref{fig:ex-checks-gvc0}---the \ttt{\_id} field is initialized with the value of a global integer \ttt{instCntr} that uniquely identifies the instance. The call to library function \ttt{addStructAcc} on line \ref{ilw-gvc0-accalloc} performs this functionality and then increments \ttt{instCntr}. It also adds all fields in the struct instance (\eg \ttt{l->val:(l->\_id,0)}, \ttt{l->next:(l->\_id,1)}, and \ttt{l->\_id:(l->\_id,2)}) to \ttt{\_ownedFields} and marks them as owned. The only other times \ttt{\_ownedFields} can change are at method/function calls. Methods, like \ttt{insertLast} and \ttt{insertLastWrapper}, may add or drop owned fields during their executions. They may also contain run-time checks, such as the one for \ttt{acyclic(l)} on line \ref{ilw-gvc0-acyclicl}, that need owned fields for verification. So, \gvc adds an additional parameter to the their declarations (\eg line \ref{ilw-gvc0-ownedfields-param}, \fig~\ref{fig:ex-checks-gvc0}) to accept, initialize, and then modify \ttt{\_ownedFields} in their contexts. A callee's pre- and postcondition controls what owned fields are passed to and from the callee via this new parameter. When a method's precondition is imprecise\footnote{Here, a formula is also imprecise if it contains predicates that expose $\qm$ when fully unrolled---an equi-recursive treatment.}, then any caller will pass all of its owned fields to the method, as on line \ref{ilw-gvc0-ownedfields-pass} for the call to \ttt{insertLast}. 
After execution, the callee method returns all of its owned fields to the caller. When a method's precondition is precise, then any caller only passes its owned fields specified by the precondition to the method. If the method's postcondition is imprecise, then after execution the callee method returns all of its owned fields as before; otherwise, only the owned fields specified by the postcondition are returned. Finally, as an optimization, in precisely specified methods (no external---pre- and postconditions---or internal---loop invariants, unfolds, folds, etc.---specifications contain imprecision and no run-time checks are required), \gvc does not implement any \ttt{\_ownedFields} tracking. In this case, \gvc uses the callee's pre- and postcondition to modularly update \ttt{\_ownedFields} in the caller.

\paragraph{Verifying accessibility predicates and separating conjunctions with owned fields.}
Now, \ttt{\_ownedFields} tracking is used to verify accessibility predicates and uses of the separating conjunction. Run-time checks for accessibility predicates are turned into assertions that ensure the presence of their heap location in \ttt{\_ownedFields}. For example, $\phiAcc{\ttt{l->val}}$ looks like \ttt{assertAcc(\_ownedFields,l->\_id,0);} in C0 code, where $0$ is the index for \ttt{val} in the \ttt{Node} struct. The \ttt{assertAcc} library function indexes into \ttt{\_ownedFields} using \ttt{l->\_id} and $0$ and ensures a boolean entry is in there and is \ttt{true}; otherwise, \ttt{assertAcc} throws an error. Wherever \gvc must check separation of heap locations (as indicated in run-time checks from \gviper via a flag\footnote{Note, this flag is not formalized in this paper for simplicity.}),---such as for the nodes in list \ttt{l} at lines \ref{ilw-gvc0-acyclicl}-\ref{ilw-gvc0-sep1}---it creates (with the library method \ttt{initOwnedFields}) an auxiliary data structure \ttt{\_tempFields} of type \ttt{OwnedFields}.  We check that heap cells are disjoint by adding them one at a time to \ttt{\_tempFields} and failing if the cell is already there. \gvc generates a \ttt{sep\_X} method for each predicate \ttt{X} to actually perform the separation check; and when done, discards \ttt{\_tempFields}, as its purpose was only to check separation.  Similar checks are created for the \ttt{acyclic(\textbackslash result)} check on lines \ref{ilw-gvc0-acyclicresult}-\ref{ilw-gvc0-sep2}.


\section{Empirical Evaluation}
\label{sec:empirical}
\input{contents/empirical}

\section{Related Work}
Much of the closely related work, particularly work on gradual verification \cite{bader2018gradual, wise2020gradual,zimmerman2024sound}, gradual typing \cite{siek2015refined, siek2006gradual, hermanAl:hosc10, takikawa16}, and static verification \cite{reynolds2002separation, smans2009implicit, parkinson2005separation, MuellerSchwerhoffSummers16}, has been discussed throughout the paper. Note, this paper and \citet{zimmerman2024sound}'s paper have distinct contributions.  Our work includes the original development of symbolic execution based gradual verification and describes the first implementation of the same, as well as related empirical results. \citet{zimmerman2024sound}'s paper proves that the approach is sound. Now, we discuss additional related work.

\paragraph{Gradual typing}
Additional related work in gradual typing includes richer type systems such as gradual refinement types~\cite{10.1145/3093333.3009856} and gradual dependent types~\cite{10.1145/3341692,lennonAl:toplas2022}. These systems focus on pure functional programming, while \gco targets imperative programs. There is an extensive body of work on optimizing run-time checks in gradual type systems. \citet{10.1145/3133880} show that in languages with nominal type systems, such as Java, gradual typing does not exhibit the usual slowdowns induced by structural types. \citet{10.1145/3276503} reduce run-time overhead from redundant contract checking by contract wrappers. They eliminate unnecessary contract checking by determining---across multiple contract checking boundaries for some datatype or function call---whether some of the contracts being checked imply others. While the results in \S\ref{sec:empirical} are promising, we may be able to draw from the extensive body of work in gradual typing to achieve further performance gains. 

\paragraph{Static verification}
Work in formal verification contains approaches that try to reduce the specification burden of users---a goal of \gco. \citet{furia2010inferring} infer loop invariants with heuristics that weaken postconditions into invariants. When that approach fails, verification also fails because invariants are missing.  Similarly, several tools (Smallfoot \cite{berdine2006smallfoot}, jStar~\cite{distefano2008jstar}, and Chalice~\cite{leino2009verification}) use heuristics to infer fold and unfold statements for verification. In contrast, \gco does not fail solely because invariants, folds, or unfolds are missing; imprecision begets optimism. However, \gco may benefit from similar heuristic approaches by leveraging additional static information to further reduce run-time overhead. 

\emph{Abductive reasoning} (\emph{abductive inference}) tries to find an explanatory hypothesis for a desired outcome \cite{dillig2012automated}. In static verification, the desired outcome is a proof obligation ($O$), facts ($F$) are invariants derived from the program and specifications using some analysis, and the explanatory hypothesis ($E$) are invariants that do not contradict the derived facts ($\text{SAT}(F \wedge E)$) and are required to discharge the proof obligation ($F \wedge E \models O$). Ideally, $F$ should be sufficient to discharge $O$, but missing or insufficient specifications often results in $F$ being too weak to prove $O$ leading to false positives (alarms) in tools.
So, work in applying abductive reasoning to static verification \cite{blackshear2013almost,calcagno2009compositional,dillig2012automated,das2015angelic,chandra2009snugglebug} aims to compute $E$ in order to prioritize---with minimal human intervention---verification failures caused by bugs in a program and de-emphasize false positives (alarms) caused by missing or incomplete specifications. 
In angelic verification \cite{blackshear2013almost, das2015angelic} and \citet{calcagno2009compositional}'s work, entire specifications, such as preconditions, postconditions, and loop invariants are generated as explanatory hypotheses. \citet{dillig2012automated} instead compute smaller, intermediate formulas as explanatory hypotheses.

Similar to prior abductive reasoning work \cite{blackshear2013almost,calcagno2009compositional,dillig2012automated,das2015angelic}, \gco's static system reasons around missing or incomplete specifications to compute facts $F$ as part of imprecise formulas $\qm~\wedge~F$. At proof obligations, we approximate the weakest formula that can replace $\qm$ in $\qm~\wedge~F \models O$ and $\text{SAT}(\qm~\wedge~F)$ successfully.
So, like \citet{dillig2012automated} we compute intermediate explanatory hypotheses rather than whole specifications like \citet{blackshear2013almost}, \citet{calcagno2009compositional}, and \citet{das2015angelic}. But, rather than relying on users to validate generated hypotheses \cite{blackshear2013almost,calcagno2009compositional,dillig2012automated,das2015angelic}, we check their correctness at run time. This significantly simplifies their computation---since they do not need to be human readable and can statically mark code as unreachable---and allows \gco to be sound (prior abduction work is not). 

\paragraph{Dynamic verification}
\citet{meyer1988eiffel} introduced the Eiffel language, which automatically performs dynamic verification of pre- and postconditions and class invariants in first order logic. \citet{nguyen2008runtime} extended dynamic verification to support separation logic assertions. More recently, \citet{agten2015modularverif} applied dynamic checking at the boundaries between statically verified and unverified code to guarantee that no assertion failures or invalid memory accesses occur at run time in any verified code. Their approach improved on \citet{nguyen2008runtime}'s approach in terms of performance by allowing unverified code to read arbitrary memory. Further, unlike \citet{nguyen2008runtime}, \citet{agten2015modularverif}’s approach only needs access to verified code rather than the entire codebase. As with \citet{nguyen2008runtime}'s work, \gco supports dynamic verification of ownership and first order logic. \gco additionally supports run-time checking of recursive predicates. Similarly to \citet{agten2015modularverif}, \gco applies dynamic checking at the boundaries between verified and unverified code to protect verified code. However, in \gco unverified code must be accessible to the verifier as it is gradually verified as well. Future work in gradual verification should incorporate insights from \citet{agten2015modularverif}'s work to avoid requiring entire codebases for verification and to improve verification performance.

\paragraph{Hybrid verification}
Another closely related work is soft contract verification~\cite{10.1145/2628136.2628156}, which verifies dynamic contracts statically where possible and dynamically where necessary by utilizing symbolic execution. This hybrid technique does not rely on a notion of {\em precision}, which is central to gradual approaches and their metatheory~\cite{siek2015refined}. \citet{10.1145/2628136.2628156} use symbolic execution results directly to discharge proof obligations where possible, while \gco strengthens symbolic execution results to discharge proof obligations adhering to the theory of imprecise formulas from \citet{wise2020gradual}. Further, \citet{10.1145/2628136.2628156}'s work is targeted at dynamic functional languages, while our work focuses on imperative languages. We also build in memory safety as a default, while \citet{10.1145/2628136.2628156} do not.

\section{Conclusion}
Gradual verification is a promising approach to supporting incrementality and enhance adoptability of program verification. Users can focus on specifying and verifying the most important properties and components of their systems and get immediate feedback about the consistency of their specifications and the correctness of their code. By relying on symbolic execution, \gco overcomes several limitations of prior work on gradual verification of heap-manipulating programs. The experimental results show that our approach can reduce overhead significantly compared to purely dynamic checking and confirms performance trends speculated in prior work. While more work remains to extend gradual verification (and \gco) to the expressiveness of state-of-the-art static program verifiers, we believe the symbolic-execution approach to gradual verification shows promise to make verification more adoptable in software development practice.
\bibliography{references}
\clearpage

\appendix
\section{Appendix}
\FloatBarrier

\begin{figure}[!ht]
\scriptsize\ttfamily
\disableTttResize
\begin{alignat*}{2}
& \pcadd{\pi}{t} = \ttt{Let}~ (id,~bc,~pcs)::\emph{suffix}~ \ttt{match}~ \pi \\
  &\quad (id,~bc,~pcs\cup\{t\})::\emph{suffix} \\
& \pcpush{\pi}{id}{bc} =  (id,~bc,~\emptyset)::\pi \\
& \pcall{\pi} = \foldl{\pi}{\emptyset}{(\lambda~ (id_i,~bc_i,~pcs_i),~all_i ~.~ all_i \cup \{bc_i\} \cup pcs_i)}
\end{alignat*}
\caption{Path condition helper functions}
\label{fig:pc-funcs}
\end{figure}

\begin{figure}[!ht]
\scriptsize\ttfamily
\disableTttResize
\begin{alignat*}{2}  
  &\evalp{\sigma}{t}{Q}
    &&= Q(\sigma,~t) \\
  &\evalp{\sigma}{x}{Q}
    &&= Q(\sigma,~\sigma.\gamma(x)) \\
  &\evalp{\sigma_1}{op(\overline{e})}{Q}
    &&=  \evalp{\sigma_1}{\overline{e}}{(\lambda~ \sigma_2, ~\overline{t} ~.~ Q(\sigma_2,~ op'(\overline{t})))}\\
  &\evalp{\sigma_1}{e.f}{Q}
    &&=  \evalp{\sigma_1}{e}{(\lambda~ \sigma_2, ~t ~.~ \\
      &&&\quad \texttt{if}~~ (\exists~ \chunk{f}{r}{\delta} \in \sigma_2.h ~.~ \checkv{\sigma_2.\pi}{r = t}) ~~\ttt{then}~~\\
      &&&\qquad Q(\sigma_2,~\delta)\\
      &&&\quad \texttt{else if}~~ (\exists~ \chunk{f}{r}{\delta} \in \sigma_2.\oh ~.~ \checkv{\sigma_2.\pi}{r = t}) ~~\ttt{then}~~\\
      &&&\qquad Q(\sigma_2,~\delta)\\
      &&&\quad \texttt{else} ~~\texttt{if}~~ (\sigma_2.\isimp) ~~\ttt{then}~~\\
      &&&\qquad \mlightblue{\ttt{if}~~ (\sigma_2.\rchecks.\origin = (\_,~\sUnfold{\phiAcc{\_}},~\_)) ~~\ttt{then}}~~\\
      &&&\qquad\quad \mlightblue{e_t \assign \translate{\sigma_2}{t}}\\
      &&&\qquad\quad \mlightblue{\rchecks' \assign \addcheck{\sigma_2.\rchecks}{e.f}{\phiAcc{e_t.f}}}\\
      &&&\qquad \mlightblue{\texttt{else}}\\
      &&&\qquad\quad \mlightblue{\rchecks' \assign \sigma_2.\rchecks} \\
      &&&\qquad \delta \assign \fresh \\
      &&&\qquad Q(\sigma_2\{~ \oh \assign \sigma_2.\oh \cup \chunk{f}{t}{\delta},~ \pi \assign \pcadd{\sigma_2.\pi}{\{t \neq \enull\}},~ \rchecks \assign \rchecks' ~\},~\delta)\\
      &&&\quad \texttt{else}~ \failure}
\end{alignat*}
\caption{Rules for symbolically executing expressions without introducing run-time checks (except for a special case for unfold)}
\label{fig:eval-p-rules}
\end{figure}

\noindent In \predicate{eval-p} (\fig~\ref{fig:eval-p-rules}), a special case (highlighted in blue) for unfold statements is added that creates run-time checks for field accesses in the unfolded predicate's body. This case ensures soundness when introducing branch condition variables in C0 programs during run-time verification. In our implementation of \gco, these checks are optimized further as they are only produced for branch conditions in the predicate body rather than for the whole body.

\clearpage

\begin{figure}[!ht]
\scriptsize\ttfamily
\disableTttResize
\begin{alignat*}{2}  
  &\evalpc{\sigma}{t}{Q}
    &&= Q(\sigma,~t) \\
  &\evalpc{\sigma}{x}{Q}
    &&= Q(\sigma,~\sigma.\gamma(x)) \\
  &\evalpc{\sigma_1}{op(\overline{e})}{Q}
    &&=  \evalpc{\sigma_1}{\overline{e}}{(\lambda~ \sigma_2, ~\overline{t} ~.~ Q(\sigma_2,~ op'(\overline{t})))}\\
  &\evalpc{\sigma_1}{e.f}{Q}
    &&=  \evalpc{\sigma_1}{e}{(\lambda~ \sigma_2, ~t ~.~ \\
      &&&\quad \texttt{if}~~ (\exists~ \chunk{f}{r}{\delta} \in \sigma_2.h ~.~ \checkv{\sigma_2.\pi}{r = t}) ~~\ttt{then}~~\\
      &&&\qquad Q(\sigma_2,~\delta)\\
      &&&\quad \texttt{else if}~~ (\exists~ \chunk{f}{r}{\delta} \in \sigma_2.\oh ~.~ \checkv{\sigma_2.\pi}{r = t}) ~~\ttt{then}~~\\
      &&&\qquad Q(\sigma_2,~\delta)\\
      &&&\quad \texttt{else} ~~\texttt{if}~~ (\sigma_2.\isimp) ~~\ttt{then}~~\\
      &&&\qquad res,~\_ \assign \assertv{\sigma_2.\isimp}{\sigma_2.\pi}{t \neq \enull} \\
      &&&\qquad e_t \assign \translate{\sigma_2}{t}\\
      &&&\qquad \rchecks' \assign \addcheck{\sigma_2.\rchecks}{e.f}{\phiAcc{e_t.f}}\\
      &&&\qquad res \wedge Q(\sigma_2\{~ \rchecks \assign \rchecks' ~\},~\fresh)\\
      &&&\quad \texttt{else}~ \failure}
\end{alignat*}
\caption{Rules for symbolically executing expressions without modifying the optimistic heap and path condition}
\label{fig:eval-c-rules}
\end{figure}

\subsection{Diff and Translate}
\label{sec:appendix-diff-translate}

\begin{figure}[!ht]
\small\ttfamily
\disableTttResize
\centering
\begin{algorithm}[H]
  \caption{Generating minimal checks}
  \begin{algorithmic}[1]
    \Statex
    \Function{Diff}{$\phi$}
      \Let{$conjuncts$}{$CNF(\phi)$}
      \Let{$\phi'$}{$\emptyset$}
      \For{$c \gets conjuncts$}
        \If{$!check(c)$}
            \Let{$\phi'$}{$\phi' + c$}
        \EndIf
      \EndFor
      \State \Return{$\phi'$}
    \EndFunction
  \end{algorithmic}
\end{algorithm}
\caption{Algorithm for computing the \predicate{diff} between two symbolic values}
\label{fig:algorithm-diff}
\end{figure}

\begin{figure}[!ht]
\small\ttfamily
\disableTttResize
\centering
\centering
\begin{algorithm}[H]
  \caption{Variable resolution procedure}
  \begin{algorithmic}[1]
    \Function{Translate-var}{$s, v$}
      \Let{$store$}{$\emptyset$}
        \If{$s.oldStore$}
            \Let{$store$}{$s.oldStore$}
        \Else
            \Let{$store$}{$s.store$}
        \EndIf
        \Let{$aliasList$}{$aliases(v, s.pathConditions ++ s.heap ++ s.optimisticHeap)$}
        \Let{$heap$}{$s.heap ++ s.optimisticHeap$}
        \Let{$outputs$}{$\emptyset$}
      \For{$v \gets aliasList$}
        \If{$c \gets store.lookup(v)$}
            \Let{$outputs$}{$outputs + c$}
        \Else
            \If{$h \gets heap.lookup(v) \&\& c \gets store.lookup(h)$}
                \Let{$outputs$}{$outputs + c$}
            \EndIf
        \EndIf
      \EndFor
      \State \Return{$selectLongest(outputs)$}
    \EndFunction
  \end{algorithmic}
\end{algorithm}
\caption{\textsc{Translate}'s procedure for resolving variables}
\label{fig:algorithm-translate}
\end{figure}

The \textsc{diff} (\fig~\ref{fig:algorithm-diff}) function finds a minimal run-time check from an optimistically asserted formula containing statically known information. It accomplishes this by first performing a standard transformation to conjunctive normal form (CNF) on the optimistically asserted formula, to extract the maximal number of top level conjuncts. It then attempts to call \textit{check()} on each conjunct; it accumulates each conjunct for which the call does not succeed. The set of conjuncts which could not be statically discharged are returned as the final check.

The \textsc{translate} (\fig~\ref{fig:algorithm-translate}) function lifts symbolic values to concrete values. Most symbolic values are directly translated to their concrete counterparts via recursive descent; the exception is variables, whose concrete values must be reconstructed by searching the program state known by the verifier. This is done by retrieving the states of the symbolic store, which contains mappings from concrete variables to symbolic variables, and the heap, which contains field and predicate permissions. When \textsc{translate} encounters a symbolic variable, it first retrieves all possible aliasing information from \gviper's state. This includes all variables known to be equivalent to the translation target according to the path condition and the heap. If the translation target or one of its aliases exists as a value in the symbolic store, then the translator finds a key corresponding to it in the store and returns it. Note that multiple valid keys may exist for a particular symbolic variable, because \gviper may have determined that multiple concrete values are equivalent at a particular program point. If the translation target is a field, then only the top level receiver (the variable on which fields are being accessed) or one of its aliases will exist in the store. The fields being accessed are resolved by mapping their corresponding heap entries, or any aliased heap entries, to a value in the symbolic store, and resolving the store entry as described. In particular contexts, \textsc{translate} may be asked to translate a precondition for a method call, or a predicate body for an (un)fold statement. In these cases, an old store attached to the current symbolic state as described in \ref{sec:gviper-exec} is retrieved, and its symbolic store and heap are used for translation. This causes variables in a precondition or predicate to be resolved to their concrete values at the call site, or site of unfolding. This enables run-time checks produced via translate to be straightforwardly emitted to the frontend. The portion of translate related to translating variables is shown in \fig~\ref{fig:algorithm-translate}.

\subsection{Symbolic production of formulas}
\label{sec:appendix-produce}
The rules for produce are given in \fig~\ref{fig:produce-rules}. Essentially, produce takes a formula and snapshot $\delta$ (mirroring the structure of the formula) and adds the information in the formula to the symbolic state, which is then returned to the continuation $Q$.
An imprecise formula $\withqm{\phi}$ has its static part $\phi$ produced into the current state $\sigma$ alongside $\second{\delta}$. Note the snapshot $\delta$ for an imprecise formula looks like $(unit, \second{\delta})$ where $unit$ is the snapshot for $\qm$ and $\second{\delta}$ is the snapshot for $\phi$. An imprecise formula also turns $\sigma$ imprecise to produce the unknown information represented by $\qm$ into $\sigma$. For example, if the state is represented by the formula $\theta$, then this rule results in $\withqm{\theta \phiAnd \phi}$.
A symbolic value $t$ is produced into the path condition of the current state $\sigma$. Also, the snapshot $\delta$ for $t$ must be $unit$, so this fact is also stored in $\sigma$'s path condition. Then, $\sigma$ is passed to $Q$.

The produce rule for expression $e$, first evaluates $e$ to its symbolic value $t$ using eval-p. Then, $t$ is produced into the path condition of the current state $\sigma_2$ using the aforementioned symbolic value rule. Imprecision in the symbolic state can always provide accessibility predicates for fields also in the state. Therefore, when fields in $e$ are added to an imprecise state, heap chunks for those fields do not have to already be in the state, \eg the state $\withqm{\phiTrue}$ becomes $\withqm{\phiCons{\phiTrue}{e}}$. This functionality is permitted by eval-p. Similarly, an imprecise formula always provides accessibility predicates for fields in its static part, \eg the state $\phiTrue$ and produced formula $\withqm{e}$ results in the state $\withqm{\phiCons{\phiTrue}{e}}$.
The goal of produce is not to assert information in the state, but rather add information to the state. So we reduce run-time overhead by ensuring no run-time checks are produced by produce even for verifying field accesses.

The rules for producing field and predicate accessibility predicates into the state $\sigma_1$ operate in a very similar manner. Thus, we will focus on the rule for fields only. The field $e.f$ in $\phiAcc{e.f}$ first has its receiver $e$ evaluated to $t$ by eval-pc, resulting in $\sigma_2$. Then, using the parameter $\delta$ a fresh heap chunk $\chunk{f}{t}{\delta}$ is created and added to $\sigma_2$'s heap $h$, which represents $\phiAcc{e.f}$ in the state. Note, the disjoint union $\uplus$ ensures $\chunk{f}{t}{\delta}$ is not already in the heap before adding $\chunk{f}{t}{\delta}$ in there. If the chunk is in the heap, then verification will fail. Further, $\phiAcc{e.f}$ implies $e \neq \enull$ and so that fact is recorded in $\sigma_2$'s path condition as $t \neq \enull$. 

When the separating conjunction $\phiCons{\phi_1}{\phi_2}$ is produced, $\phi_1$ is first produced and then afterwards $\phi_2$ is produced into the resulting symbolic state. Note that the snapshot $\delta$ is split between the two formulas using $\first{\delta}$ and $\second{\delta}$.
Finally, to produce a conditional, \gviper branches on the symbolic value $t$ for the condition $e$ splitting execution along two different paths. Along one path only the true branch $\phi_1$ is produced into the state, and along the other path only the false branch $\phi_2$ is produced. Both paths follow the continuation to the end of its execution. More details about branching are provided next, as we describe \gviper's branch function.  

The branch function in \fig~\ref{fig:branch-func} is used to split the symbolic execution into two paths in a number of places in our algorithm: during the production or consumption of logical conditionals and during the execution of if statements. One path ($Q_t$) is taken under the assumption that the parameter $t$ is true, and the other ($Q_{\neg t}$) is taken under the assumption that $t$ is false. For each path, a branch condition corresponding to the assumption made is added to $\sigma.\rchecks$, as highlighted in blue. Additionally, paths may be pruned using check when \gviper knows for certain a path is infeasible (the assumption about $t$ would contradict the current path conditions). Now, normally, if either of the two paths fail verification, then branch marks verification as failed ($\wedge$ the results). This is still true when $\sigma$ (the current state) is precise. However, when $\sigma$ is imprecise, branch can be more permissive as highlighted in yellow. If verification fails on one of two paths only (one success, one failure), then branch returns success ($\vee$ the results). In this case, a run-time check (highlighted in blue) is added to $\rchecksall$ to force run-time execution down the success path only. Of course, two failures result in failure and two successes result in success ($\vee$ the results). No run-time checks are produced in these cases, as neither path can be soundly taken or both paths can be soundly taken at run time respectively. Note that $\gviper$ being flexible in the aforementioned way is critical to adhering to the gradual guarantee at branch points. 

\subsection{\textbf{Symbolic consumption of formulas}}
\label{sec:appendix-consume}~

\begin{figure}[!ht]
\scriptsize\ttfamily
\disableTttResize
\begin{alignat*}{2}
  &\consume{\sigma_1}{\theta}{Q}
    &&= \mlightyellow{\sigma_2 \assign \sigma_1 \{~ h,~ \pi \assign \consolidate{\sigma_1.h}{\sigma_1.\pi} ~\} } \\
    &&&\quad \consumep{\sigma_2}{\mlightyellow{\sigma_2.\isimp}}{\mlightyellow{\sigma_2.\oh}}{\sigma_2.h}{\theta}{(\lambda~ \sigma_3,~\mlightyellow{\oh'},~h_1,~\delta_1 ~.~ \\
    &&&\qquad Q(\sigma_3 \{ \mlightyellow{\oh \assign \oh'},~ h \assign h_1 \},~\delta_1))} \\
  &\mlightyellow{\consume{\sigma_1}{\withqm{\phi}}{Q}}
    &&= \mlightyellow{\sigma_2 \assign \sigma_1 \{~ h,~ \pi \assign \consolidate{\sigma_1.h}{\sigma_1.\pi} ~\} } \\
    &&&\quad \consumep{\sigma_2}{\mlightyellow{\phiTrue}}{\mlightyellow{\sigma_2.\oh}}{\sigma_2.h}{\phi}{(\lambda~ \sigma_3,~\mlightyellow{\oh'},~h_1,~\delta_1 ~.~ \\
    &&&\qquad Q(\sigma_3 \{ \mlightyellow{\isimp \assign \phiTrue,~ \oh \assign \emptyset,~ h \assign \emptyset} \},~\mlightyellow{pair(unit,~\delta_1)}))}
\end{alignat*}

\fbox{\begin{tabular}{llll}
\textcolor{light-yellow}{$\blacksquare$} & \small{Handles imprecision} &
\textcolor{light-blue}{$\blacksquare$} & \small{Handles run-time check generation and collection}
\end{tabular}}

\caption{Rules for symbolically consuming formulas (1/3)}
\label{fig:consume-rules-1}
\end{figure}

\begin{figure}[!ht]
\scriptsize\ttfamily
\disableTttResize
\ContinuedFloat
\begin{alignat*}{2}  
  &\consumep{\sigma}{\mlightyellow{f_{\qm}}}{\mlightyellow{\oh}}{h}{(\mlightblue{e},t)}{Q} 
    &&= res,~\mlightblue{\overline{t}} \assign \mlightyellow{\assertv{\sigma.\isimp}{\sigma.\pi}{t}} \\
      &&&\quad \mlightblue{\rchecks' \assign \addcheck{\sigma.\rchecks}{e}{\translate{\sigma}{\overline{t}}}} \\
      &&&\quad res \wedge Q(\sigma\{\mlightblue{\rchecks \assign \rchecks'}\},~\mlightyellow{\oh},~h,~unit) \\
  &\consumep{\sigma_1}{\mlightyellow{f_{\qm}}}{\mlightyellow{\oh}}{h}{e}{Q}
    &&= \hevalpc{\sigma_1 \{\mlightyellow{\isimp \assign f_{\qm}} \}}{e}{(\lambda~ \sigma_2,~t ~.~ \\
      &&&\hspace{-0.1cm} \consumep{\sigma_2 \{\mlightyellow{\isimp \assign \sigma_1.\isimp} \}}{\mlightyellow{f_{\qm}}}{\mlightyellow{\oh}}{h}{(\mlightblue{e},t)}{Q})}\\
  &\consumep{\sigma_1}{\mlightyellow{f_{\qm}}}{\mlightyellow{\oh}}{h}{\phiAcc{p(\overline{e})}}{Q} 
    &&= \hevalpc{\sigma_1 \{\mlightyellow{\isimp \assign f_{\qm}} \}}{\overline{e}}{(\lambda~ \sigma_2,~\overline{t} ~.~ \\
      &&&\quad \mlightyellow{\sigma_3 \assign \sigma_2 \{\isimp \assign \sigma_1.\isimp \} } \\
      &&&\quad \mlightyellow{(h_1,~\delta_1,~b_1) \assign \heapremp{\sigma_3.\isimp}{h}{\sigma_3.\pi}{p(\overline{t})}} \\
      &&&\quad \mlightyellow{\ttt{if}~~ (\sigma_3.\isimp) ~~\ttt{then}} \\
      &&&\qquad \mlightyellow{(\oh',~\delta_2,~b_2) \assign \heapremp{\sigma_3.\isimp}{\oh}{\sigma_3.\pi}{p(\overline{t})}} \\
      &&&\qquad \mlightblue{\ttt{if}~~(b_1 = b_2 = \phiFalse)~~\ttt{then}} \\
      &&&\qquad\quad \mlightblue{\rchecks' \assign \addcheck{\sigma_3.\rchecks}{\phiAcc{p(\overline{e})}}{\phiAcc{p(\overline{e})}}} \\
      &&&\qquad \mlightblue{\ttt{else}~~\rchecks' \assign \sigma_3.\rchecks} \\
      &&&\qquad \mlightyellow{Q(\sigma_3\{\mlightblue{\rchecks \assign \rchecks'}\},~\oh',~h_1,~(\ttt{if}~~(b_1)~~\ttt{then}~~\delta_1~~\ttt{else}~~\delta_2))} \\
      &&&\quad \mlightyellow{\ttt{else}~~\ttt{if}~~ (b_1) ~~\ttt{then}~~Q(\sigma_3,~\sigma_3.\oh,~h_1,~\delta_1)} \\
      &&&\quad \mlightyellow{\ttt{else}~~ \failure}
      )}\\
  &\consumep{\sigma_1}{\mlightyellow{f_{\qm}}}{\mlightyellow{h_{\qm}}}{h}{\phiAcc{e.f}}{Q} 
    &&= \hevalpc{\sigma_1 \{\mlightyellow{\isimp \assign f_{\qm} }\}}{e}{(\lambda~ \sigma_2,~t ~.~ \\
      &&&\quad \mlightyellow{\sigma_3 \assign \sigma_2 \{\isimp \assign \sigma_1.\isimp \}} \\
      &&&\quad res,~\mlightblue{\overline{t}} \assign \mlightyellow{\assertv{\sigma_3.\isimp}{\sigma_3.\pi}{t \neq \enull}} \\
      &&&\quad res ~~\wedge~~ (\\
      &&&\quad \mlightblue{\rchecks' \assign \addcheck{\sigma_3.\rchecks}{\phiAcc{e.f}}{\translate{\sigma_3}{\overline{t}}}} \\
      &&&\quad \mlightyellow{(h_1,~\delta_1,~b_1) \assign \heaprema{\sigma_3.\isimp}{h}{\sigma_3.\pi}{f(t)}} \\
      &&&\quad \mlightyellow{\ttt{if}~~ (\sigma_3.\isimp) ~~\ttt{then}} \\
      &&&\qquad \mlightyellow{(\oh',~\delta_2,~b_2) \assign \heaprema{\sigma_3.\isimp}{\oh}{\sigma_3.\pi}{f(t)}} \\
      &&&\qquad \mlightblue{\ttt{if}~~(b_1 = b_2 = \phiFalse)~~\ttt{then}} \\
      &&&\qquad\quad \mlightblue{\rchecks'' \assign \addcheck{\rchecks'}{\phiAcc{e.f}}{\phiAcc{\translate{\sigma_3}{t}.f}}} \\
      &&&\qquad \mlightblue{\ttt{else}~~\rchecks'' \assign \rchecks'} \\
      &&&\qquad \mlightyellow{Q(\sigma_3\{\mlightblue{\rchecks \assign \rchecks''}\},~h_{\qm}',~h_1,~(\ttt{if}~~(b_1)~~\ttt{then}~~\delta_1~~\ttt{else}~~\delta_2))} \\
      &&&\quad \mlightyellow{\ttt{else}~~\ttt{if}~~ (b_1) ~~\ttt{then}~~ Q(\sigma_3\{\mlightblue{\rchecks \assign \rchecks'}\},~\sigma_3.\oh,~h_1,~\delta_1)} \\
      &&&\quad \mlightyellow{\ttt{else}~~ \failure}
      ))} 
\end{alignat*}

\fbox{\begin{tabular}{llll}
\textcolor{light-yellow}{$\blacksquare$} & \small{Handles imprecision} &
\textcolor{light-blue}{$\blacksquare$} & \small{Handles run-time check generation and collection}
\end{tabular}}

\caption{Rules for symbolically consuming formulas (2/3)}
\label{fig:consume-rules-2}
\end{figure}

\begin{figure}[!ht]
\scriptsize\ttfamily
\disableTttResize
\ContinuedFloat
\begin{alignat*}{2}  
  &\consumep{\sigma_1}{\mlightyellow{f_{\qm}}}{\mlightyellow{\oh}}{h}{\phiCons{\phi_1}{\phi_2}}{Q} 
    &&= \consumep{\sigma_1}{\mlightyellow{f_{\qm}}}{\mlightyellow{\oh}}{h}{\phi_1}{(\lambda~ \sigma_2,~\mlightyellow{\oh'},~h',~\delta_1 ~.~\\
    &&&\quad \consumep{\sigma_2}{\mlightyellow{f_{\qm}}}{\mlightyellow{h_{\qm}'}}{h'}{\phi_2}{(\lambda~ \sigma_3,~\mlightyellow{h_{\qm}''},~h'',~\delta_2 ~.~\\
    &&&\qquad Q(\sigma_3,~\mlightyellow{h_{\qm}''},~h'',~pair(\delta_1,\delta_2)))})} \\
  &\consumep{\sigma_1}{\mlightyellow{f_{\qm}}}{\mlightyellow{\oh}}{h}{\phiCond{e}{\phi_1}{\phi_2}}{Q} 
    &&= \hevalpc{\sigma_1 \{\mlightyellow{\isimp \assign f_{\qm}} \}}{e}{(\lambda~ \sigma_2,~t ~.~ \\
    &&&\quad \mlightyellow{\sigma_3 \assign \sigma_2 \{\isimp \assign \sigma_1.\isimp \}} \\
    &&&\quad \hbranch{\sigma_3}{e}{t}{\\
    &&&\qquad (\lambda~ \sigma_4 ~.~ \consumep{\sigma_4}{\mlightyellow{f_{\qm}}}{\mlightyellow{h_{\qm}}}{h}{\phi_1}{Q})}{\\
    &&&\qquad (\lambda~ \sigma_4 ~.~ \consumep{\sigma_4}{\mlightyellow{f_{\qm}}}{\mlightyellow{h_{\qm}}}{h}{\phi_2}{Q})}
    )} 
\end{alignat*}

\fbox{\begin{tabular}{llll}
\textcolor{light-yellow}{$\blacksquare$} & \small{Handles imprecision} &
\textcolor{light-blue}{$\blacksquare$} & \small{Handles run-time check generation and collection}
\end{tabular}}

\caption{Rules for symbolically consuming formulas (3/3)}
\label{fig:consume-rules-3}
\end{figure}
The goals of \predicate{consume} are 3-fold: 1) given a symbolic state $\sigma$ and formula $\gphi$ check whether $\gphi$ is established by $\sigma$, \ie $\gphiImplies{\gphi_{\sigma}}{\gphi}$ where $\gphi_{\sigma}$ is the formula which represents the state $\sigma$, 2) produce and collect run-time checks that are minimally sufficient for $\sigma$ to establish $\gphi$ soundly, and 3) remove accessibility predicates and predicates that are asserted in $\gphi$ from $\sigma$. Note that $\gphiImplies{}{}$ is the consistent implication formally defined by \citet{wise2020gradual}. 
The rules for \predicate{consume} are given in \fig~\ref{fig:consume-rules-1}.

The \predicate{consume} function always begins by consolidating information across the given heap $\sigma_1.h$ and path condition $\sigma_1.\pi$. The invariant on the heap $\sigma_1.h$ ensures all heap chunks in $\sigma_1.h$ are separated in memory, \eg $\chunk{f}{x}{\delta_1} \in \sigma_1.h$ and $\chunk{f}{y}{\delta_2} \in \sigma_1.h$ implies $x \neq y$. Similarly, $\chunk{f}{x}{\delta_1} \in \sigma_1.h$ implies $x \neq \enull$. Therefore, such information is added to the path condition $\sigma_1.\pi$ during consolidation. Further, \predicate{consolidate} ensures $\sigma_1.h$ and $\sigma_1.\pi$ are consistent, \ie do not contain contradictory information.  We use the definition of \predicate{consolidate} from \cite{schwerhoff2016advancing}, without repeating it here.

After consolidation, \predicate{consume} calls a helper function \predicate{consume'}, which performs the major functionality of \predicate{consume}. Along with the state $\sigma_2$ from consolidation, \predicate{consume'} accepts a boolean flag, optimistic heap $\sigma_2.\oh$, regular heap $\sigma_2.h$, the formula to be consumed $\gphi$, and a continuation. The boolean flag sent to \predicate{consume'} controls how $\sigma_2$ provides access to fields in $\gphi$. When $\gphi$ is precise (is $\theta$), then $\sigma_2$ provides access to fields in $\theta$ through heap chunks or imprecision where applicable. Therefore, in this case, the boolean flag is set to $\sigma_2.\isimp$. However, when $\gphi$ is imprecise (\ie $\withqm{\phi}$), then the boolean flag is set to $\phiTrue$ so access to fields in $\gphi$ is always justified: first by $\sigma_2$ if applicable and second by imprecision in $\gphi$. Copies of the optimistic heap $\sigma_2.\oh$ and regular heap $\sigma_2.h$ are sent to \predicate{consume'} where heap chunks from $\gphi$ are removed from them. If \predicate{consume'} succeeds, then when $\gphi$ is precise execution continues with the residual heap chunks. When $\gphi$ is imprecise execution continues with empty heaps, because $\gphi$ may require and assert any heap chunk in $\sigma_2$. Residual heap chunks are instead represented by imprecision, \ie execution continues with an imprecise state. Finally, \predicate{consume'} also sends snapshots collected for removed heap chunks to the continuation.

Rules for \predicate{consume'} can also be found in \fig~\ref{fig:consume-rules-2}. Cases for expressions $e$, the separating conjunction $\phiCons{\phi_1}{\phi_2}$, and logical conditionals $\phiCond{e}{\phi_1}{\phi_2}$ 
are straightforward. Expressions are evaluated to symbolic values that are then consumed with the corresponding rule. In a separating conjunction, $\phi_1$ is consumed first, then afterward $\phi_2$ is consumed. The rule for logical conditionals evaluates the condition $e$ to a symbolic value, and then uses the \predicate{branch} function to consume $\phi_1$ and $\phi_2$ along different execution paths. The case for $\phiAcc{p(\overline{e})}$ is also 
very similar to the case for $\phiAcc{e.f}$ that we discuss later in this section.

When a symbolic value $t$ is consumed, the current state $\sigma$ must establish $t$, \ie $\gphiImplies{\sigma}{t}$, or verification fails. The \predicate{assert} function (defined in \fig~\ref{fig:check-assert-funcs}) implements this functionality. In particular, \predicate{assert} returns $\success$ when $\pi$ can statically prove $t$ or when $\sigma$ is imprecise and $t$ does not contradict constraints in $\pi$---here, $t$ is optimistically assumed to be true. Otherwise, \predicate{assert} returns $\failure$. When \predicate{assert} succeeds, it also returns a set of symbolic values $\overline{t}$ that are residuals of $t$ that cannot be proved statically by $\pi$. If $t$ is proven entirely statically, then \predicate{assert} returns $\phiTrue$. A run-time check is created for the residuals $\overline{t}$ and is added to $\sigma$ to be passed to the continuation $Q$. Note that \predicate{translate} is used to create an expression from $\overline{t}$ that can be evaluated at run time. Further, the location $e$ is the expression that evaluates to $t$ and is passed to \predicate{consume'} alongside $t$. The heaps $\oh$ and $h$ are passed unmodified to $Q$ alongside the snapshot $unit$.

The \predicate{consume'} rule for accessibility predicates $\phiAcc{e.f}$, first evaluates the receiver $e$ to $t$ using \predicate{eval-c}, the current state $\sigma_1$, and the parameter $f_\qm$. The parameter $f_\qm$ is the boolean flag mentioned previously. Assigning $f_\qm$ to $\sigma_1.\isimp$ during evaluation allows $f_\qm$ to control whether or not imprecision verifies field accesses. This occurs in all of the \predicate{consume'} rules where expressions and thus fields are evaluated. After evaluation, the $\isimp$ field is reset resulting in $\sigma_3$, and \predicate{assert} is used to ensure the receiver $t$ is non-null. If $t \neq \enull$ is optimistically true, a run-time check for $t \neq \enull$ at location $\phiAcc{e.f}$ is created and added to $\sigma_3.\rchecks$.
Next, \predicate{heap-rem-acc} is used to remove the heap chunks from heap $h$ that overlap with or may potentially overlap with $\phiAcc{e.f}$ in memory. The \predicate{heap-rem-acc} function is formally defined alongside a similar function for predicates (\predicate{heap-rem-pred}) in the \fig~\ref{fig:heap-add-rem}. If a field chunk is not statically proven to be disjoint from $\phiAcc{e.f}$, then it is removed. Further, since predicates are opaque, \gviper cannot tell whether or not their predicate bodies overlap with $\phiAcc{e.f}$. Therefore, predicate chunks are almost always considered to potentially overlap with $\phiAcc{e.f}$. The only time this is not the case is if they both exist in the heap $h$, which ensures its heap chunks do not overlap in memory. The \predicate{heap-rem-acc} function also checks that $\phiAcc{e.f}$ has a corresponding heap chunk in $h$. If so, its snapshot $\delta_1$ is returned and $b_1$ is assigned $\phiTrue$. Otherwise, a fresh snapshot is returned with $\phiFalse$. If the current state $\sigma_3$ is imprecise, then heap chunks are similarly removed from $\oh$ and $\phiAcc{e.f}$ is checked for existence in $\oh$. If a field chunk for $\phiAcc{e.f}$ is not found in either heap, then a run-time check is generated for it and passed to the continuation $Q$ alongside the two heaps after removal and $\phiAcc{e.f}$'s snapshot. Without imprecision, \predicate{consume'} will fail when a field chunk for $\phiAcc{e.f}$ is not found in $h$.

\begin{figure}[!ht]
\scriptsize\ttfamily
\disableTttResize
\begin{alignat*}{2}
  &\heapremp{\isimp}{h}{\pi}{p(\overline{t})}
   &&= \ttt{if}~~ \exists~ (\chunk{p}{\overline{r}}{\delta} \in h ~.~ \checkv{\pi}{\bigwedge \overline{t=r}}) ~~\ttt{then} \\
     &&&\qquad (h \setminus \{ \chunk{p}{\overline{r}}{\delta} \},~\delta,~\phiTrue) \\
     &&&\quad \ttt{else}~~ (\emptyset,~\fresh,~\phiFalse) \\
  &\heaprema{\isimp}{h}{\pi}{f(t)}
   &&= h' \assign \foldl{h}{\emptyset}{(\lambda~ \chunk{f_{src}}{\overline{r}}{\delta},~h_{dst} ~.~ \\
     &&&\qquad \ttt{if}~~ (\neg (|\overline{r}|=1) \parallel \neg (f = f_{src}) \parallel \neg \checkqm{\isimp}{\pi}{t = r} )~~\ttt{then} \\
     &&&\qquad\quad h_{dst} \cup \chunk{f_{src}}{\overline{r}}{\delta} \\
     &&&\qquad \ttt{else}~~ h_{dst})
     } \\
     &&&\quad \ttt{if}~~ \exists~ \chunk{f}{r}{\delta} \in h ~.~ \checkv{\pi}{t = r} ~~\ttt{then} \\
     &&&\qquad (h',~\delta,~\phiTrue) \\
     &&&\quad \ttt{else}~~ \\
     &&&\qquad h' \assign \foldl{h'}{\emptyset}{(\lambda~ \chunk{f_{src}}{\overline{r}}{\delta},~h_{dst} ~.~ \\
     &&&\qquad\quad \ttt{if}~~ (\chunk{f_{src}}{\overline{r}}{\delta}~~\ttt{is a field chunk})~~\ttt{then} \\
     &&&\qquad\qquad h_{dst} \cup \chunk{f_{src}}{\overline{r}}{\delta} \\
     &&&\qquad\quad \ttt{else}~~ h_{dst})
     } \\
     &&&\qquad (h',~\fresh,~\phiFalse)
\end{alignat*}
\caption{Heap remove function definitions}
\label{fig:heap-add-rem}
\end{figure}

\begin{figure}[!ht]
\scriptsize\ttfamily
\disableTttResize
\begin{alignat*}{2}
  &\checkv{\pi}{t} = \phiImplies{\pcall{\pi}}{t}\\
  &\checkqm{\mlightyellow{\isimp}}{\pi}{t} =     
    \begin{cases}
	\phiTrue,~ \mlightblue{\phiTrue} & \texttt{if } \checkv{\pi}{t} \\
	\mlightyellow{\phiTrue,~ \mlightblue{\diff(\pcall{\pi},t)}} & \mlightyellow{\ttt{if}~ (\isimp \wedge (\bigwedge \pcall{\pi} \wedge t)_{\ttt{SAT}})} \\
	\phiFalse,~ \mlightblue{\emptyset} & \texttt{otherwise} \\		
    \end{cases}\\
  &\assertv{\mlightyellow{\isimp}}{\pi}{t} = 
    \begin{cases}
	\success,~ \mlightblue{\overline{t}} & \mlightyellow{\texttt{if } (b=\phiTrue)~~\ttt{where}~~b,~\mlightblue{\overline{t}} \assign \checkqm{\isimp}{\pi}{t}} \\
	\failure,~ \mlightblue{\emptyset} & \texttt{otherwise} \\		
    \end{cases}
\end{alignat*}

\fbox{\begin{tabular}{llll}
\textcolor{light-yellow}{$\blacksquare$} & \small{Handles imprecision} &
\textcolor{light-blue}{$\blacksquare$} & \small{Handles run-time check generation and collection}
\end{tabular}}

\caption{Check and assert function definitions}
\label{fig:check-assert-funcs}
\end{figure}

\subsection{\textbf{Symbolic execution of statements}}
\label{sec:appendix-exec}

\begin{figure}[!ht]
\scriptsize\ttfamily
\disableTttResize
\begin{alignat*}{2}
  &\exec{\sigma_1}{\sSeq{s_1}{s_2}}{Q}
    &&= \exec{\sigma_1}{s_1}{(\lambda~ \sigma_2 ~.~ \exec{\sigma_2}{s_2}{Q})} \\
  &\exec{\sigma}{\sDeclare{T}{x}}{Q}
    &&=  Q(\sigma \{ \gamma \assign \havoc{\sigma.\gamma}{x} \})\\
  &\exec{\sigma_1}{\sVarAssign{x}{e}}{Q}
    &&=  \heval{\sigma_1}{e}{(\lambda~ \sigma_2,~t ~.~ Q(\sigma_2 \{ \gamma \assign \sigma_2.\gamma[x \mapsto t]\}))} \\
  &\exec{\sigma_1}{\sFieldAssign{x}{f}{e}}{Q}
    &&= \heval{\sigma_1}{e}{(\lambda~ \sigma_2,~t ~.~
       \hconsume{\sigma_2}{\phiAcc{x.f}}{(\lambda~ \sigma_3,~\_ ~.~ \\
      &&&\qquad \hproduce{\sigma_3}{\phiCons{\phiAcc{x.f}}{x.f = t}}{pair(\fresh,~unit)}{Q})})}\\
  &\exec{\sigma}{\sAlloc{x}{\overline{f}}}{Q}
     &&= \hproduce{\sigma \{ \gamma \assign \havoc{\sigma.\gamma}{x} \}}{\overline{\phiAcc{x.f}}}{\fresh}{Q}\\
  &\exec{\sigma_1}{\sCall{\overline{z}}{m}{\overline{e}}}{Q}
    &&=  \heval{\sigma_1}{\overline{e}}{(\lambda~ \sigma_2,~\overline{t} ~.~ \\
      &&&\qquad \mlightblue{\rchecks' \assign \sigma_2.\rchecks\{ \origin \assign (\sigma_2,~\sCall{\overline{z}}{m}{\overline{e}},~ \overline{t}) \}} \\
      &&&\qquad \hconsume{\sigma_2\{\mlightblue{\rchecks \assign \rchecks'}\}}{meth_{pre}[\overline{meth_{args} \mapsto t}]}{(\lambda~ \sigma_3,~\delta ~.~ \\
      &&&\qquad\quad \mlightyellow{\ttt{if}~~(\iseimp{meth_{pre}})~~\ttt{then}} \\
      &&&\qquad\qquad \mlightyellow{\sigma_4 \assign \sigma_3 \{ \isimp \assign \phiTrue,~ \oh \assign \emptyset,~ h \assign \emptyset,}\\
      &&&\qquad\qquad\qquad\qquad\quad\mlightyellow{\gamma \assign \havoc{\sigma_3.\gamma}{\overline{z}} \}} \\
      &&&\qquad\quad \mlightyellow{\ttt{else}}~~\sigma_4 \assign \sigma_3 \{ \gamma \assign \havoc{\sigma_3.\gamma}{\overline{z}} \} \\
      &&&\qquad\quad \hproduce{\sigma_4}{meth_{post}[\overline{meth_{args} \mapsto t}][\overline{meth_{ret} \mapsto z}]}{\fresh}{ \\
      &&&\qquad\qquad (\lambda~ \sigma_5 ~.~ Q(\sigma_5\{\mlightblue{\rchecks \assign \sigma_5.\rchecks\{ \origin \assign \none \}} \})
      )})})}\\
  &\exec{\sigma_1}{\sAssert{\phi}}{Q}
    &&=  \hconsume{\sigma_1}{\phi}{(\lambda~ \sigma_2,~\delta ~.~ \\
    &&&\qquad \hwellformed{\sigma_2}{\mlightyellow{\qm ~\&\&~ \phi}}{\delta}{(\lambda~ \sigma_3 ~.~ \\
    &&&\qquad\quad Q(\sigma_1 \{\mlightyellow{ \pi \assign \sigma_3.\pi},~ \mlightblue{\rchecks \assign \sigma_3.\rchecks}\}))})}\\
  &\exec{\sigma_1}{\sFold{\phiAcc{p(\overline{e})}}}{Q}
    &&=  \heval{\sigma_1}{\overline{e}}{(\lambda~ \sigma_2,~\overline{t} ~.~ \\
      &&&\qquad \mlightblue{\rchecks' \assign \sigma_2.\rchecks\{ \origin \assign (\sigma_2,~\sFold{\phiAcc{p(\overline{e})}},~ \overline{t}) \}} \\
      &&&\qquad \hconsume{\sigma_2\{ \mlightblue{\rchecks \assign \rchecks'} \}}{pred_{body}[\overline{pred_{args} \mapsto t}]}{(\lambda~ \sigma_3,~\delta ~.~ \\
      &&&\qquad\quad \hproduce{\sigma_3\{\mlightblue{\rchecks \assign \sigma_3.\rchecks\{ \origin \assign \none \}}\}}{\phiAcc{p(\overline{t})}}{\delta}{\\
      &&&\qquad\qquad Q})}))}\\
  &\exec{\sigma_1}{\sUnfold{\phiAcc{p(\overline{e})}}}{Q}
    &&=  \heval{\sigma_1}{\overline{e}}{(\lambda~ \sigma_2,~\overline{t} ~.~ \\
      &&&\qquad \mlightblue{\rchecks' \assign \sigma_2.\rchecks\{ \origin \assign (\sigma_2,~\sUnfold{\phiAcc{p(\overline{e})}},~ \overline{t}) \}} \\
      &&&\qquad \hconsume{\sigma_2\{ \mlightblue{\rchecks \assign \rchecks'} \}}{\phiAcc{p(\overline{t})}}{(\lambda~ \sigma_3,~\delta ~.~ \\
      &&&\qquad\quad \hproduce{\sigma_3}{pred_{body}[\overline{pred_{args} \mapsto t}]}{\delta}{(\lambda~\sigma_4 ~.~\\
      &&&\qquad\qquad Q(\sigma_4\{\mlightblue{\rchecks \assign \sigma_4.\rchecks\{ \origin \assign \none \}}\}))})}))}\\
  &\exec{\sigma_1}{\sIf{e}{stmt_1}{stmt_2}}{Q}
    &&=  \heval{\sigma_1}{e}{(\lambda~ \sigma_2,~t ~.~ \\
      &&& \hspace{-2.5cm} \quad \hbranch{\sigma_2}{e}{t}{ 
      (\lambda~ \sigma_3 ~.~ \exec{\sigma_3}{stmt_1}{Q})}{
      (\lambda~ \sigma_3 ~.~ \exec{\sigma_3}{stmt_2}{Q})
      })}\\
\end{alignat*}

\fbox{\begin{tabular}{llll}
\textcolor{light-yellow}{$\blacksquare$} & \small{Handles imprecision} &
\textcolor{light-blue}{$\blacksquare$} & \small{Handles run-time check generation and collection}
\end{tabular}}

\caption{Rules for symbolically executing program statements (1/2)}
\label{fig:exec-rules}
\end{figure}

\begin{figure}[!ht]
\scriptsize\ttfamily
\disableTttResize
\ContinuedFloat
\begin{alignat*}{2}
   &\exec{\sigma_1}{\sWhile{e}{\mlightyellow{\gphi}}{stmt}}{Q}
    &&=\quad \gamma_2 \assign \havoc{\sigma_1.\gamma}{\overline{x}} \\
      &&&\hspace{-3cm} \mlightblue{res_{body}} \assign \hwellformed{\\
      &&& \hspace{-3cm} \quad \sigma_1 \{ \mlightyellow{\isimp \assign \phiFalse,~ h_{\qm} \assign \emptyset},~ h \assign \emptyset,~ \gamma \assign \gamma_2,~ \\
      &&&\hspace{-3cm} \qquad \mlightblue{\rchecks \assign \sigma_1.\rchecks\{ \origin \assign (\sigma_1,~ \sWhile{e}{\gphi}{stmt},~ \texttt{beginning}) \}} \}}{\\
      &&&\hspace{-3cm} \quad \mlightyellow{\phiCons{\gphi}{e}}}{\fresh}{ (\lambda~ \mlightblue{\sigma_3\{\rchecks \assign \sigma_3.\rchecks\{ \origin \assign \none \}\}} ~.~\\
      &&&\hspace{-3cm} \qquad \exec{\sigma_3}{stmt}{(\lambda~ \sigma_4 ~.~\\
      &&&\hspace{-3cm} \qquad\quad \heval{\mlightblue{\sigma_4\{\rchecks \assign \sigma_4.\rchecks\{ \origin \assign (\sigma_4,~ \sWhile{e}{\gphi}{stmt},~ \texttt{end}) \}\}}}{\\
      &&&\hspace{-3cm} \qquad\qquad e}{(\lambda~ \sigma_e,~\_ ~.\\
      &&&\hspace{-3cm} \qquad\qquad\qquad \hconsume{\mlightblue{\sigma_4\{\rchecks \assign \sigma_4.\rchecks\{ \origin \assign \sigma_e.\rchecks.\origin,~\rcs \assign \sigma_e.\rchecks.\rcs\}\}}}{\\
      &&&\hspace{-3cm} \qquad\qquad\qquad\quad \mlightyellow{\gphi}}{(\lambda~ \sigma_5,~\_ ~.~ \mlightblue{\rchecksall \assign \rchecksall \cup \sigma_5.\rchecks.\rcs} ~;~ \success)})})})} \\
      &&&\hspace{-3cm} \mlightblue{res_{after}} \assign \heval{\mlightblue{\sigma_1\{\rchecks \assign \sigma_1.\rchecks\{ \origin \assign (\sigma_1,~ \sWhile{e}{\gphi}{stmt},~ \texttt{before}) \} \}}}{ \\ 
      &&&\hspace{-3cm} \quad e}{(\lambda~ \sigma_e,~\_~.~\hconsume{ \\
      &&&\hspace{-3cm} \qquad \mlightblue{\sigma_1\{\rchecks \assign \sigma_1.\rchecks\{ \origin \assign \sigma_e.\rchecks.\origin,~ \rcs \assign \sigma_e.\rchecks.\rcs \} \}}}{\\
      &&&\hspace{-3cm} \qquad \mlightyellow{\gphi}}{(\lambda~ \sigma_2,~ \_ ~.~\\
      &&&\hspace{-3cm} \qquad\quad \mlightyellow{\ttt{if}~~(\iseimp{\gphi})~~\ttt{then}} \\
      &&&\hspace{-3cm} \qquad\qquad \mlightyellow{\sigma_3 \assign \sigma_2 \{ \isimp \assign \phiTrue,~ \oh \assign \emptyset,~ h \assign \emptyset,~ \gamma \assign \gamma_2 \}} \\
      &&&\hspace{-3cm} \qquad\quad \mlightyellow{\ttt{else}}~~\sigma_3 \assign \sigma_2 \{ \gamma \assign \gamma_2 \} \\
      &&&\hspace{-3cm} \qquad\quad \hproduce{\sigma_3 \{ \mlightblue{\sigma_3.\rchecks\{ \origin \assign (\sigma_3,~\sWhile{e}{\gphi}{stmt},~ \texttt{after}) \}} \}}{\\
      &&&\hspace{-3cm} \qquad\qquad \mlightyellow{\phiCons{\gphi}{!e}}}{\fresh}{Q})})}\\
      &&&\hspace{-3cm}\mlightyellow{\texttt{if ($\sigma_1$.isImprecise) then}}\\
      &&&\hspace{-3cm} \quad \mlightblue{\texttt{if ($\neg res_{body} \wedge res_{after}$) then}}\\
      &&&\hspace{-3cm} \qquad \mlightblue{\texttt{$\rchecks ' \assign \addcheck{\sigma_1.\rchecks}{e}{\neg e}$}}\\
      &&&\hspace{-3cm} \qquad \mlightblue{\texttt{$\rchecksall \assign \rchecksall \cup \rchecks'.\rcs.\predicate{last}$}}\\
      &&&\hspace{-3cm} \quad \mlightyellow{\texttt{($\neg res_{body} \vee res_{after}) \wedge (res_{body} \vee res_{after}$)}}\\
      &&&\hspace{-3cm}\texttt{else}\\
      &&&\hspace{-3cm} \quad \texttt{$res_{body} \wedge res_{after}$}\\
    &&&\hspace{-4cm} \ttt{where $\overline{x}$ are variables modified by the loop body}
\end{alignat*}

\fbox{\begin{tabular}{llll}
\textcolor{light-yellow}{$\blacksquare$} & \small{Handles imprecision} &
\textcolor{light-blue}{$\blacksquare$} & \small{Handles run-time check generation and collection}
\end{tabular}}

\caption{Rules for symbolically executing program statements (2/2)}
\label{fig:rules-exec-while}
\end{figure}

\begin{figure}[!ht]
\scriptsize\ttfamily
\disableTttResize
\begin{alignat*}{2}
&\iseimp{\withqm{\phi}}
  &&=  \phiTrue\\
&\iseimp{\phiCons{\phi_1}{\phi_2}}
  &&= \iseimp{\phi_1} \lor \iseimp{\phi_2} \\
&\iseimp{\phiCond{e}{\phi_1}{\phi_2}}
  &&= \iseimp{\phi_1} \lor \iseimp{\phi_2} \\
&\iseimp{\phiAcc{p(\overline{e})}}
  &&= \ttt{if}~~(p \in \ttt{VisitedPreds})~~\ttt{then} \\
    &&&\qquad \phiFalse \\
    &&&\quad \ttt{else}\\
    &&&\qquad \ttt{VisitedPreds} \assign \ttt{VisitedPreds} \cup p \\
    &&& \qquad b \assign \iseimp{pred_{body}}\\
    &&&\qquad \ttt{VisitedPreds} \assign \ttt{VisitedPreds} \setminus p \\
    &&&\qquad b \\
&\iseimp{\_}
  &&= \phiFalse \\
\end{alignat*}
\caption{Boolean function determining if a gradual formula is equi-recursively imprecise or not}
\label{fig:equi-imprecise}
\end{figure}


The \predicate{exec} rules for sequence statements, variable declarations and assignments, allocations, and if statements are pretty much unchanged from Viper. The only difference is that \gviper's versions of \predicate{eval}, \predicate{produce}, \predicate{branch}, and \predicate{consume} (defined previously) are used instead of Viper's. Statements in a sequence are executed one after another, and variable declarations introduce a fresh symbolic value for the variable into the state. Variable assignments evaluate the right-hand side to a symbolic value and update the variable in the symbolic store with the result. Allocations produce fresh heap chunks for fields into the state. Finally, if statements have their condition evaluated and then branch is used to split execution along two paths to symbolically execute the true and false branches.

Symbolic execution of field assignments first evaluates the right-hand side expression $e$ to the symbolic value $t$ with the current state $\sigma_1$ and \predicate{eval}. Any field reads in $e$ are either directly or optimistically verified using $\sigma_1$. Then, the resulting state $\sigma_2$ must establish write access to $x.f$ in \predicate{consume}, \ie $\gphiImplies{\sigma_2}{\phiAcc{x.f}}$. The call to \predicate{consume} also removes the field chunk for $\phiAcc{x.f}$ from $\sigma_2$ (if it is in there) resulting in $\sigma_3$. Therefore, the call to \predicate{produce} can safely add a fresh field chunk for $\phiAcc{x.f}$ alongside $x.f = t$ to $\sigma_3$ before it is passed to the continuation $Q$. Under the hood, run-time checks are collected where required for soundness and passed to $Q$.

The \predicate{exec} rule for method calls similarly uses \predicate{eval} to evaluate the given args $\overline{e}$ to symbolic values $\overline{t}$, asserts the method's precondition $meth_{pre}$ holds in the current state, consumes the heap chunks in the precondition, and produces the method's postcondition $meth_{post}$ into the continuation. Run-time checks are also collected where necessary (under the hood) and passed to the continuation. \gviper makes an exception when consuming preconditions at method calls (and loop invariants before entering loops), which can be seen in the \ttt{if-then} in the method call rule. If \gviper determines the precondition (invariant) is equi-recursively imprecise, then it will conservatively remove all the heap chunks from both symbolic heaps after the consume. This exception ensures the static verification semantics in \gviper lines up with the equi-recursive, dynamic verification semantics encoded by \gvc in \S\ref{sec:c0-runtime-checks} such that \gco is sound. Note that the $\origin$ field of $\rchecks$ is set to $\sCall{\overline{z}}{m}{\overline{e}}$ before consuming $meth_{pre}$ and reset to $\none$ after producing $meth_{post}$. Setting the origin indicates that run-time checks or branch conditions for $meth_{pre}$ or $meth_{post}$ should be attached to the method call statement rather than where they are declared. The origin arguments $\sigma_2$ and $\overline{t}$ are used to reverse the substitution $[\overline{meth_{args} \mapsto t}]$ in run-time checks and branch conditions for $meth_{pre}$ and $meth_{post}$. The rule for (un)folding predicates operates the same as for method calls where $meth_{pre}$ is the predicate body (predicate instance) and $meth_{post}$ is the predicate instance (predicate body). The $\origin$ is set to $\sFold{\phiAcc{p(\overline{e})}}$ and $\sUnfold{\phiAcc{p(\overline{e})}}$ respectively.

In contrast, $\phi$ in $\sAssert{\phi}$ maintains a $\none$ origin field, because $\phi$'s use and declaration align at the same program location $\sAssert{\phi}$. The assert rule relies on \predicate{consume} to assert $\phi$ holds in the current state $\sigma_1$. If the consume succeeds, the state $\sigma_1$ is passed to the continuation nearly unmodified. Path condition constraints from $\phi$ hold in $\sigma_1$ either directly or optimistically. Therefore, these constraints are added to $\sigma_1$ to avoid producing run-time checks for them in later program statements. Run-time checks from the consume are also passed to the continuation.
Note that $\phi$ is checked for well-formedness here (\fig~\ref{fig:wellformed-func}). A formula is well-formed if it contains $\qm$ or accessibility predicates that verify access to the formula's fields (\emph{self-framing}). Additionally, the formula cannot contain duplicate accessibility predicates or predicate instances. Finally, \predicate{well-formed} adds the formula's information to the given symbolic state. Here, $\phi$ does not need to be self-framed, and so it is joined with $\qm$ in the call to \predicate{well-formed}. $\qm$ verifies access to all of $\phi$'s fields.

\begin{figure}[!ht]
\scriptsize\ttfamily
\disableTttResize
\begin{alignat*}{2}
    & \wellformed{\sigma_1}{\gphi}{\delta}{Q}
    &&= \produce{\sigma_1}{\gphi}{\delta}{(\lambda~ \sigma_2 ~.~\\
    &&&\qquad \produce{\sigma_1\{~ \pi \assign \sigma_2.\pi ~\}}{\gphi}{\delta}{Q})}
\end{alignat*}
\caption{Well-formed formula function definition}
\label{fig:wellformed-func}
\end{figure}

Finally, while the while loop rule is the largest rule and looks fairly complex, it just combines ideas from other rules that are discussed in great detail in this section and from the branch rule described in \S\ref{sec:appendix-produce}.

\subsection{\textbf{Valid program}}
\label{sec:gviper-validprogram}
A \gviper program is valid if all of its method and predicate declarations are verified successfully as defined in \fig~\ref{fig:valid-program}. 
In particular, a method $m$'s declaration is verified first by checking well-formedness of $m$'s precondition $meth_{pre}$ and postcondition $meth_{post}$ using the empty state $\sigma_0$ (well-formedness is described in \S\ref{sec:appendix-exec}). Note, fresh symbolic values are created and added to $\sigma_0$ for $m$'s argument variables $\overline{x}$ and return variables $\overline{y}$. If $meth_{pre}$ and $meth_{post}$ are well-formed, then the body of $m$ ($meth_{body}$) is symbolically executed (\S\ref{sec:appendix-exec}) starting with the symbolic state $\sigma_1$ containing $meth_{pre}$. Recall, \predicate{well-formed} additionally produces the formula that is being checked into the symbolic state. The symbolic state $\sigma_2$ is produced after the symbolic execution of $meth_{body}$. Then, $meth_{post}$ is checked for validity against $\sigma_2$, \ie $\sigma_2$ must establish $meth_{post}$ (\S\ref{sec:appendix-consume}). If $meth_{post}$ is established, then verification succeeds; and as a result, the run-time checks collected during verification are added to $\rchecksall$ (highlighted in blue).
A valid predicate $p$ is simply valid if $p$'s body $pred_{body}$ is well-formed. As before, fresh symbolic values are created for $p$'s argument variables $\overline{x}$. Note, no run-time checks are added to $\rchecksall$ here, because well-formedness checks do not produce any run-time checks.

\end{document}

%% file: definitions.tex
\newunicodechar{≠}{=\llap{/}}
\newunicodechar{≔}{:\raisebox{-0.18ex}[\height][\depth]{=}}
\newunicodechar{≤}{\rlap{\begingroup \fontsize{13pt}{13pt}\selectfont \raisebox{-0.58ex}[\height][\depth]{-} \endgroup}\shifttext{0.15ex}{\raisebox{0.18ex}[\height][\depth]{<}}}
\newunicodechar{≥}{\rlap{\begingroup \fontsize{13pt}{13pt}\selectfont \raisebox{-0.58ex}[\height][\depth]{-} \endgroup}\shifttext{0.15ex}{\raisebox{0.18ex}[\height][\depth]{>}}}
\newunicodechar{∧}{\shifttext{1ex}{\begingroup \fontsize{8pt}{8pt}\selectfont {\shifttext{0.4ex}{/}\shifttext{-0.4ex}{\textbackslash}} \endgroup}}
\newunicodechar{∨}{\shifttext{1ex}{\begingroup \fontsize{8pt}{8pt}\selectfont {\shifttext{0.4ex}{\textbackslash}\shifttext{-0.4ex}{/}} \endgroup}}

\makeatletter
\newcommand*{\shifttext}[2]{%
	\settowidth{\@tempdima}{#2}%
	\makebox[\@tempdima]{\hspace*{#1}#2}%
}
\makeatother

\newcommand{\ie}{\emph{i.e.}\xspace}

\newcommand{\eg}{\emph{e.g.}\xspace}
\newcommand{\fig}{Fig.\xspace}
\usepackage[leftmargin=3em,vskip=0em]{quoting}
  
\definecolor{light-gray}{gray}{0.87}
\newcommand{\lightgray}[1]{\colorbox{light-gray}{#1}}
\definecolor{gray}{gray}{0.75}

\definecolor{drk-gray}{RGB}{169,169,169}
\newcommand{\drkgray}[1]{\colorbox{drk-gray}{#1}}

\definecolor{light-purple}{RGB}{229,204,255}
\newcommand{\lightpurple}[1]{\colorbox{light-purple}{#1}}

\definecolor{light-yellow}{RGB}{255,228,181}
\newcommand{\lightyellow}[1]{\colorbox{light-yellow}{#1}}
\newcommand{\mlightyellow}[1]{\mcolor{light-yellow}{#1}}

\definecolor{light-blue}{RGB}{189,215,238}
\newcommand{\lightblue}[1]{\colorbox{light-blue}{#1}}
\newcommand{\mlightblue}[1]{\mcolor{light-blue}{#1}}

\definecolor{light-green}{RGB}{152,251,152}
\newcommand{\lightgreen}[1]{\colorbox{light-green}{#1}}

\definecolor{light-red}{RGB}{255,204,204}
\newcommand{\lightred}[1]{\colorbox{light-red}{#1}}

\definecolor{rred}{RGB}{255,153,153}

\definecolor{light-pink}{RGB}{255,204,255}
\newcommand{\lightpink}[1]{\colorbox{light-pink}{#1}}
\newcommand{\mlightpink}[1]{\mcolor{light-pink}{#1}}

\definecolor{light-orange}{RGB}{255,204,153}
\newcommand{\lightorange}[1]{\colorbox{light-orange}{#1}}
\newcommand{\mlightorange}[1]{\mcolor{light-orange}{#1}}

\definecolor{neon-blue}{RGB}{153,255,255}
\newcommand{\neonblue}[1]{\colorbox{neon-blue}{#1}}

\definecolor{neon-yellow}{RGB}{255,255,153}
\newcommand{\neonyellow}[1]{\colorbox{neon-yellow}{#1}}

\newcommand{\mcolor}[2]{\colorbox{#1}{$\displaystyle #2$}}

\newcommand{\grad}[1]{\widetilde{#1}}

\newcommand{\gphi}{\grad{\phi}}
\newcommand{\sphi}{\theta}

\newcommand{\std}{\textrm}
\newcommand{\ttt}[1]{\textup{\texttt{\small{#1}}}}
\newcommand{\disableTttResize}[0]{\renewcommand{\small}[1]{##1}}

\newcommand{\predicate}[1]{\textup{\textmd{\textsf{#1}}}}

\newcommand{\gco}{\std{Gradual C0}\xspace}
\newcommand{\gviper}{\std{Gradual Viper}\xspace}
\newcommand{\gvc}{\std{GVC0}\xspace}

\newcommand{\eresult}{\ttt{result}}
\newcommand{\phiTrue}[0]{\ttt{true}}
\newcommand{\phiFalse}[0]{\ttt{false}}

\newcommand{\qm}{\ttt{?}}

\newcommand{\diff}{\predicate{diff}}

\newcommand{\sVarAssign}[2]{{#1}~\ttt{≔}~{#2}}

\newcommand{\sAssert}[1]{\ttt{assert}~#1}
\newcommand{\sDeclare}[2]{\ttt{var}~#2:#1}
\newcommand{\sSeq}[2]{{#1}\ttt{;}~{#2}}
\newcommand{\type}[1]{\ttt{#1}}
\newcommand{\Tint}{\type{Int}}
\newcommand{\Tbool}{\type{Bool}}
\newcommand{\Tref}{\type{Ref}}

\newcommand{\pred}[3]{{#1}\ttt{(}{#2}\ttt{)}~\ttt{\{}~{#3}~\ttt{\}}}
\newcommand{\contract}[2]{\ttt{requires}~{#1}~\ttt{ensures}~{#2}}

\newcommand{\enull}{\ttt{null}}
\newcommand{\phiAcc}[1]{\ttt{\textbf{acc}}(#1)}
\newcommand{\phiCons}[2]{{#1}\:\&\&\:{#2}}
\newcommand{\phiAnd}{~\&\&~}
\newcommand{\phiCond}[3]{#1~\ttt{?}~#2~\ttt{:}~#3}
\newcommand{\withqm}[1]{\phiCons{\qm}{\ensuremath{#1}}}

\newcommand{\sFieldAssign}[3]{#1\ttt{.}#2~\ttt{≔}~#3}
\newcommand{\sAlloc}[2]{#1~\ttt{≔}~\ttt{new}(#2)}
\newcommand{\sCall}[3]{#1~\ttt{≔}~#2\ttt{(}#3\ttt{)}}
\newcommand{\sWhile}[3]{\ttt{while}~(#1)~\ttt{invariant}~#2~\{~#3~\}}
\newcommand{\sWhilegvc}[3]{\ttt{while}~(#1)~\ttt{//@loop\_invariant}~#2~\{~#3~\}}

\newcommand{\sForgvc}[5]{\ttt{for}~(#1;~#2;~#3)~\ttt{//@loop\_invariant}~#4~\{~#5~\}}
\newcommand{\sIf}[3]{\ttt{if}~(#1)~\{~#2~\}~\ttt{else}~\{~#3~\}}
\newcommand{\sFold}[1]{\ttt{fold}~{#1}}
\newcommand{\sUnfold}[1]{\ttt{unfold}~{#1}}

\newcommand{\phiImplies}[2]{{#1} \,\,{\Rightarrow}\,\, {#2}}

\newcommand{\gphiImplies}[2]{{#1} \,\,{\grad{\Rightarrow}}\,\, {#2}}

\newcommand{\exec}[3]{\predicate{exec}\ttt{(}#1,~#2,~#3\ttt{)}}
\newcommand{\hexec}[3]{\lightyellow{\predicate{exec}}\ttt{(}#1,~#2,~#3\ttt{)}}
\newcommand{\produce}[4]{\predicate{produce}\ttt{(}#1,~#2,~#3,~#4\ttt{)}}
\newcommand{\hproduce}[4]{\lightyellow{\predicate{produce}}\ttt{(}#1,~#2,~#3,~#4\ttt{)}}
\newcommand{\consume}[3]{\predicate{consume}\ttt{(}#1,~#2,~#3\ttt{)}}
\newcommand{\hconsume}[3]{\lightyellow{\predicate{consume}}\ttt{(}#1,~#2,~#3\ttt{)}}
\newcommand{\consumep}[6]{\predicate{consume'}\ttt{(}#1,~#2,~#3,~#4,~#5,~#6\ttt{)}}
\newcommand{\eval}[3]{\predicate{eval}\ttt{(}#1,~#2,~#3\ttt{)}}
\newcommand{\heval}[3]{\lightyellow{\predicate{eval}}\ttt{(}#1,~#2,~#3\ttt{)}}

\newcommand{\evalpc}[3]{\predicate{eval-c}\ttt{(}#1,~#2,~#3\ttt{)}}
\newcommand{\evalp}[3]{\predicate{eval-p}\ttt{(}#1,~#2,~#3\ttt{)}}
\newcommand{\hevalpc}[3]{\lightyellow{\predicate{eval-c}}\ttt{(}#1,~#2,~#3\ttt{)}}
\newcommand{\hevalp}[3]{\lightyellow{\predicate{eval-p}}\ttt{(}#1,~#2,~#3\ttt{)}}

\newcommand{\verify}[1]{\predicate{verify}\ttt{(}#1\ttt{)}}
\newcommand{\wellformed}[4]{\predicate{well-formed}\ttt{(}#1,~#2,~#3,~#4\ttt{)}}
\newcommand{\hwellformed}[4]{\lightyellow{\predicate{well-formed}}\ttt{(}#1,~#2,~#3,~#4\ttt{)}}

\newcommand{\rchecksall}{\mathfrak{R}}
\newcommand{\rchecks}{\mathcal{R}}
\newcommand{\bcs}{\ttt{bcs}}
\newcommand{\rcs}{\ttt{rcs}}
\newcommand{\origin}{\ttt{origin}}
\newcommand{\none}{\ttt{none}}
\newcommand{\location}{\ttt{location}}
\newcommand{\addcheck}[3]{\predicate{addcheck}(#1,~#2,~#3)}
\newcommand{\addbc}[3]{\predicate{addbc}(#1,~#2,~#3)}

\newcommand{\translate}[2]{\predicate{translate}(#1,~#2)}

\newcommand{\havoc}[2]{\predicate{havoc}\ttt{(}#1,~#2\ttt{)}}

\newcommand{\pcadd}[2]{\predicate{pc-add}\ttt{(}#1,~#2\ttt{)}}
\newcommand{\pcall}[1]{\predicate{pc-all}\ttt{(}#1\ttt{)}}
\newcommand{\pcpush}[3]{\predicate{pc-push}\ttt{(}#1,~#2,~#3\ttt{)}}

\newcommand{\foldl}[3]{\predicate{foldl}\ttt{(}#1,~#2,~#3\ttt{)}}

\newcommand{\heapremp}[4]{\predicate{heap-rem-pred}\ttt{(}#1,~#2,~#3,~#4\ttt{)}}
\newcommand{\heaprema}[4]{\predicate{heap-rem-acc}\ttt{(}#1,~#2,~#3,~#4\ttt{)}}

\newcommand{\consolidate}[2]{\predicate{consolidate}\ttt{(}#1,~#2\ttt{)}}

\newcommand{\checkv}[2]{\predicate{check}\ttt{(}#1,~#2\ttt{)}}
\newcommand{\checkqm}[3]{\predicate{check}\ttt{(}#1,~#2,~#3\ttt{)}}
\newcommand{\assertv}[3]{\predicate{assert}\ttt{(}#1,~#2,~#3\ttt{)}}

\newcommand{\branch}[5]{\predicate{branch}\ttt{(}#1,~#2,~#3,~#4,~#5\ttt{)}}
\newcommand{\hbranch}[5]{\lightyellow{\predicate{branch}}\ttt{(}#1,~#2,~#3,~#4,~#5\ttt{)}}

\newcommand{\iseimp}[1]{\predicate{equi-imp}\ttt{(}#1\ttt{)}}

\newcommand{\first}[1]{\predicate{first}\ttt{(}#1\ttt{)}}
\newcommand{\second}[1]{\predicate{second}\ttt{(}#1\ttt{)}}

\newcommand{\chunk}[3]{#1\ttt{(}#2;~#3\ttt{)}}
\newcommand{\success}{\predicate{success}\ttt{(}\ttt{)}}
\newcommand{\failure}{\predicate{failure}\ttt{(}\ttt{)}}
\newcommand{\fresh}{\predicate{fresh}}
\newcommand{\assign}{~\ttt{≔}~}
\newcommand{\oh}{h_{\qm}}
\newcommand{\isimp}{\ttt{isImprecise}}

%% file: contents/empirical.tex
The seminal work on gradual typing~\cite{siek2006gradual} selectively inserts run-time casts in support of optimistic static checking: for instance, whenever a function application is deemed well-typed only because of imprecision---such as passing an argument of the unknown type to a function that expects an integer---the type-directed cast insertion procedure inserts a run-time check. But if the application is definitely well-typed, no cast is inserted. This approach ensures that a fully-precise program does not incur any overhead related to run-time type checking. While it is tempting to assume that more precision necessarily results in better performance, the reality has been shown to be more subtle: both the nature of the inserted checks (such as higher-order function wrappers) as well as when/how often they are executed is of utmost importance~\cite{takikawa16,10.1145/3133880}, and anticipating the performance impact of precision is challenging~\cite{camporaAl:icfp2018}.

The performance of gradual verification has never been studied until now, due to the lack of a working gradual verifier. Here, we explore the relation between minimizing dynamic check insertion with statically available information and observed run-time performance in gradual verification with \gco.
Specifically, we explore the performance characteristics of \gco for thousands of partial specifications generated from four data structures, as inspired by \citet{takikawa16}'s work in gradual typing.
In particular, we observe how adding or removing individual atomic formulas and $\qm$ within a specification impacts the degree of static and dynamic verification and, as a result, the run-time overhead of the program. Additionally, we compare the run-time performance of \gco to a fully dynamic approach, as readily available in C0.
The aforementioned ideas are captured in the following research questions:
\begin{itemize}
    \item [\textbf{RQ1}:] As specifications are made more precise, can more verification conditions be eliminated statically?
    \item [\textbf{RQ2}:] Does gradual verification result in less run-time overhead than a fully dynamic approach?
    \item [\textbf{RQ3}:] Are there particular types of specification elements that have significant impact in run-time overhead, and can high overhead be avoided?
\end{itemize}

\subsection{Creating Performance Lattices}
\label{sec:empirical-lattices}
We define a \textit{complete} specification as being statically verifiable when all \ttt{?}s are removed, and then a \textit{partial} specification as a subset of formulas from a complete specification that are joined with $\qm$. 
Like \citet{takikawa16}, we model the gradual verification process as a series of steps from an unspecified program to a statically verifiable specification where, at each step, an \textit{element} is added to the current, partial specification. An element is an atomic conjunct (excluding boolean primitives) in any type of method contract, assertion, or loop invariant. 
We form a lattice of partial specifications by varying which elements of the complete specification are included. We also similarly vary the presence of $\qm$ in formulas that are complete---contain the same elements as their counterparts in the statically verifiable specification---and have related fold and unfold statements in the partial specification. Otherwise, $\qm$ is always added to incomplete formulas. This strategy creates lattices where the bottom entry is an empty specification containing only $\qm$s and the top entry is a statically verifiable specification. A \textit{path} through a lattice is the set of specifications created by appending $n$ elements or removing $\qm$s one at a time from the bottom to the top of the lattice. The large array of partial specifications created in each lattice closely approximates the positive specifications supported by the gradual guarantee \cite{wise2020gradual}, which are less precise variants of successfully verified programs. For reference, we give a more formal statement of the gradual guarantee: 

\begin{quote}
\emph{Let $p_1$ and $p_2$ be \gco programs where $p_1 \sqsubseteq p_2$ (\ie the formulas in $p_1$ are more precise than those in $p_2$). If $p_1$ statically verifies, then $p_2$ statically verifies. Additionally, $p_2$ must execute at least as far as $p_1$ executes at run time.}
\end{quote}

\noindent Now, to illustrate the aforementioned approach, consider the following loop invariant:
\begin{center}
\begin{tabular}{c}
\begin{lstlisting}[label={fig:invar_ex},numbers=none]
//@ loop_invariant sortedSeg(list, curr, curr->val) && curr->val <= val;
\end{lstlisting}
\end{tabular}
\end{center}
The invariant is made of two elements: the \ttt{sortedSeg} predicate instance and the boolean expression \ttt{curr->val <= val;}. 
The lattice generated for a program with this invariant has five unique specifications: four contain a combination of the two elements joined with \ttt{?}, and the fifth is the complete invariant above.

\subsection{Data Structures}
To apply this methodology, we implemented and fully specified four recursive heap data structures with \gco: binary search tree, sorted linked list, composite tree, and AVL tree. We chose these data structures because complete static specifications exist for them in prior work and they are interesting use cases for gradual verification. Linked list is implemented with a while loop rather than recursion. Binary search tree is a more complex data structure with a more complex property (BST property) than a linked list and uses recursion. Composite tree implements a structure where modifications do not have to start at the root, but can be applied directly to any node in the tree. Its invariant also applies to any node in the tree. Finally, AVL tree implements the most complex invariant (the balanced property) and data structure with many interdependent functions and predicates related to tree rotations. Each data structure has a test program that contains its implementation and a main function that adds elements to the structure based on a workload parameter $\omega$. We design the test programs to incur as little run-time overhead as possible outside of structure size and run-time checks. For each example and corresponding test program, Table $\ref{tbl:ex-desc}$ displays the distribution of elements in the complete specification, as well as the run-time complexity of the test program and the number of unique partial specifications generated by our benchmarking tool.
\begin{table}[t]
{\small
\centering
\input{contents/table_desc}}

\disableTttResize
\captionsetup{font=footnotesize}
\caption[font=footnotesize]{Description of benchmark examples. For each example, the table shows the complexity of the test program without verification, the number of sampled partial specifications, and the distribution of specification elements for the complete specification. Element counts are formatted as ``\textit{Accessibility Predicate}/\textit{Predicate Instance}/\textit{Boolean Expression}/\textit{Imprecision}"}
\label{tbl:ex-desc}
\vspace{-1.5em}
\end{table}

\paragraph{Binary Search Tree (BST)}
The implementation of the binary search tree is typical; each node contains a value and pointers to left and right nodes. We statically specify memory safety and preservation of the binary search tree property---that is, any node's value is greater than any value in its left subtree and less than any value in its right subtree.
The test program creates a root node with value $\omega$ and sequentially adds and removes a set of $\omega$ values in the range $[0, 2\omega]$. Note that values are removed in the same order they were added.

\paragraph{Linked List}
We implement a linked list with insertion similar to the one given in \fig~\ref{ex:ll-insert}. Insertion is statically specified for memory safety as well as preservation of list sortedness. Its test program creates a new list and inserts $\omega$ arbitrary elements.

\paragraph{Composite}
The composite data structure is a binary tree where each node tracks the size of its subtree---this is verified by its specification along with memory safety. Its test program starts with a root node and builds a tree of size $\omega$ by randomly descending from the root until a node without a left or right subtree is reached. A new node is added in the empty position, and then traversal backtracks to the root.

\paragraph{AVL Tree} The implementation of AVL tree with insertion is standard except that the height of the left and right subtrees is stored in each node (instead of the overall height of the tree). This allows us to easily state the AVL balanced property---for every node in the tree the height difference between its left and right children is at most 1---without using functions or ghost variables, which Gradual C0 does not currently support. In addition to specifying the AVL balanced property for insertion, we also specify memory safety. 
The AVL test program starts with a root node and builds a tree of size $\omega$ by inserting randomly valued nodes into the tree using balanced insertion.

\subsection{Experimental Setup}

With upwards of 100 elements in the specifications for each data structure, it is combinatorially infeasible to fully explore every partial specification. Therefore, unlike \citet{takikawa16}, we proceed by {\em sampling} a subset of partial specifications in a lattice, rather than executing them all. Specifically, we sample 16 unique paths through the lattice from randomized orderings of specification elements. 
We chose partial specifications along lattice paths to explore trends in migration from no specifications to complete specifications, which is how we imagine developers may use our tool. We also randomly sampled paths, rather than using a another heuristic for selecting paths, to be prescriptive to users of \gco. We wanted to find and recommend new specification patterns that users should apply or avoid depending on their performance.
Every step is executed with three workloads chosen arbitrarily to ensure observable differences in timing. Each timing measurement is the median of 10 iterations. Programs were executed on four physical Intel Core i5-4250U 1.3GHz Cores with 16 GB of RAM.

We introduce two baseline verifiers to compare \gco against. The \textit{dynamic verifier} transforms every specification into a run-time check and inserts accessibility predicate checks for field dereferences---thereby emulating a fully dynamic verifier. The \textit{framing verifier} only performs the accessibility predicate checks, and therefore represents the minimal dynamic checks that must be performed in a language that checks ownership.\footnote{These framing checks could fail, for example, if some function lower in the call stack owns data that is accessed by the currently-executing function.} We implement the baseline verifiers ourselves using \citet{wise2020gradual}'s dynamic semantics, which checks everything at run time, as a guide. \citet{wise2020gradual}'s work is the only work that we are aware of that handles run-time checking of both ownership and recursive predicates.

\subsection{Evaluation}
\begin{figure}[t]
    \centering
    \textbf{\small Mean Verification Conditions}\\
    \vspace{1em}
    \includegraphics[width=.24\textwidth]{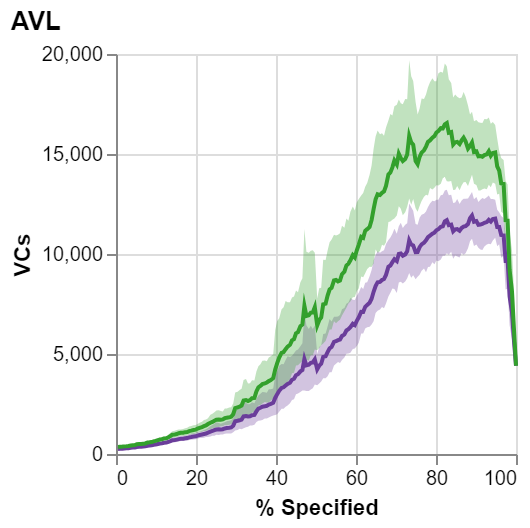}
    \includegraphics[width=.24\textwidth]{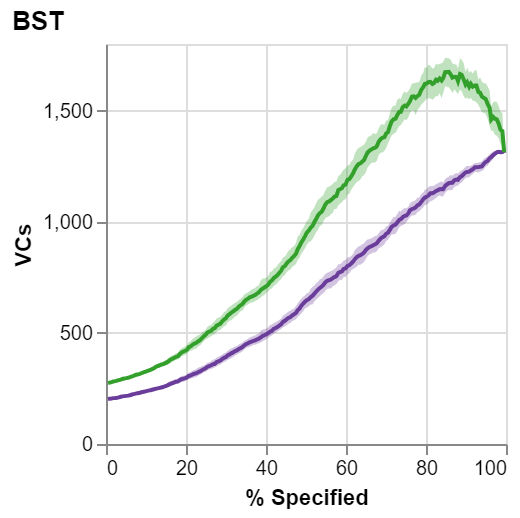}
    \includegraphics[width=.24\textwidth]{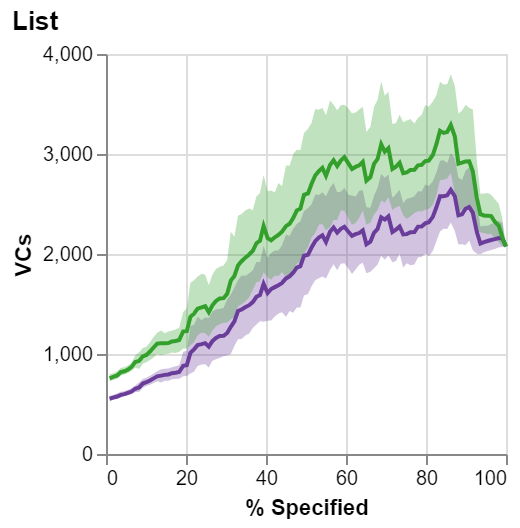}
    \includegraphics[width=.24\textwidth]{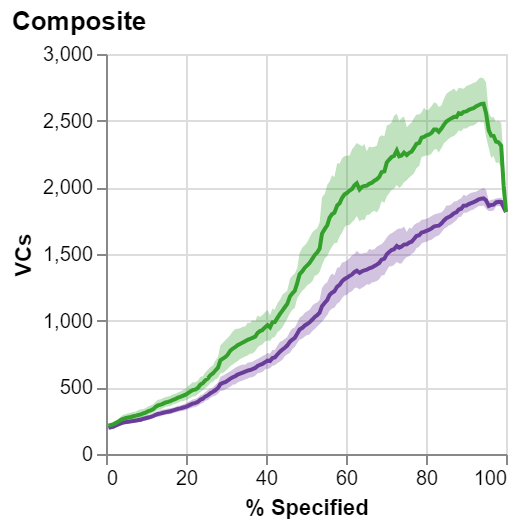}
    \scriptsize
        \cbox{legend_green} Total Verification Conditions \quad
        \cbox{legend_purple} Statically Eliminated Verification Conditions
    
    \disableTttResize
    \caption{For each example, the average quantity of verification conditions and the subset that were eliminated statically at each level of specification completeness across all paths sampled. Shading indicates the standard deviation.}
    \label{fig:vcs-plots}
\end{figure}

\fig~\ref{fig:vcs-plots} shows how the total number of verification conditions (proof obligations) changes as more of each benchmark is specified (green curve). The figure also similarly shows the number of verification conditions that are statically verified as each benchmark is specified (purple curve). From the green curve, we see that even when there are no specifications, there are verification conditions, \eg before a field is accessed, the object reference must be non-null and the field must be owned. Some of these verification conditions can be verified statically as illustrated by the purple curve. As more of a benchmark is specified, there are more verification conditions (green curve); but also, more of these verification conditions are discharged statically and do not have to be checked dynamically (purple curve). Towards the right end of the plots, the two curves converge until they meet when all the verification conditions are discharged statically. As a result, the answer to \textbf{RQ1} is \emph{yes}. Note, the number of verification conditions does decrease when enough of the benchmark is specified. This is due to being able to prune symbolic execution paths with new static information.

\begin{figure}[t]
    \centering
    \textbf{\small Mean Execution Time Over All Paths}\\
    \vspace{1em}
    \includegraphics[trim=0em 0em 0em 0em, width=0.25\textwidth]{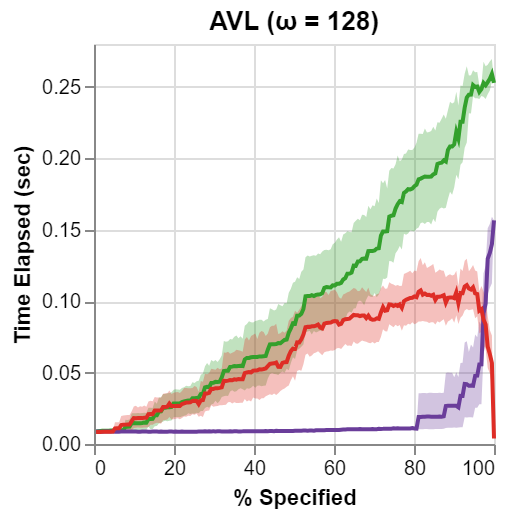}
    \includegraphics[trim=0em 0em 0em 0em, width=0.24\textwidth]{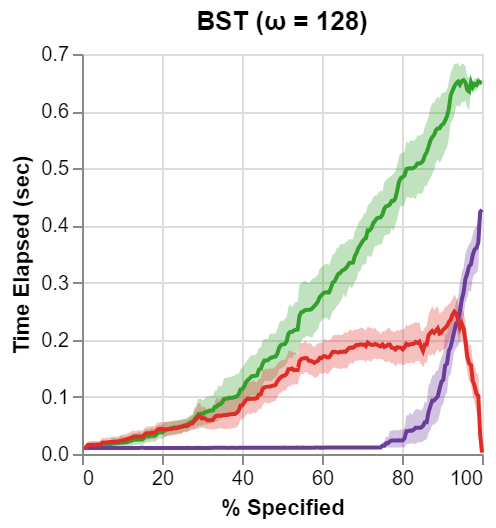}
    \includegraphics[trim=0em 0em 0em 0em, width=0.245\textwidth]{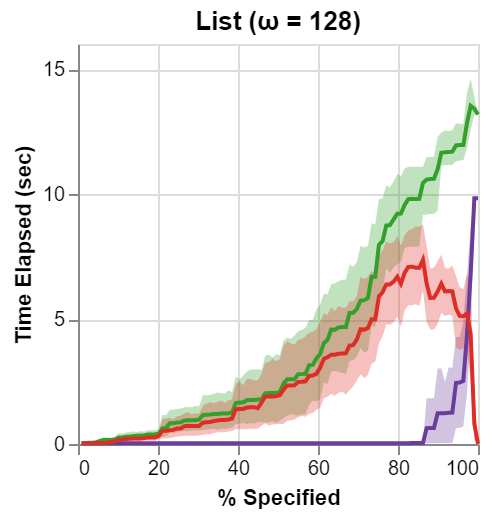}
    \includegraphics[trim=0em 0em 0em 0em, width=0.24\textwidth]{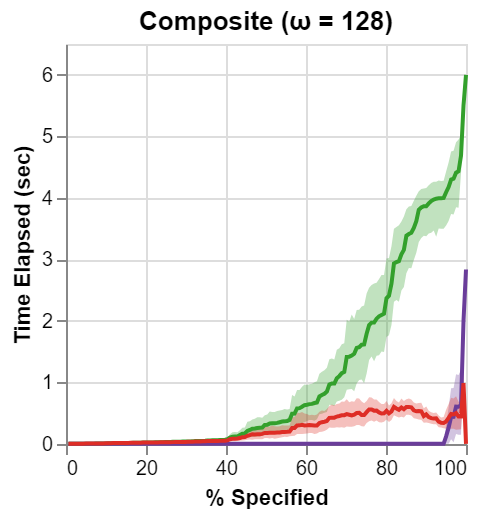}
    \\
    \scriptsize
        \cbox{legend_green} Dynamic Verification \quad \cbox{legend_purple} Only Framing Checks \quad \cbox{legend_red} Gradual Verification

    \disableTttResize
    \caption{The mean time elapsed at each step over the 16 paths sampled. Shading indicates the confidence interval of the mean for each verification type.}
    \label{fig:bmw}

\end{figure}

The plots in \fig~\ref{fig:bmw} display the run-time performance (in red) of dynamically checking the verification conditions from \fig~\ref{fig:vcs-plots}. The plots also show how the run-time performance of the dynamic verifier (in green) and framing verifier (in purple) change as more of each benchmark is specified. The green lines show that as more properties are specified, the cost of run-time verification increases. With \gco, some of these properties can be checked statically; therefore, the run-time cost of gradual verification, shown in red, starts equivalent but eventually ends up significantly lower than the cost of pure run-time verification. 

Notably, the purple lines are significantly lower than the red and greens ones until they exhibit a dramatic increase starting at around 80\% specified all the way to 100\%. Eventually (after about 95\% specified), the purple lines end above the red ones (but below the green ones) where running time is orders of magnitude higher than at the start of the incline. The framing verifier (in purple) only checks that heap accesses are safe---\ie they are owned and their receivers are non-null. 
So unsurprisingly, the dynamic and gradual verifiers, which check more properties like heap separation, nearly always have significantly higher run-time verification overhead than the framing verifier. Eventually, \gco outperforms the framing verifier when enough properties, including framing, are checked statically.

The dramatic increase in the framing verifier's run-time performance is caused by the owned fields passing strategy employed at method boundaries (described in \S\ref{sec:gvc0-accpreds}) to verify memory safety at run time. To respect precondition abstractions, only owned fields specified by a callee's precondition are passed by the caller to the callee when the precondition is precise. Similarly, when a callee's postcondition is precise, then only the owned fields specified by the postcondition are passed back to the caller. Computing owned fields from precise contracts is costly; and even more-so for contracts containing recursive predicates like in our benchmarks. Further, our benchmarks call such methods frequently during execution. As a result, execution time increases significantly at each path step where one of the aforementioned methods gets a precise pre or postcondition from $\qm$ removal. This, of course, happens more frequently as more of a benchmark is specified. At 100\% specified every method contract is precise, and so the owned fields passing strategy is used at every method call and return leading to the highest run-time costs for the framing verifier. In contrast, \gco checks fully-specified methods completely statically and does not use the owned field passing strategy for calls to these methods. As a result, looking at the red lines, \gco is not heavily affected by this phenomena---we see slight increases starting at 90\% specified but they are significantly less costly. Additionally, once a critical mass of specifications have been written, \gco's run-time verification cost decreases until reaching zero---which is the same as running the raw C0 version of the benchmark.
If the spikes around 90\% specified are too costly, production gradual verifiers can reduce them by employing more optimal permission passing strategies.
In general, according to the red lines, \gco's performance increases gradually as more proof obligations are specified but are not yet statically verified; and thus, must be checked at run time. When a critical mass of specifications are written that allows more and more of these proof obligations to be proven statically, run-time performance starts to decrease until reaching the spikes around 90\% specified caused by owned fields passing. After the spikes, performance decreases to the benchmark's raw baseline. This trend is consistent with speculations made in \citet{wise2020gradual}'s work, and also confirms that increasing precision in gradual verification does not always correspond with decreased run-time overhead from dynamic verification.

\begin{table}[t]
{\small
\centering
\input{figures_11-16-23/table}}

\disableTttResize
\captionsetup{font=footnotesize}
\caption{Summary statistics for the performance of each example over 16 paths at selected workloads ($\omega$), comparing gradual verification (GV) against dynamic verification (DV). The grouped column ``\% in $\Delta t$, GV. vs. DV." displays summary statistics for the percent decrease in time elapsed for each step when using GV versus DV. The column ``\% Steps GV<DV for Paths DV<GV" shows the distribution of steps that performed best under GV that were part of paths containing steps that performed better under DV. The final column shows the percentage of paths in which every step performed better under GV.}
\label{tbl:perf_summary}
\vspace{-1.5em}
\end{table}

Table \ref{tbl:perf_summary} displays summary statistics for \gco's performance on every sampled partial specification compared to the dynamic verification baseline. Depending on the workload and example, on average \gco reduces run-time overhead by 11.5-34.4\% (Table \ref{tbl:perf_summary}, Column 3) compared to the dynamic verifier. Note, the speed-ups are consistent as $\omega$ increases: -14.4\%, -24.7\%, -18.0\%, and -34.4\% at the lowest $\omega$ values compared to -11.5\%, -21.7\%, -20.7\%, and -30.1\% at the largest. While \gco generally improves performance, there are some outliers in the data (Table \ref{tbl:perf_summary}, Column 5) where \gco is slower than dynamic verification by 138.9-603.3\%. Fortunately, for lattice paths that produce these poor-performing specifications, gradual verification still outperforms dynamic verification (on average) for 74.2-85.8\% (Table \ref{tbl:perf_summary}, Column 7) of all steps. Further, these outliers appear under 20\% specified where the bookkeeping we insert to track conditionals, which is unoptimized and could be improved, and measurement error are the cause of such outcomes. 

\fig~\ref{fig:bmw} displays the average run-time cost across all paths under each of our benchmarks and verifiers. In all the plots, for some early parts of the path the cost of \gco is comparable to or exceeds the cost of the dynamic verifier, but after 50\% completion, static optimization kicks in and \gco begins to significantly outperform it. Further, Table \ref{tbl:perf_summary} shows that on average \gco reduces run-time overhead by 11.5-34.4\% compared to the dynamic verifier. Therefore, the answer to \textbf{RQ2} is \textit{yes}. 

\begin{figure}[t]
\centering
\textbf{\small 99th Percentile Changes in Run-time Overhead}\\
\vspace{1em}
\includegraphics[width=.325\textwidth]{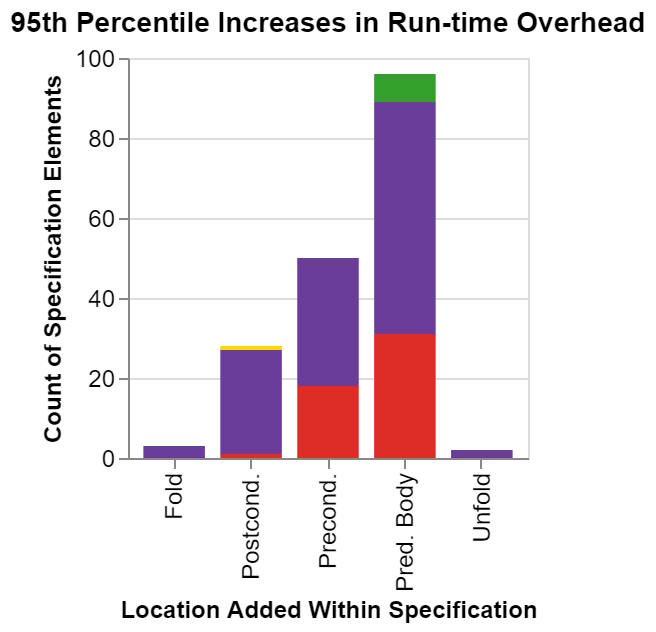} \quad
\includegraphics[width=.337\textwidth]{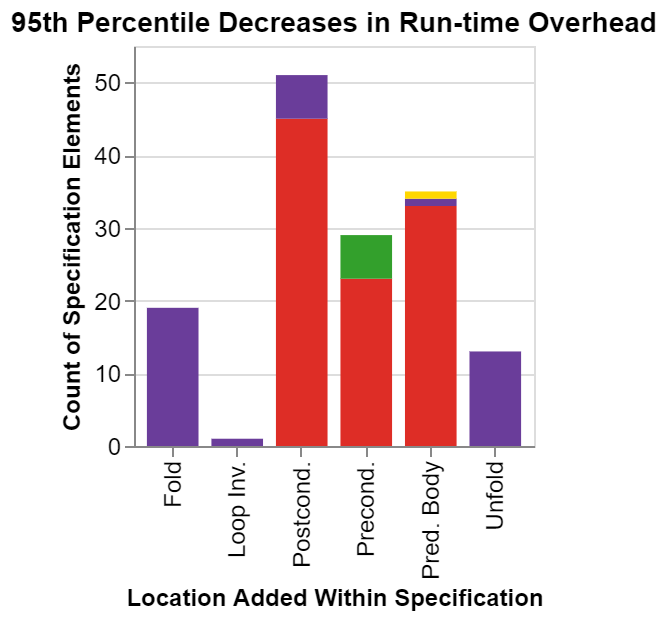}\\
\vspace{1em}
\scriptsize
\cbox{legend_green} Accessibility Predicate \quad
\cbox{legend_purple} Predicate Instance     \quad
\cbox{legend_yellow} Boolean Expression \quad
\cbox{legend_red} $\qm$ Removed    

\disableTttResize
\caption{The quantity of specification elements, grouped by type and location, that caused the highest ($P_{99\%}$) increases and decreases in time elapsed out of every path sampled}
\label{fig:99th-percentile}
\end{figure}

\fig~\ref{fig:99th-percentile} captures the impact that different types of specification elements (accessibility predicates, predicates, and boolean expressions) have on \gco's run-time performance when specified in different locations. It also captures the impact removing $\qm$ from a formula has on performance. Elements that when added or $\qm$ that when removed from one step in a lattice path to another increase run-time overhead significantly (in the top 1\%) are counted in the left sub-figure, and ones that decrease run-time overhead significantly (top 1\%) are counted in the right sub-figure. The count for accessibility predicates is colored in green, predicates in purple, boolean expressions in yellow, and $\qm$ removal in red.

Adding recursive predicates to preconditions, postconditions, and predicate bodies is the most frequent cause (67.6\%) of dramatic increases in run-time verification overhead during the specification process. When these predicates are added to preconditions and postconditions they create additional proof obligations for them in callers and callees (respectively) that are frequently checked at run time. Similarly, when they are added to predicate bodies any proof obligations for the enclosing predicate that are checked at run time become far more expensive. Fortunately, folding or unfolding a predicate can decrease run-time cost when doing so discharges such proof obligations statically (as seen in the right sub-figure). Therefore, users of \gco may consider specifying proofs of recursive predicates in frequently-executed code to significantly reduce checking costs. 

Removing $\qm$ from preconditions, postconditions, and predicate bodies when the costly owned fields passing strategy is still required in corresponding methods is the second most frequent cause (27.9\%) of increases in \gco's run-time overhead. This corresponds with the spikes at 90\% specified in \fig~\ref{fig:bmw} for \gco: removal of $\qm$ in the aforementioned locations leads to precise pre- and postconditions that trigger the use of this costly strategy. 
Eventually, a critical mass of specifications are written so that when $\qm$s are removed further this costly strategy is no longer necessary (\ie when callee methods are full statically verified) and so run-time performance improves dramatically---the downward trends seen prior to full static specification in \fig~\ref{fig:bmw}. This is reflected in the right sub-figure in \fig\ref{fig:99th-percentile}, where removing $\qm$ from preconditions, postconditions, and predicate bodies is the most frequent cause (68.2\%) of significant decreases in run-time overhead.
This suggests a strategy for avoiding high checking costs: specify frequently-executed code in critical-mass chunks that are fully statically verifiable, leaving boundaries between statically and dynamically verified code in places that are executed less frequently.

Overall, the answer to \textbf{RQ3} is \emph{yes}; we have identified some key contributors to run-time overhead, whose optimization is a promising direction for future work, and we have also identified strategies for minimizing run time overhead in practice.

Finally, all of the partial specification evaluated in our study were successfully verified by \gco. Since they originated from complete and correct specifications on code, we can conclude \gco adheres to the gradual guarantee for these partial specifications and likely adheres to the gradual guarantee for common use cases of \gco.

\paragraph{Threats to Validity}  
Our test programs were executed on multiple devices, each with the same CPU and memory configuration. However, we did not otherwise control for differences in performance between devices. While the test programs we used are of sufficient complexity to demonstrate interesting empirical trends, they are not representative of all software. Further, the baseline we used for dynamic verification is entirely unoptimized as we naively insert a check for each written element of a specification. Finally, due to computational constraints, only a small subset of over $2^{100}$ possible imprecise specifications were sampled, and we did not use a formal criteria to choose our workload values. 
As such, while our results reveal interesting trends, including significant performance improvements by \gco over dynamic verification, more work is needed to validate the robustness of those trends.

\subsection{Qualitative Experience with AVL Tree}
Notably, it was our experience that the incrementality of gradual verification was very helpful for developing a complete specification of the AVL tree example. In particular, a run-time verification error from a partial specification helped us realize the contract for the \ttt{rotateRight} helper function was not general enough. We fully specified \ttt{rotateRight} and proved it correct. However, \ttt{insert}'s pre- and postconditions were left as $\qm$, and so static verification could not show us that the contract proved for \ttt{rotateRight} was insufficiently general. Nevertheless, we ran the program; gradual verification inserted run-time checks, and the precondition for \ttt{rotateRight} failed. This early notification allowed us to identify the problem with the specification and fix it immediately. Otherwise, we would have had to get deep into the static verification of \ttt{insert}---a complicated function, 50 lines long, with lots of tricky logic and invariants---before discovering the error, and a lot of verification work built on the faulty specification would have had to be redone.
Interestingly, it is conventional wisdom that one of the benefits of static checking is that you get feedback early, when it is easier to correct mistakes. Here, we encountered a scenario where gradual verification had a similar benefit over static verification! We found an error (in a specification) earlier than we would have otherwise, presumably saving time.

%% file: contents/table_desc.tex


    \scriptsize
    
    \begin{tabular}{l|c|c|cccccc|}
    \cline{2-9}
    \multirow{2}{*}{\textbf{Example}}       & \multirow{2}{*}{\textbf{Unverified Complexity}}                              & \multirow{2}{*}{\textbf{\# Specs}} & \multicolumn{6}{c|}{\textbf{Contents of Complete Spec}}                                                                                                                       \\ \cline{4-9} 
                                & \multicolumn{1}{c|}{} &   \multicolumn{1}{c|}{} &   \multicolumn{1}{c|}{\textit{Fold}} & \multicolumn{1}{c|}{\textit{Unfold}} & \multicolumn{1}{c|}{\textit{Pre.}}    & \multicolumn{1}{c|}{\textit{Post.}}  & \multicolumn{1}{c|}{\textit{Pred. Body}} & \textit{Loop Inv.} \\ \hline\hline
    \multicolumn{1}{c|}{Binary Search Tree} & \multicolumn{1}{c|}{$O(n~log(n))$} &
    3473   & \multicolumn{1}{c|}{43}   & \multicolumn{1}{c|}{23}     & \multicolumn{1}{c|}{0/20/21/24} & \multicolumn{1}{c|}{0/22/6/24} & \multicolumn{1}{c|}{6/6/7/4}      & 0/2/4/2     \\ \hline
    Linked List                             & \multicolumn{1}{c|}{$O(n)$} &
    1745                       & \multicolumn{1}{c|}{17}   & \multicolumn{1}{c|}{10}     & \multicolumn{1}{c|}{8/6/15/5}   & \multicolumn{1}{c|}{4/5/6/5}  & \multicolumn{1}{c|}{4/3/4/3}      & 4/3/5/2    \\ \hline
    Composite                               & \multicolumn{1}{c|}{$O(n~log(n))$} &
     2577                      & \multicolumn{1}{c|}{28}   & \multicolumn{1}{c|}{15}     & \multicolumn{1}{c|}{0/10/2/12} & \multicolumn{1}{c|}{0/11/1/12} & \multicolumn{1}{c|}{32/9/17/3}    &  0/3/2/3     \\ \hline
    AVL                      & \multicolumn{1}{c|}{$O(n~log(n))$} &
                           3057 & \multicolumn{1}{c|}{25}   & \multicolumn{1}{c|}{14}     & \multicolumn{1}{c|}{3/4/5/9}  & \multicolumn{1}{c|}{3/6/9/9} & \multicolumn{1}{c|}{25/8/21/3}    &  1/1/2/1     \\ \hline
    \end{tabular}

%% file: figures_11-16-23/table.tex
\scriptsize
\begin{tabular}{c|c|cccc|cccc|c|}
\cline{2-11}
\multicolumn{1}{l|}{\multirow{2}{*}{Example}} & \multirow{2}{*}{$\omega$} & \multicolumn{4}{c|}{\% $\Delta t$, GV vs. DV}                                                   & \multicolumn{4}{c|}{\% Steps GV < DV  for Paths DV < GV}                  & \multirow{2}{*}{\% Paths GV < DV} \\ \cline{3-10}
\multicolumn{1}{l|}{}                         &                           & \multicolumn{1}{c|}{Mean}  & \multicolumn{1}{c|}{St. Dev.} & \multicolumn{1}{c|}{Max}  & Min    & \multicolumn{1}{c|}{Mean} & \multicolumn{1}{c|}{St. Dev.} & \multicolumn{1}{c|}{Max}   & Min. &                                                          \\ \hline\hline
\multirow{3}{*}{AVL}                          & 32                         & \multicolumn{1}{c|}{-14.4}  & \multicolumn{1}{c|}{29.4}     & \multicolumn{1}{c|}{170.0} & -88.8 & \multicolumn{1}{c|}{80.6} & \multicolumn{1}{c|}{14.8}   & \multicolumn{1}{c|}{95.3}  & 50.8 & 0.0                                                        \\ \cline{2-11} 
                                              & 64                         & \multicolumn{1}{c|}{-16.1} & \multicolumn{1}{c|}{50.5}     & \multicolumn{1}{c|}{256.3} & -96.0 & \multicolumn{1}{c|}{84.3} & \multicolumn{1}{c|}{15.5}      & \multicolumn{1}{c|}{98.4} & 55.5 & 0.0                                                      \\ \cline{2-11} 
                                              & 128                        & \multicolumn{1}{c|}{-11.5} & \multicolumn{1}{c|}{74.0}     & \multicolumn{1}{c|}{384.7} & -98.6  & \multicolumn{1}{c|}{82.2} & \multicolumn{1}{c|}{16.5}      & \multicolumn{1}{c|}{99.5} & 55.0 & 0.0                                                     \\ \hline
\multirow{3}{*}{BST}                          & 32                         & \multicolumn{1}{c|}{-24.7} & \multicolumn{1}{c|}{35.0}     & \multicolumn{1}{c|}{144.5} & -93.8  & \multicolumn{1}{c|}{74.2} & \multicolumn{1}{c|}{9.0}      & \multicolumn{1}{c|}{89.4}  & 60.8 & 0.0                                                      \\ \cline{2-11} 
                                              & 64                         & \multicolumn{1}{c|}{-23.9} & \multicolumn{1}{c|}{46.8}     & \multicolumn{1}{c|}{301.3} & -98.5  & \multicolumn{1}{c|}{74.7} & \multicolumn{1}{c|}{9.3}      & \multicolumn{1}{c|}{92.2} & 59.4 & 0.0                                                     \\ \cline{2-11} 
                                              & 128                        & \multicolumn{1}{c|}{-21.7} & \multicolumn{1}{c|}{55.6}       & \multicolumn{1}{c|}{436.1} & -99.6  & \multicolumn{1}{c|}{74.7} & \multicolumn{1}{c|}{11.9}      & \multicolumn{1}{c|}{96.3} & 51.6 & 0.0                                                     \\ \hline
\multirow{3}{*}{Linked List}                  & 32                         & \multicolumn{1}{c|}{-18.0} & \multicolumn{1}{c|}{27.7}     & \multicolumn{1}{c|}{138.9} & -95.7  & \multicolumn{1}{c|}{78.8} & \multicolumn{1}{c|}{14.8}     & \multicolumn{1}{c|}{96.3}  & 34.9 & 0.0                                                        \\ \cline{2-11} 
                                              & 64                        & \multicolumn{1}{c|}{-21.0} & \multicolumn{1}{c|}{48.6}     & \multicolumn{1}{c|}{389.0} & -99.8  & \multicolumn{1}{c|}{85.2} & \multicolumn{1}{c|}{15.1}     & \multicolumn{1}{c|}{100.0}  & 37.6 & 6.3                                                       \\ \cline{2-11} 
                                              & 128                        & \multicolumn{1}{c|}{-20.7} & \multicolumn{1}{c|}{59.7}     & \multicolumn{1}{c|}{603.3} & -100.0  & \multicolumn{1}{c|}{85.8} & \multicolumn{1}{c|}{15.0}     & \multicolumn{1}{c|}{99.1}  & 42.2 & 0.0                                                        \\ \hline
\multirow{3}{*}{Composite}                    & 32                         & \multicolumn{1}{c|}{-34.4} & \multicolumn{1}{c|}{40.1}     & \multicolumn{1}{c|}{141.7} & -99.1  & \multicolumn{1}{c|}{80.3} & \multicolumn{1}{c|}{10.7}      & \multicolumn{1}{c|}{94.4} & 60.9 & 0.0                                                     \\ \cline{2-11} 
                                              & 64                         & \multicolumn{1}{c|}{-33.1} & \multicolumn{1}{c|}{50.1}     & \multicolumn{1}{c|}{258.6} & -99.8  & \multicolumn{1}{c|}{80.0} & \multicolumn{1}{c|}{12.2}      & \multicolumn{1}{c|}{96.3} & 50.9 & 0.0                                                     \\ \cline{2-11} 
                                              & 128                        & \multicolumn{1}{c|}{-30.1} & \multicolumn{1}{c|}{63.9}     & \multicolumn{1}{c|}{419.0} & -100.0 & \multicolumn{1}{c|}{80.4} & \multicolumn{1}{c|}{13.0}      & \multicolumn{1}{c|}{96.3} & 48.4 & 0.0                                                     \\ \hline
\end{tabular}